\documentclass[usenatbib,useAMS,letterpaper]{mnras}
\usepackage{float}
\usepackage{graphicx}
\usepackage{epstopdf}
\usepackage{ textcomp }
\usepackage{multirow}
\usepackage{ctable}
\usepackage{url}
\usepackage{times}
\usepackage{amsmath}
\usepackage{amssymb}
\usepackage{xspace}
\usepackage{mathrsfs} 
\usepackage{tabularx}
\usepackage{fixltx2e}
\usepackage{color}
\usepackage{placeins}
\usepackage{comment}

\newcommand{\acknowledgments}{\begin{small}\section*{Acknowledgements}\end{small}}

\defcitealias{2019MNRAS.487.4393S}{Paper {\small I}}
\defcitealias{2020MNRAS.491.1190S}{Paper {\small II}}
\newcommand\sref[1]{\hyperref[#1]{\S~\ref*{#1}}}
\newcommand\fref[1]{\hyperref[#1]{Fig.~\ref*{#1}}}
\newcommand\Eqref[1]{Eq.~(\hyperref[#1]{\ref*{#1}})}
\newcommand\eeqref[1]{Eq.~\hyperref[#1]{\ref*{#1}}}

\newcommand\tref[1]{\hyperref[#1]{Table~\ref*{#1}}}
\newcommand\aref[1]{\hyperref[#1]{Appendix~\ref*{#1}}}

\newcommand{\pfh}[1]{\textcolor{black}{#1}}
\newcommand{\hlt}[1]{\textcolor{blue}{#1}}

\newcommand{\mr}[1]{\multirow{2}{*}{#1}}

\newcommand{\mrrr}[1]{\multirow{4}{*}{#1}}

\newcommand{\oneline}[1]{%
  \newdimen{\namewidth}%
  \setlength{\namewidth}{\widthof{#1}}%
  \ifthenelse{\lengthtest{\namewidth < \textwidth}}%
  {#1}
  {\resizebox{\textwidth}{!}{#1}}
}
 
\title[What Types of Jet Quench?]{Which AGN Jets Quench Star Formation in Massive Galaxies?}
\author[]{
\parbox[t]{\textwidth}{
Kung-Yi Su$^{1,2,3}$\thanks{E-mail: k.su@columbia.edu}, Philip F. Hopkins$^3$, Greg L. Bryan$^1$, Rachel S. Somerville$^2$, Christopher C. Hayward$^2$, Daniel Angl\'{e}s-Alc\'{a}zar$^{4,2}$, Claude-Andr\'{e} Faucher-Gigu\`{e}re$^5$, Sarah Wellons$^5$, Jonathan Stern$^5$,  Bryan A. Terrazas$^6$, T. K. Chan$^{7,8}$, Matthew E. Orr$^{9,3}$,  Cameron Hummels$^{3}$, Robert Feldmann$^{10}$,  Du\v{s}an Kere\v{s}$^8$
}
\vspace*{6pt} \\
$^1$Department of Astronomy, Columbia University, 550 West 120th Street, New York, NY 10027, USA\\
$^2$Center for Computational Astrophysics, Flatiron Institute, 162 Fifth Avenue, New York, NY 10010, USA\\
$^3$TAPIR 350-17, California Institute of Technology, 1200 E. California Boulevard, Pasadena, CA 91125, USA\\
$^4$Department of Physics, University of Connecticut, 196 Auditorium Road, U-3046, Storrs, CT 06269-3046, USA\\
$^5$Department of Physics \& Astronomy and CIERA, Northwestern University, 1800 Sherman Ave, Evanston, IL 60201, USA\\
$^6$Harvard-Smithsonian Center for Astrophysics, Cambridge, MA 02138, USA\\
$^7$Institute for Computational Cosmology, Durham University, South Road, Durham DH1 3LE, UK\\
$^8$Department of Physics and Center for Astrophysics and Space Science, University of California at San Diego, 9500 Gilman Drive, La Jolla, CA 92093, USA\\
$^9$Department of Physics and Astronomy, Rutgers, The State University of New Jersey, 136 Frelinghuysen Rd, Piscataway, NJ 08854, USA\\
$^{10}$Institute for Computational Science, University of Zurich, Winterthurerstrasse 190, CH-8057 Zurich, Switzerland
}
\usepackage[all]{hypcap}

\begin{document}
\long\def\/*#1*/{}
\date{Submitted to MNRAS}

\pagerange{\pageref{firstpage}--\pageref{lastpage}} \pubyear{2021}

\maketitle

\label{firstpage}

\begin{abstract}
In the absence of additional heating, radiative cooling of gas in the halos of massive galaxies (Milky Way and above) produces cold gas or stars in excess of that observed. Previous work has suggested that a contribution to this heating from AGN jets is likely required, however the form of jet energy required to quench remains unclear. This is particularly challenging for galaxy simulations, in which the resolution is orders of magnitude coarser than necessary to form and evolve the jet. On such scales, the uncertain parameter space includes: jet energy form (kinetic, thermal, and cosmic ray energy), energy flux, momentum flux, mass flux, magnetic field strength and geometry, jet precession angle and period, jet opening-angle, and duty cycle. We investigate all of these parameters in a $10^{14}\,{\rm M}_{\odot}$ halo using high-resolution non-cosmological MHD simulations with the FIRE-2 (Feedback In Realistic Environments) stellar feedback model, conduction, and viscosity. We explore which scenarios match observational constraints and show that cosmic ray-dominated jets can most efficiently quench the central galaxy through a combination of cosmic ray pressure support and a modification of the thermal instability. Jets with most of their energy in mildly relativistic ($\sim$ MeV or $\sim10^{10}$ K) thermal plasma can also work, but require a factor $\sim 10$ larger energy input. 
For a fixed energy flux, jets with lower mass loading (higher specific energy, hence longer cooling times) quench more effectively.   For this halo size,  kinetic jets are less efficient in quenching unless they have wide opening or precession angles; however, if the jet becomes too wide, it produces a large, low-density core, in tension with observations.  Magnetic fields, while they may be critical for jet acceleration near the black hole horizon, also play a relatively minor role except when the magnetic flux reaches $\gtrsim 10^{44}$ erg s$^{-1}$ in a kinetic jet model, which causes the jet cocoon to significantly widen, and the quenching to become explosive. We conclude that the criteria for a jet model to be successful are an optimal energy flux and a sufficiently wide jet cocoon with long enough cooling time at the cooling radius.

\end{abstract}


\section{Introduction}
\label{S:intro}
A major outstanding problem in galaxy formation for decades has been how to ``quench'' {\it massive} galaxies (stellar masses $\gtrsim 10^{11}\,{\rm M}_{\odot}$ or above $\sim L_{\ast}$ in the galaxy luminosity function) and keep them ``red and dead'' over a large fraction of cosmic time \citep[see e.g.,][]{2003ApJS..149..289B,2003MNRAS.341...54K,2003MNRAS.343..871M,2004ApJ...600..681B,2005MNRAS.363....2K,2005ApJ...629..143B,2006MNRAS.368....2D,2009MNRAS.396.2332K,2010A&A...523A..13P,2012MNRAS.424..232W,2015MNRAS.446.1939F,2015Natur.519..203V}. The difficulty lies in the classic ``cooling flow'' problem --- X-ray observations have found significant radiative cooling in the hot gas of elliptical galaxies and clusters, indicating cooling times shorter than a Hubble time \citep{1994ApJ...436L..63F,2006PhR...427....1P,2019MNRAS.488.2549S}. However, compared to the inferred cooling flow (reaching up to $\sim 1000\,{\rm M}_\odot {\rm yr}^{-1}$ in clusters), neither sufficient cold gas  from H{\scriptsize I} and CO observations  \citep{2011ApJ...731...33M,2013ApJ...767..153W} nor sufficient star formation  \citep{2001A&A...365L..87T,2008ApJ...681.1035O,2008ApJ...687..899R} has been observed in galaxies. Simulations and semi-analytic models which do not suppress the cooling flows, and simply allow gas to cool into the galactic core, typically predict over an order of magnitude higher star formation rates (SFRs) than observed \citep[for recent examples see, e.g., the weak/no feedback runs in][]{2007MNRAS.380..877S,somerville:2008,2009MNRAS.398...53B,2015MNRAS.449.4105C,2015ApJ...811...73L,2017MNRAS.472L.109A}.

Some heat source or pressure support must be present to compensate for the observed cooling. Moreover, the heating must still preserve the cool core structure (e.g., density and entropy profiles) observed in the majority of galaxies \citep{1998MNRAS.298..416P,2009A&A...501..835M}.  One way to achieve this is to suppress the cooling flow and maintain a very-low-SFR, stable cool-core (CC) cluster. Another possibility is that clusters undergo cool-core---non-cool-core (NCC) cycles: a stronger episode of feedback overturns the cooling flows, resulting in a  non-cool-core cluster, which gradually recovers to a cool-core cluster and starts another episode of feedback. 

The various non-AGN solutions to the cooling flow problem  proposed in the literature generally belong to the former case,  including: stellar feedback from shock-heated AGB winds \citep{2015ApJ...803...77C}, Type Ia supernovae (SNe) \citep[e.g.][and references therein]{2012MNRAS.420.3174S}, SNe-injected cosmic rays (CRs) \citep{2017ApJ...834..208R,2017MNRAS.465.4500P,Buts18,Farb18,2018MNRAS.475..570J}, magnetic fields \citep{1990ApJ...348...73S,1996ARA&A..34..155B,2012MNRAS.422.2152B} and thermal conduction \citep{1981ApJ...247..464B,1983ApJ...267..547T,2002MNRAS.335L...7V,2002MNRAS.335L..71F,2003ApJ...582..162Z} in the circum-galactic medium (CGM) or intra-cluster medium (ICM), or ``morphological quenching'' via altering the galaxy morphology and gravitational stability properties \citep{2009ApJ...707..250M,2009ApJ...703..785D}. Although these processes can slightly suppress star formation, or help suppress the cooling flows, most previous studies, including our own exhaustive survey studying each of these in simulations similar to those presented here \citep[][hereafter \citetalias{2019MNRAS.487.4393S}]{2019MNRAS.487.4393S}, have shown that they do not fundamentally alter the classic cooling flow picture. In the end, the star formation is still regulated by cooling flows, and the star formation rate is orders of magnitude too high.

Consequently, AGN feedback seems to be the most promising candidate to solve the cooling flow problem, and there has been a tremendous amount of theoretical work on the topic (for recent studies, see the reference in later paragraphs for the AGN jet and e.g., \citealt[][]{2017ApJ...837..149G,2017MNRAS.468..751E,2018MNRAS.479.4056W,2018ApJ...866...70L,2018ApJ...856..115P,2018ApJ...864....6Y} for other type of AGN feedback; also see e.g., \citealt{1998A&A...331L...1S,1999MNRAS.308L..39F,2001ApJ...551..131C,2005ApJ...630..705H,2006ApJS..163....1H,2006MNRAS.365...11C,2009ApJ...699...89C,2012ApJ...754..125C} for earlier works). 
Observational studies also infer that the available energy budget from AGN can match the cooling rate \citep{2004ApJ...607..800B}. There are also observations of un-ambiguous cases of AGN expelling gas from galaxies, injecting thermal energy via shocks or sound waves, or via photo-ionization and Compton heating, or via ``stirring'' the CGM and ICM, and creating ``bubbles'' of hot plasma with non-negligible relativistic components which are ubiquitous around massive galaxies \citep[see, e.g.,][for a detailed review]{2012ARA&A..50..455F,2018ARA&A..56..625H}.

However, despite its plausibility and the extensive work above, the detailed physics of AGN feedback remain uncertain, as do the relevant ``input parameters.'' Several studies also suggested certain categories of AGN feedback models struggle to stably quench the star formation, self-regulate themselves, or meet some of the observational constraints \citep[e.g.,][]{2004ApJ...607..800B,2006ApJ...645...83V,2020arXiv200400021G,2020MNRAS.491.1190S}. Therefore, a broad systematic exploration of AGN feedback models can be useful to understand which, if any, are more plausible for solving the cooling flow problem.
In \cite{2020MNRAS.491.1190S} (here after \citetalias{2020MNRAS.491.1190S}), we explored various idealized AGN “toy models” with energy injection in different forms (e.g., direct isotropic momentum injection, turbulent stirring, thermal heating, cosmic-ray injection). We found that turbulent stirring within a radius of order the halo scale radius, or cosmic ray injection (with appropriate energetics) were able to maintain a stable, cool-core, low-SFR halo for extended periods, across halos with mass $10^{12}-10^{14} {\rm M}_\odot$, without obviously violating observational constraints on halo gas properties or exceeding plausible energy budgets for low luminosity AGN in massive galaxies. But in that study, we did not attempt to model realistic jets or AGN outflows; instead, we intentionally considered energy input or ``stirring'' rates distributed according to an arbitrary spatial kernel, without considering how that energy would actually propagate from a collimated geometry, or how turbulence would actually be produced. Given that AGN jets can be a dominant source of cosmic rays and an important mechanism to stir  turbulence in the CGM, we move a step forward in this work to study the effects of a wide range of more realistic jet models in cooling flows.

Extensive studies have shown that various AGN jet models are, in principle, capable of quenching a galaxy and stopping the cooling flows in galaxy-scale simulations \citep[e.g.,][]{2010MNRAS.409..985D,2012MNRAS.424..190G,2012MNRAS.427.1614Y,2014ApJ...789...54L,2015ApJ...811...73L,2015ApJ...811..108P,2016ApJ...818..181Y,2017ApJ...834..208R,2017MNRAS.472.4707B}. However, in such simulations, AGN jets are launched from the smallest resolved scale, acting as a sort of inner boundary condition, instead of being generated self-consistently. Due to the uncertainties of the jet properties at these scales, the details of how the jet is launched are highly model-dependent, spanning a vast parameter space. 

AGN jets  most likely physically consist of relativistic particles at the black hole horizon scale, powered by magnetic fields  through the Blandford-Znajek process \citep{1977MNRAS.179..433B,2011MNRAS.418L..79T,2019ARA&A..57..467B}, where the magnetic energy is supplied by the black hole spin. Recent developments in GRMHD simulations have made it possible to self-consistently follow the formation and evolution of the jet in simulations resolving the black hole horizon scale and accretion disc \citep[e.g.,][]{2004PThPS.155..132H,2011MNRAS.418L..79T,2012MNRAS.423.3083M,2019ApJ...874..168W}, and the fields carried with the jet can reach $\sim$ Gauss at scales $ \ll 1 {\rm pc}$ \citep{2014ApJ...781...48G}.
 However, at the finest resolvable scale ($\gtrsim 10$ pc) in galaxy simulations, the jet velocity and magnetic field strength evolve radically through interactions with the surrounding gas. Depending on the model and the sub-resolution environment around the black hole, part of the kinetic energy can be transformed into thermal or cosmic ray energy. The balance between thermal, kinetic, magnetic, and cosmic ray energy at the scales where jets begin to interact with resolvable galaxy scales (the key for quenching models) therefore remains highly uncertain.
 
 Momentum and kinetic energy can be directly transferred to the gas, suppressing inflows. The fast-moving jets can also shock heat the surrounding gas. Many models have invoked kinetic jets to suppress cooling flows and SFRs in massive halos \citep[e.g.,][]{2010MNRAS.409..985D,2012MNRAS.424..190G,2014ApJ...789...54L,2015ApJ...811..108P,2016ApJ...818..181Y}. 
 Many models in the literature also invoke the idea that AGN can effectively drive strong pressure-driven outflows and offset cooling if a large fraction of the accretion energy is thermalized \citep{2004cbhg.symp..374B,2005MNRAS.361..776S,2005Natur.433..604D,hopkins:qso.all,hopkins:red.galaxies,hopkins:bhfp.theory,hopkins:groups.qso,hopkins:twostage.feedback,2009ApJ...690..802J,2010ApJ...722..642O,2012MNRAS.425..605F,2013MNRAS.428.2885D,2014MNRAS.437.1456B,2017MNRAS.465.3291W,2018MNRAS.473.4077P,2018MNRAS.474.3673R,2018MNRAS.478.3100R,2020MNRAS.497.5292T}. Physically, as the jet propagates, part of the kinetic energy can thermalize through shocks.  Some studies have argued that the heat from those weak shocks can suppress cooling flows and SFRs in massive halos  \citep{2016ApJ...829...90Y,2017ApJ...847..106L,2019MNRAS.483.2465M}.  
 The magnetic fields carried by the jet at its launch might also help  suppress cooling flows by providing additional pressure support \citep{1990ApJ...348...73S,1996ARA&A..34..155B,2012MNRAS.422.2152B}, although our studies find that they have limited effects on global star formation properties of sub-$L_{\ast}$ galaxies \citep{2017MNRAS.471..144S}\footnote{Even if magnetic fields are dynamically important on large scales, they can still be critical on scales near the black hole that we do not resolve.}. 
 Finally, CRs arise generically from processes that occur in fast shocks, so could come from shocked winds or outflows, but are particularly associated with relativistic jets from AGN (where they can make up the bulk of the jet energy; \citealt{2006PhRvD..74d3005B,2017ApJ...844...13R}) and hot, relativistic plasma-filled ``bubbles'' or ``cavities'' (perhaps inflated by jets in the first place) around AGN. Different authors have argued that they could help suppress cooling flows by providing additional pressure support to the gas, driving pressurized outflows in the galaxy or CGM, or via heating the CGM/ICM directly via collisional (hadronic \&\ Coulomb) and streaming-instability losses \citep{2008MNRAS.384..251G,2010ApJ...720..652S,2011A&A...527A..99E,2011ApJ...738..182F,2013MNRAS.434.2209W,2013MNRAS.432.1434F,2013ApJ...779...10P,2017MNRAS.465.4500P,2017ApJ...834..208R,2017ApJ...844...13R,2017MNRAS.467.1449J,2017MNRAS.467.1478J,2018MNRAS.475..570J}.

 The direction and geometry of the jet at these scales are also uncertain. The width of the jet can change substantially with time or distance. Although there is still a debate as to whether jets precess or not, several proposed mechanisms including  self-induced warping of an irradiated accretion disc \citep{1996MNRAS.281..357P,1997MNRAS.292..136P}, torn accretion discs \citep{2012MNRAS.422.2547N,2012ApJ...757L..24N} due to the Lense-Thirring effect \citep{1918PhyZ...19..156L}, massive black hole binaries \citep{1980Natur.287..307B}, or simply the widely varying angular momentum direction of gas accreting from larger scales on short time scales ($<0.1$Myr) \citep{2012MNRAS.425.1121H,2020arXiv200812303A}, can plausibly alter the angular momentum direction of the accretion disc, causing jet precession. Multiple observations also suggest jet precession occurs \cite[e.g., ][]{2006MNRAS.366..758D,2011A&A...533A.111M,2013ApJ...768...11B,2016A&A...590A..73A}. Reflecting such uncertainties, AGN feedback models have adopted energy injection with opening-angles ranging from small or negligible \citep[e.g.][]{2014ApJ...789...54L,2014ApJ...789..153L,2017MNRAS.470.4530W,2019MNRAS.483.2465M}, to much wider opening-angles \citep[e.g.,][]{2015ApJ...811..108P,2017ApJ...845...91H,2018RAA....18...81H} to isotropic \citep[e.g.,][]{2015ApJ...815...41R,2020MNRAS.491.1190S,2020MNRAS.497.5292T}. Several models also consider jet precession \citep[e.g.][]{2014ApJ...789...54L,2014ApJ...789..153L,2016ApJ...818..181Y,2017MNRAS.472.4707B,2019MNRAS.483.2465M}.
 
 In particular, jet precession might more efficiently drive turbulence in the CGM/ICM through changing bulk motion or secondary instabilities \citep[e.g.][]{2014ApJ...789...54L,2016ApJ...818..181Y,2016ApJ...817..110Z,2017MNRAS.472.4707B,2018PASJ...70....9H,2019MNRAS.483.2465M}. The enhanced turbulence can also suppress cooling flows by providing direct pressure support to the gas \citep{2012MNRAS.419L..29P}, or by heating the gas ``directly'' via viscous dissipation \citep{2014MNRAS.443..687B,2014Natur.515...85Z}, effectively conducting heat from the outer hot halo to the inner cool core \citep{2014MNRAS.443..687B}, or mixing cold structures back into the hot gas in a thermally unstable medium and thereby efficiently re-distributing heat \citep[e.g.][]{2003ApJ...596L.139K,2004MNRAS.347.1130V,2006ApJ...645...83V,2010ApJ...712L.194P,2010ApJ...713.1332R,2014MNRAS.443..687B}. In \citetalias{2020MNRAS.491.1190S}, we showed that an explicit externally driven turbulence could very effectively quench the galaxy. Here we further test whether a precessing jet can drive such turbulence and thereby quench the galaxy.
 
 Finally, AGN feedback is generally episodic. A period of strong feedback can shut down the cooling flow and the black hole accretion, which subsequently turns off feedback. During the time without feedback, the cooling flows and accretion can be reestablished, starting another episode of feedback, as may have occurred in the Phoenix cluster \citep{2019MNRAS.488.2549S}. However, the duty cycle and the duration of each episode are highly dependent on both the accretion and feedback models.
Most of the literature above only studied a limited part of this large parameter space. In order to narrow down the parameter space of jet launching, in this study we conduct the most extensive set of simulations to date surveying AGN jet launching parameters including energy form (kinetic, thermal, and cosmic ray energy), energy flux, mass flux, magnetic field strength and geometry, jet precession angle and period, jet opening-angle, and jet duty cycles. We will test for what part of the parameter space jets can quench  galaxies without violating observational constraints on the CGM density and entropy profiles. For viable models, we will also study how and why those models work and what is the required energy. 

All of these questions have been studied to a varying extent in the literature already (see references above). However, this work expands on these previous studies in at least four important ways. 
{\bf (a)} We attempt a broader and more comprehensive survey, across a wide variety of parameters characterizing jet injection, using an otherwise identical set of physics and numerics, to enable fair comparisons. 
{\bf (b)} We implement all of these in global simulations that self-consistently (and simultaneously) treat the entire halo and star-forming galactic disk. We also reach higher resolution than most previous work, which allows us to resolve more detailed sub-structure in the CGM and galactic disk.
{\bf (c)} We include explicit, detailed treatments of radiative cooling, the multi-phase ISM and CGM, star formation, and stellar feedback following the FIRE-2\footnote{FIRE project website: \href{http://fire.northwestern.edu}{\textit{http://fire.northwestern.edu}}} simulation physics \citep{2014MNRAS.445..581H,2017arXiv170206148H}, in order to more robustly model both the gas dynamics and the response of galactic star formation rates to cooling flows.
{\bf (d)} We test our jet model in simulations with MHD, conduction, viscosity, and explicit cosmic-ray transport and dynamics (from AGN) to capture any interaction between the jet and fluid microphysics.

In \sref{S:methods}, we summarize our initial conditions (ICs) and the AGN jet parameters we survey and describe our numerical simulations.  We present our results and describe the observational properties of the more successful runs in \sref{S:results}. 
We discuss the effects of each model in turn, and explain why it works or does not, in \sref{s:energy_form} - \sref{dis:pres}. We summarize in \sref{sec:conclusions}. 
We include some observational properties of all the successful and unsuccessful runs in \aref{a:d_t}.

\section{Methodology} \label{S:methods}
We perform simulations of isolated galaxies with a halo mass of $\sim10^{14}{\rm M}_\odot$.
We set up the initial conditions according to the observed profiles of cool-core clusters at low redshift, as detailed in \sref{S:ic}. Without any AGN feedback, although the galaxies have initial properties consistent with observations, their cooling flow rates, and SFRs quickly run away, exceeding the observational values by orders of magnitude (\citetalias{2019MNRAS.487.4393S} and \citetalias{2020MNRAS.491.1190S}). We evolve the simulations with various AGN jet models and test to what extent (if any) they suppress the cooling flow and whether they can maintain stably quenched galaxies. 

We note that while there are more constraints from X-ray observations for rich clusters of mass $\sim 10^{15}{\rm M}_\odot$, we focus on a galaxy with a halo mass of $10^{14}{\rm M}_\odot$. The reason is that a  halo of this mass already has most of the cooling flow properties of the more massive clusters, and will experience a major cooling catastrophe unless properly quenched, but requires less computational expense. We will consider how jet models scale with halo mass in future work.

Our simulations use {\sc GIZMO}\footnote{A public version of this code is available at \href{http://www.tapir.caltech.edu/~phopkins/Site/GIZMO.html}{\textit{http://www.tapir.caltech.edu/$\sim$phopkins/Site/GIZMO.html}}}  \citep{2015MNRAS.450...53H}, in its meshless finite mass (MFM) mode, which is a Lagrangian mesh-free Godunov method, capturing advantages of grid-based and smoothed-particle hydrodynamics (SPH) methods. Numerical implementation details and extensive tests are presented in a series of methods papers for, e.g.,\ the hydrodynamics and self-gravity \citep{2015MNRAS.450...53H}, magnetohydrodynamics \citep[MHD;][]{2016MNRAS.455...51H,2015arXiv150907877H}, anisotropic conduction and viscosity \citep{2017MNRAS.466.3387H,2017MNRAS.471..144S}, and cosmic rays \citep{chan:2018.cosmicray.fire.gammaray}.

All of our simulations have the FIRE-2 implementation of the Feedback In Realistic Environments (FIRE) physical treatments of the ISM, star formation, and stellar feedback, the details of which are given in \citet{hopkins:sne.methods,2017arXiv170206148H} along with extensive numerical tests.  Cooling is followed from $10-10^{10}$K, including the effects of photo-electric and photo-ionization heating, collisional, Compton, fine structure, recombination, atomic, and molecular cooling. 

Star formation is treated via a sink particle method, allowed only in  molecular, self-shielding, locally self-gravitating gas, above a density $n>100\,{\rm cm^{-3}}$ \citep{2013MNRAS.432.2647H}. Star particles, once formed, are treated as a single stellar population with metallicity inherited from their parent gas particle at formation. All feedback rates (SNe and mass-loss rates, spectra, etc.) and strengths are IMF-averaged values calculated from {\small STARBURST99} \citep{1999ApJS..123....3L} with a \citet{2002Sci...295...82K} IMF. The stellar feedback model includes: (1) Radiative feedback including photo-ionization and photo-electric heating, as well as single and multiple-scattering radiation pressure tracked in five bands  (ionizing, FUV, NUV, optical-NIR, IR), (2) OB and AGB winds, resulting in continuous stellar mass loss and injection of mass,  metals, energy, and momentum  (3) Type II and Ia SNe (including both prompt and delayed populations) occuring according to tabulated rates, and injecting the appropriate mass, metals, momentum, and energy to the surrounding gas. All the simulations except the `B0' run also include MHD, and fully anisotropic conduction, and viscosity with the Spitzer-Braginski coefficients.

\subsection{Initial Conditions}
\label{S:ic}

The initial conditions studied here are presented and described in detail in \citetalias{2019MNRAS.487.4393S}. The ICs are designed to be similar to observed cool-core systems of similar mass wherever possible at $z\sim0$ \citep[see e.g.,][]{2012ApJ...748...11H,2013MNRAS.436.2879H,2013ApJ...775...89S,2015ApJ...805..104S,2017A&A...603A..80M}. Their properties are summarized in \tref{tab:ic}. In this paper, we focus on the {\bf m14} halo from \citetalias{2019MNRAS.487.4393S}, which has the most dramatic (massive) cooling flow. The dark matter (DM) halo, bulge, black hole, and gas+stellar disk are initialized following  \cite{1999MNRAS.307..162S} and \cite{2000MNRAS.312..859S}.
We assume a spherical, isotropic, \citet{1996ApJ...462..563N} profile DM halo; a \cite{1990ApJ...356..359H} profile stellar bulge ($2\times10^{12} {\rm M}_\odot$); an exponential, rotation-supported disk of gas and stars ($10^{10}$ and $2\times10^{10} {\rm M}_\odot$, respectively) initialized with Toomre $Q\approx1$; a BH with mass $1/300$ of the bulge mass \citep[e.g.,][]{2004ApJ...604L..89H}; and an extended spherical, hydrostatic gas halo with a $\beta$-profile ($\beta=1/2$) and rotation at twice the net DM spin (so $\sim 10-15\%$ of the support against gravity comes from rotation, and most of the support from thermal pressure as expected in a massive halo). All the components of the initial conditions are `live'. The initial metallicity of the CGM/ICM drops from solar ($Z=0.02$) to $Z=0.001$ with radius as $Z(r)=0.02\,(0.05+0.95/(1+(r/20\,{\rm kpc})^{1.5}))$. Initial magnetic fields are azimuthal with a seed vlue of $|{\bf B}|=0.3\,\mu{\rm G}/(1+(r/20\,{\rm kpc})^{0.375})$ (which will later be amplified) extending throughout the ICM, and the initial CR energy density is in equipartition with the local initial magnetic energy density. The ICs are run adiabatically (no cooling or star formation) to relax any initial transients. 

 Our {\bf m14} halo has an initial cooling rate of $\sim 8\times10^{43}{\rm erg\,s}^{-1}$, with $\sim3 \times10^{43}{\rm erg\,s}^{-1}$ radiated in the X-ray band (0.5-7 kev). 

A resolution study is included in the appendix of \citetalias{2019MNRAS.487.4393S}. 
To achieve better convergence, we use a hierarchical super-Lagrangian refinement scheme (\citetalias{2019MNRAS.487.4393S} and \citetalias{2020MNRAS.491.1190S}) to reach $\sim 3 \times 10^{4}\,{\rm M}_{\sun}$ mass resolution in the core region and around the z-axis where the jet is launched, much higher than many previous global studies. The mass resolution decreases as a function of both radius ($r_{\rm 3d}$) and distance from the z-axis ($r_{\rm 2d}$), roughly proportional to $r_{\rm 3d}$ and $2^{r_{\rm 2d}/10 {\rm kpc}}$, whichever is smaller, down to $2\times 10^6 {\rm M}_{\sun}$. The highest resolution region is where either $r_{\rm 3d}$ or $r_{\rm 2d}$ is smaller than 10 kpc.

\begin{table*}
\begin{center}
 \caption{Properties of Initial Conditions for the Simulations/Halos Studied In This Paper}
 \label{tab:ic}
 \begin{tabular*}{\textwidth}{@{\extracolsep{\fill}}lccccccccccccccc}
 \hline
\hline
&\multicolumn{2}{c}{\underline{Resolution}}&\multicolumn{3}{c}{\underline{DM halo}}&&\multicolumn{2}{c}{\underline{Stellar Bulge}}&\multicolumn{2}{c}{\underline{Stellar Disc}}&\multicolumn{2}{c}{\underline{Gas Disc}}&\multicolumn{2}{c}{\underline{Gas Halo}}  \\
$\,\,\,\,$Model  &$\epsilon_g$ &$m_g$        &$M_{\rm halo}$   &$r_{\rm dh}$            &$V_{\rm Max}$    &$M_{\rm baryon}$    &$M_b$ 
                 &$a$          &$M_d$        & $r_d$             &$M_{\rm gd}$       &$r_{\rm gd}$         &$M_{\rm gh}$         &$r_{\rm gh}$     \\
                 &(pc)         &(M$_\odot$)  &(M$_\odot$)      & (kpc)           &(km/s)           &(M$_\odot$)      &(M$_\odot$) 
                  &(kpc)        &(M$_\odot$)  &(kpc)            &(M$_\odot$)    &(kpc)            &(M$_\odot$)      &(kpc)\\
\hline
{\bf $\,\,\,\,$m14}  &{\bf1}  &{\bf3e4}      &{\bf8.5e13}      &{\bf 220}       &{\bf600}         &{\bf1.5e13}        &{\bf2.0e11}
                     &{\bf3.9}     &{\bf2.0e10}          &{\bf3.9}              &{\bf1e10}           &{\bf3.9}              &{\bf1.5e13}           &{\bf 22}            \\          
\hline 
\hline
\end{tabular*}
\end{center}
\begin{flushleft}
Parameters of the galaxy/halo model studied in this paper (\sref{S:ic}): 
(1) Model name. The number following `m' labels the approximate logarithmic halo mass. 
(2) $\epsilon_g$: Minimum gravitational force softening for gas (the softening for gas in all simulations is adaptive, and matched to the hydrodynamic resolution; here, we quote the minimum Plummer equivalent softening).
(3) $m_g$: Gas mass (resolution element). There is a resolution gradient for m14, so its $m_g$ is the mass of the highest resolution elements.
(4) $M_{\rm halo}$: Halo mass. 
(5) $r_{\rm dh}$: NFW halo scale radius (the corresponding concentration of m14 is $c=5.5$).
(6) $V_{\rm max}$: Halo maximum circular velocity.
(7) $M_{\rm baryon}$: Total baryonic mass. 
(8) $M_b$: Bulge mass.
(9) $a$: Bulge Hernquist-profile scale-length.
(10) $M_d$ : Stellar disc mass.
(11) $r_d$ : Stellar disc exponential scale-length.
(12) $M_{\rm gd}$: Gas disc mass. 
(13) $r_{\rm gd}$: Gas disc exponential scale-length.
(14) $M_{\rm gh}$: Hydrostatic gas halo mass. 
(15) $r_{\rm gh}$: Hydrostatic gas halo $\beta=1/2$ profile scale-length.
\end{flushleft}
\end{table*}

\subsection{AGN Jet Models}
\label{S:physics}
In this paper, we focus on the effects of a given AGN jet.  All the jet models are run with a preset mass, energy, and momentum flux: we do not attempt to simultaneously model BH accretion from $\sim10-100$ pc to the event horizon. We systematically vary the jet velocity, energy composition (kinetic, thermal, magnetic, and cosmic ray energy), mass flux, opening-angle, procession, and duty cycle. We note that such variations reflect the uncertainties from the nature of AGN jets and the sub-resolution ($<$10 pc) physics around the black hole, which affects the balance of different energy forms and other jet parameters. A full list of simulations can be found in \tref{tab:run}. We emphasize that the parameters in the table reflect the jet parameters at our launch scale. The jet properties will continuously evolve as it interacts with the surrounding gas. 

We launch the jet with a particle spawning method, which creates new gas cells (resolution elements) from the central black hole. With this method, we have better control of the jet properties as the launching is less dependent on the neighbor-finding results. We can also enforce a higher resolution for the jet elements.  The numerical method in this paper is similar to \cite{2020MNRAS.497.5292T}, which studied the effects of BAL wind feedback on disk galaxies.
The spawned gas particles have a mass resolution of 5000 ${\rm M}_\odot$ and are forbidden to de-refine (merge into a regular gas element) before they decelerate to 10\% of the launch velocity. Two particles are spawned in opposite z-directions at the same time when the accumulated jet mass flux reaches twice the target spawned particle mass, so linear momentum is always exactly conserved. Initially, the spawned particle is randomly placed on a sphere with a radius of $r_0$, which is either 10 pc or half the distance between the black hole and the closest gas particle, whichever is smaller.  If the particle is initialized at a position $(r_0,\theta_0,\phi_0)$ and the jet opening-angle of a specific model is $\theta_{\rm op}$, the initial velocity direction of the jet will be set at $2\theta_{\rm op}\theta_0/\pi$ for $\theta_0<\pi/2$  and at $\pi-2\theta_{\rm op}(\pi-\theta_0)/\pi$ for $\theta_0>\pi/2$. With this, the projected paths of any two particles will not intersect.

The naming of each model starts from the {\it primary} form of energy at our injection scale (`Kin',  `Th', and  `CR' for kinetic, thermal and CR energy, respectively) and the corresponding energy flux in erg s$^{-1}$. The run with `B4' in the name means that the initial jet magnetic field is toroidal with a maximum field strength of $10^{-4}$ G. The number after the `m' label provides the mass flux in units of ${\rm M}_\odot$ yr$^{-1}$. The number after `w' and `pr' gives the initial opening-angle and precession angle in degrees. The number after `$t_p$' and `$t_d$' denotes the precession period and the duty cycle in Myr. If a specific quantity is not labeled in the name, the jet model has fiducial values of a constant mass flux of 2 ${\rm M}_\odot$ yr$^{-1}$, an initial toroidal magnetic field with a maximum field strength of $10^{-3}$, $1^{\circ}$ opening-angle, no precession, and 100\% duty cycle (i.e., we only label runs with parameters that differ from these default values).

\subsubsection{Form of Jet Energy}
\label{S:kinetic}
Each spawned jet element carries mass, velocity, thermal energy, magnetic field, and cosmic ray energy, so the energy flux of each kind is well controlled. 
In the initial conditions, we set the blackhole mass at $10^9{\rm M}_\odot$, corresponding to an Eddington luminosity  $ L_{\rm Edd}\sim 10^{47}$ erg s$^{-1}$. We systemically test the jet model with energy input in each form ranging from $\sim6\times10^{42}$ up to $6\times10^{44}$ erg s$^{-1}$, which corresponds to roughly $6\times 10^{-5}-6\times 10^{-3} L_{\rm Edd}$, around the total X-ray luminosity of the whole system and the plausible energy flux according to \citetalias{2020MNRAS.491.1190S}.

Each energy form is briefly described below:

\begin{itemize}
    \item {\bf Kinetic component}: 
    Despite the relativistic nature of jets at the black hole scale, such scales are orders of magnitude smaller than the finest scale we can resolve. Instead, we have to initiate the spawned element at $\sim10$ pc, at which point the jet has already been decelerated significantly. Moreover, we are also constrained by the Newtonian nature of our MHD solver, which cannot accurately treat relativistic velocities. Accordingly, we leave the jet element's initial spawning velocity as a free parameter varied within 3000-30000 km s$^{-1}$. This roughly spans the range from the minimum velocity required to sustain a clear bi-polar jet shape to the maximum feasible velocity, as limited by the numerical methods and computational time.
    
    \item {\bf Thermal component}:   At the scale where we initiate the jet, a significant amount of energy has probably already been thermalized, but the fraction is uncertain.  Therefore, we also leave the internal temperature of the jet plasma as a free parameter ranging from $10^7-3\times 10^{10} {\rm K}$, corresponding to roughly the same range of specific energy as the kinetic jet models we studied. We emphasize that due to the injection in jet form, this is very different from ``traditional" thermal feedback, which we will discuss later.
    
    \item {\bf Magnetic component}:  
    Given the uncertainties of the magnetic field strength and geometry at the scale where we launch our jet, we parameterize the initial magnetic fields as either purely toroidal or purely poloidal with different strengths. The toroidal magnetic fields follow:
    \begin{align}
    B_{{\rm tor},\phi} \propto r \sin \theta \exp\left(-\frac{r^2}{r^2_{\rm inj}}\right),     
    \end{align}
    where $r_{\rm inj}$ is set to 10 pc.
    The poloidal magnetic fields follow:
    \begin{align}
    &B_{{\rm pol},r} \propto \frac{r\sin\theta r\cos\theta}{r^2_{\rm inj}} \exp\left(-\frac{r^2}{r^2_{\rm inj}}\right),     \notag\\
    &B_{{\rm pol},z} \propto \left(1- \frac{r^2\sin^2\theta}{r^2_{\rm inj}}\right) \exp\left(-\frac{r^2}{r^2_{\rm inj}}\right). 
    \end{align}
    \citep[e.g.,][]{2014ApJ...781...48G}
    \footnote{Given that the spawned particle is launched at a radius of 10 pc (most of the time, since the closest neighborhood gas cell rarely goes closer to the black hole), we always set the exponential part in the poloidal expression (except for B$_{\rm pol}$3e-3 where we used the full expression) and the $r$ times exponential part in the toroidal expression to a constant. This only deviates from the above expressions during the time in a run when there is a very strong cooling flow (SFR$\gtrsim 45 {\rm M}_\odot$ yr$^{-1}$ ). During those times, the closest particle to the black hole is smaller than 10 pc and therefore the jet particles are spawned at a smaller radius. However, magnetic fields do very little in those cases anyway, and none of the conclusions should change. We also checked the B$_{\rm pol}$1e-3 case to confirm that such variation does not cause any qualitative difference.}
    
    We study models with a maximum initial magnetic field strength in the jet plasma ranging from $10^{-4}\mu G$ to $3\times10^{-3}\mu G$ (roughly the maximum feasible magnetic field strength limited by the numerical method and computational time).  We use a toroidal magnetic field configuration with the maximum field strength at $10^{-3}\mu G$ as the fiducial parameters since it is roughly the minimum value able to clearly affect the global magnetic field configurations.

    \item {\bf Cosmic Rays}: 
    We treat this component analogously to our `thermal jet' runs -- simply injecting a fixed specific cosmic ray energy with the spawned jet elements, i.e., assuming a constant fraction of the jet plasma energy is in CRs. The CR physics and numerical implementation are described in detail in \citet{chan:2018.cosmicray.fire.gammaray}. Briefly, this treats CRs including streaming (at the local Alfv\'en speed, with the appropriate streaming loss term, which thermalizes, following \citealt{Uhlig2012}, but with $v_{\rm st}=v_A$), diffusion with a fixed diffusivity $\kappa_{\rm CR}$, adiabatic energy exchange with the gas and cosmic ray pressure, and hadronic and Coulomb losses (following \citealt{2008MNRAS.384..251G}). We follow a single energy bin (i.e.,\ GeV proton CRs, which dominate the pressure), treated in the ultra-relativistic limit. Streaming and diffusion are fully-anisotropic along magnetic field lines. In \citet{chan:2018.cosmicray.fire.gammaray,hopkins:cr.mhd.fire2,2021MNRAS.501.4184H}, we showed that matching observed $\gamma$-ray luminosities in simulations with the physics above, requires $\kappa_{\rm CR}\sim 10^{29}\,{\rm cm^{2}\,s^{-1}}$, in good agreement with detailed CR transport models that include an extended gaseous halo around the Galaxy  \citep[see e.g.][]{1998ApJ...509..212S,2010ApJ...722L..58S,2011ApJ...729..106T}, so we adopt this as our fiducial value.\footnote{We caution that we do not account for the possibility of  different diffusion coefficients in different environments \citep[see e.g.,][]{2021MNRAS.501.3663H,2021MNRAS.501.4184H}.}$^{\hbox{,}}$\footnote{We also note that in runs with CR jets, CRs from SNe are not included, so we have a clean test of the impact of AGN CR jets. We showed in \citetalias{2019MNRAS.487.4393S} that CRs from stars have little effect on the cooling flows in massive galaxies.}
    
    We study models with CR energy fluxes ranging from $6\times10^{42}-6\times10^{43} {\rm erg\, s^{-1}}$, roughly the values suggested to be capable of stably quenching a $10^{14}{\rm M}_\odot$ halo in our previous study with isotropic energy injection \citepalias{2020MNRAS.491.1190S}.

\end{itemize}
\subsubsection{Mass flux and specific energy}
  All of the tested runs have a constant mass flux, unless we specify a duty cycle below unity. We explicitly test the jet models with the same energy flux in each form but with a different mass flux and, therefore, specific energy. The tested mass fluxes range from 0.02-2 ${\rm M}_\odot$ yr$^{-1}$ (0.01-1 $L_{\rm Edd}/c^2$),  roughly comparable to the values obtained in the AGN feedback models considered in the literature \citep[e.g.,][]{2012MNRAS.427.1614Y,2015ApJ...811...73L,2015ApJ...811..108P}.  Correspondingly, the kinetic, thermal, or cosmic ray energy per unit mass we tested ranged from $\sim 5\times10^{16} -5\times10^{18}$ erg g$^{-1}$, so that the total energy flux is comparable to the halo cooling rate and the required energy flux suggested in \citetalias{2020MNRAS.491.1190S}.

\subsubsection{Jet opening-angle}

Due to our jet spawning method, we are able to strictly control the initial opening-angle of the jet. We emphasize that this is the initial opening-angle, which may change as the jet expands or collimates due to external interactions.
We perform most of our runs with a default initial jet opening-angle of $1^{\circ}$; however, we explicitly tested different jet opening-angles ranging from  $1^{\circ}$ (a very narrow jet) to completely isotropic (resembling a BAL wind model; e.g., \citealt{2016MNRAS.458..816H,2020MNRAS.497.5292T}).

\subsubsection{Jet precession}

We tested jet precession with different precession angles and periods. The angles ranged from $15^{\circ}-45^{\circ}$, and the periods ranged from 10-100 Myr, a broader range than the values usually quoted in the literature ($\sim 8-15^{\circ}$, 5-10 Myr)  \citep[e.g.,][]{2014ApJ...789..153L,2016ApJ...818..181Y,2017MNRAS.472.4707B}.

\subsubsection{Jet duty cycles and episodic lifetimes}
Cycles of AGN jets are observed and can naturally occur in simulations with a self-consistently coupled AGN accretion and feedback model. Given that we do not intend to model accretion explicitly in this study, we test models with preset fixed duty cycles and episodic life times. In `Th6e44-B4-t$_d$10' the mass flux follows $\dot{M}\propto\sin^2(\pi t/2 {\rm Myr})$ ``on'' for 2 Myr and is then turned off ($\dot{M}=0$) for 8 Myr, before repeating.  In `Th6e44-B4-t$_d$100', the mass flux is constant when ``on'' at $\dot{M}=2{\rm M}_\odot$ yr $^{-1}$ for 10 Myr and then off for 90 Myr.  When the jet is on, both models have a specific thermal energy of $\sim 5\times 10^{17}$ erg g$^{-1}$; the label `6e44' in the model name refers to its energy flux at the peak, in erg s$^{-1}$. In both runs, the averaged energy flux is $6\times10^{43}$ erg s$^{-1}$, and the duty cycle (percentage of time that a jet is on) is $10-20\%$ and the recurrence time of the jet ($10-100$ Myr) is broadly within the observational range \citep[e.g.,][and references therein]{2007ARA&A..45..117M} and the range of values seen in self-regulating AGN jet simulations in the literature \citep[e.g.,][]{2015ApJ...811..108P,2015ApJ...811...73L,2016ApJ...818..181Y}.

\setlength{\tabcolsep}{4pt}
\begin{table*}
\begin{center}
 \caption{Physics variations (run at highest resolution) in our halo-{\bf m14} survey}
 \label{tab:run}
\resizebox{17.7cm}{!}{%
\begin{tabular}{cccc|cccc|ccc|cc|cc|cc}
\hline
\hline
Model           &$\Delta T$& SFR &Summary  &$\dot{E}_{\rm Kin}$ &$\dot{E}_{\rm Th}$ &$\dot{E}_{\rm Mag}$ &$\dot{E}_{\rm CR}$ &$\dot{M}$ &v &$\dot{P}$ &T & B &$\theta_{\rm op}$ & $\theta_{p}$ & T$_{p}^{\rm on}/$T$_{p}^{all}$  & T$_p$ \\
                &Gyr  &\multicolumn{2}{l|}{${\rm M}_\odot$ yr$^{-1}$}&\multicolumn{4}{c|}{erg s$^{-1}$} & ${\rm M}_\odot$ yr$^{-1}$ &km s$^{-1}$ &cgs &K &G & \multicolumn{2}{c|}{deg} & \multicolumn{2}{c}{Myr}\\
\hline
NoJet               &1.5 & 65 & strong CF &\multicolumn{4}{c|}{N/A} &\multicolumn{3}{c|}{N/A} &\multicolumn{2}{c|}{N/A} &\multicolumn{2}{c|}{N/A} &\multicolumn{2}{c}{N/A}\\ 
\hline
\multicolumn{1}{c}{\hlt{ \bf{Kinetic Energy Flux}}}&\multicolumn{3}{c|}{}&\multicolumn{4}{c|}{}&\multicolumn{3}{c|}{}&\multicolumn{2}{c|}{}&\multicolumn{2}{c|}{}&\multicolumn{2}{c}{}\\
Kin6e42               &1.0 & 44 & strong CF         &\hlt{5.8e42} &1.9e41 &8e37-8e41&0      &2.0 &3e3   &3.9e34 &1e7  &1e-3 (t) &1   &N/A &N/A &N/A \\ 
Kin6e43               &1.5 & 16 & slight $\downarrow$ &\hlt{5.8e43} &1.9e41 &1e43     &0      &2.0 &9.5e3 &3.9e34 &1e7  &1e-3 (t) &1   &N/A &N/A &N/A \\ 
Kin6e44               &0.5 & 0  & overheated         &\hlt{5.8e44} &1.9e41 &2e44-3e44&0      &2.0 &3e4   &3.9e34 &1e7  &1e-3 (t) &1   &N/A &N/A &N/A \\ 
\hline
\multicolumn{1}{c}{\bf \hlt{Thermal Energy Flux}}&\multicolumn{3}{c|}{}&\multicolumn{4}{c|}{}&\multicolumn{3}{c|}{}&\multicolumn{2}{c|}{}&\multicolumn{2}{c|}{}&\multicolumn{2}{c}{}\\
Th6e42                &1.0 & 27 & strong CF         &\hlt{5.8e42} &\hlt{5.8e42} &1e39-8e41&0      &2.0 &3e3   &3.9e34 &3e8  &1e-3 (t) &1   &N/A &N/A &N/A \\ 
{\bf Th6e43}          &{\bf 1.5} & 0.51 & {\bf quenched} &{\bf 5.8e42} &\hlt{\bf 5.8e43} &{\bf 2e42}      &{\bf 0}      &{\bf 2.0} &{\bf 3e3}   &{\bf 3.9e34} &{\bf 3e9}  &{\bf 1e-3 (t)} &{\bf 1}   &{\bf N/A} &{\bf N/A} &{\bf N/A} \\ 
Th6e44                &0.5 & 0  & overheated         &5.8e42 &\hlt{5.8e44} &2e43     &0      &2.0 &3e3   &3.9e34 &3e10 &1e-3 (t) &1   &N/A &N/A &N/A \\ 
\hline 
\multicolumn{1}{c}{\bf \hlt{CR Energy Flux}}&\multicolumn{3}{c|}{}&\multicolumn{4}{c|}{}&\multicolumn{3}{c|}{}&\multicolumn{2}{c|}{}&\multicolumn{2}{c|}{}&\multicolumn{2}{c}{}\\
{\bf CR6e42}          &{\bf 1.5} &4.9 & {\bf  strong $\downarrow$} &\hlt{\bf 5.8e42} &{\bf 1.9e41} &{\bf 6e41-7e41} &\hlt{\bf 5.8e42} &{\bf 2.0} &{\bf 3e3}   &{\bf 3.9e34} &{\bf 1e7}  &{\bf 1e-3 (t)} &{\bf 1}   &{\bf N/A} &{\bf N/A} &{\bf N/A} \\ 
{\bf CR6e43}          &{\bf 1.5} &0.25 & {\bf quenched} &{\bf 5.8e42} &{\bf 1.9e41} &{\bf 1e42}      &\hlt{\bf 5.8e43} &{\bf 2.0} &{\bf 3e3}   &{\bf 3.9e34} &{\bf 1e7}  &{\bf 1e-3 (t)} &{\bf 1}   &{\bf N/A} &{\bf N/A} &{\bf N/A} \\ 
{ CR6e44-B4}          &{ 1.0} &0 & { overheated} &{ 5.8e42} &{ 1.9e41} &{ 1e40}      &\hlt{ 5.8e44} &{ 2.0} &{ 3e3}   &{ 3.9e34} &{ 1e7}  &{ 1e-3 (t)} &{ 1}   &{ N/A} &{ N/A} &{ N/A} \\ 
\hline
\multicolumn{1}{c}{\bf \hlt{Magnetic fields}}&\multicolumn{3}{c|}{}&\multicolumn{4}{c|}{}&\multicolumn{3}{c|}{}&\multicolumn{2}{c|}{}&\multicolumn{2}{c|}{}&\multicolumn{2}{c}{}\\
B0                    &1.0 & 64 & strong CF         &\hlt{5.8e42} &1.9e41 &0        &0      &2.0 &3e3   &3.9e34 &1e7  &0        &1   &N/A &N/A &N/A \\ 
B$_{\rm tor}$1e-4         &1.0 & 54& strong CF         &\hlt{5.8e42} &1.9e41 &2e36-1e40&0      &2.0 &3e3   &3.9e34 &1e7  &1e-4 (t) &1   &N/A &N/A &N/A \\ 
B$_{\rm tor}$1e-3(Kin6e42)&1.0 & 44& strong CF         &\hlt{5.8e42} &1.9e41 &8e37-8e41&0      &2.0 &3e3   &3.9e34 &1e7  &1e-3 (t) &1   &N/A &N/A &N/A \\ 
B$_{\rm tor}$3e-3         &1.5 & 23 & strong CF         &5.8e42 &1.9e41 &\hlt{2e43}     &0      &2.0 &3e3   &3.9e34 &1e7  &3e-3 (t) &1   &N/A &N/A &N/A \\ 
B$_{\rm pol}$1e-4         &1.0 &65 & strong CF         &\hlt{5.8e42} &1.9e41 &3e36-5e39&0      &2.0 &3e3   &3.9e34 &1e7  &1e-4 (p) &1   &N/A &N/A &N/A \\ 
B$_{\rm pol}$1e-3         &1.0 &43 & strong CF         &\hlt{5.8e42} &1.9e41 &2e38-4e41&0      &2.0 &3e3   &3.9e34 &1e7  &1e-3 (p) &1   &N/A &N/A &N/A \\ 
B$_{\rm pol}$3e-3         &1.0 &35 & strong CF &\hlt{5.8e42} &1.9e41 &2e42-3e42&0      &2.0 &3e3   &3.9e34 &1e7  &3e-3 (p) &1   &N/A &N/A &N/A \\ 
\hline
\multicolumn{1}{c}{\bf \hlt{Jet width}}&\multicolumn{3}{c|}{}&\multicolumn{4}{c|}{}&\multicolumn{3}{c|}{}&\multicolumn{2}{c|}{}&\multicolumn{2}{c|}{}&\multicolumn{2}{c}{}\\
Kin6e43-w15           &1.5 &19 & slight $\downarrow$ &\hlt{5.8e43} &1.9e41 &1e43-2e43&0      &2.0 &9.5e3 &3.9e34 &1e7   &1e-3 (t) &15  &N/A &N/A &N/A \\ 
Kin6e43-w30           &1.5 &11 & slight $\downarrow$ &\hlt{5.8e43} &1.9e41 &1e43-2e43     &0      &2.0 &9.5e3 &3.9e34 &1e7   &1e-3 (t) &30  &N/A &N/A &N/A \\ 
Kin6e43-w45           &1.5 &7.1 & slight $\downarrow$ &\hlt{5.8e43} &1.9e41 &1e43-2e43     &0      &2.0 &9.5e3 &3.9e34 &1e7   &1e-3 (t) &45  &N/A &N/A &N/A \\ 
{\bf Kin6e43-wiso}    &{\bf 1.5} &0 & {\bf quenched} &\hlt{\bf 5.8e43} &{\bf 1.9e41} &{\bf 2e43}&{\bf 0}      &{\bf 2.0} &{\bf 9.5e3} &{\bf 3.9e34} &{\bf 1e7}  &{\bf 1e-3 (t)} &{\bf iso} &{\bf N/A} &{\bf N/A} &{\bf N/A} \\ 
\hline
\multicolumn{1}{c}{\bf \hlt{Jet Precession}}&\multicolumn{3}{c|}{}&\multicolumn{4}{c|}{}&\multicolumn{3}{c|}{}&\multicolumn{2}{c|}{}&\multicolumn{2}{c|}{}&\multicolumn{2}{c}{}\\
Kin6e43-pr15-t$_p$10  &1.5 &17 & slight $\downarrow$ &\hlt{5.8e43} &1.9e41 &1e43     &0      &2.0 &9.5e3 &3.9e34 &1e7   &1e-3 (t) &1   &15 &N/A &10 \\ 
Kin6e43-pr30-t$_p$10  &1.5 &12 & slight $\downarrow$ &\hlt{5.8e43} &1.9e41 &1e43     &0      &2.0 &9.5e3 &3.9e34 &1e7  &1e-3 (t) &1   &30 &N/A &10 \\ 
{\bf Kin6e43-pr45-t$_p$10}  &{\bf 1.5} &1.5       & {\bf strong $\downarrow$} &\hlt{\bf 5.8e43} &{\bf 1.9e41} &{\bf 1e43}     &{\bf 0}      &{\bf 2.0} &{\bf 9.5e3} &{\bf 3.9e34} &{\bf 1e7}  &{\bf 1e-3 (t)} &{\bf 1}   &{\bf 45} &{\bf N/A} &{\bf 10}\\  
{\bf Kin6e43-pr30-t$_p$100} &{\bf 1.5} &{\bf 3.0} & {\bf strong $\downarrow$} &\hlt{\bf 5.8e43} &{\bf 1.9e41} &{\bf 1e43}     &{\bf 0}      &{\bf 2.0} &{\bf 9.5e3} &{\bf 3.9e34} &{\bf 1e7}  &{\bf 1e-3 (t)} &{\bf 1}   &{\bf 30} &{\bf N/A} &{\bf 100}\\ 
{\bf Kin6e43-pr45-t$_p$100} &{\bf 1.5} &{\bf 1.0} & {\bf quenched} &\hlt{\bf 5.8e43} &{\bf 1.9e41} &{\bf 1e43-3e43}     &{\bf 0}      &{\bf 2.0} &{\bf 9.5e3} &{\bf 3.9e34} &{\bf 1e7}  &{\bf 1e-3 (t)} &{\bf 1}   &{\bf 45} &{\bf N/A} &{\bf 100}\\ 
\hline
\multicolumn{1}{c}{\bf \hlt{Kinetic Specific Energy}}&\multicolumn{3}{c|}{}&\multicolumn{4}{c|}{}&\multicolumn{3}{c|}{}&\multicolumn{2}{c|}{}&\multicolumn{2}{c|}{}&\multicolumn{2}{c}{}\\
Kin6e42-B4-m2e-2      &1.02  &43 & strong CF          &\hlt{5.8e42} &1.9e39 &4e34-1e41&0     &0.02 &3e4   &3.9e32 &1e7  &1e-4 (t) &1   &N/A &N/A    &N/A \\ 
Kin6e43-B4-m2e-1      &1.5   &17 & slight$\downarrow$ &\hlt{5.8e43} &1.9e40 &9e40-4e42&0     &0.2    &3e4   &3.9e33 &1e7  &1e-4 (t) &1   &N/A &N/A    &N/A \\ 
Kin6e44-B4         &1.5   &13 & slight$\downarrow$ &\hlt{5.8e44} &1.9e41 &3e42-5e42&0     &2.0    &3e4   &3.9e34 &1e7  &1e-4 (t) &1   &N/A &N/A    &N/A \\ 
\hline
\multicolumn{1}{c}{\bf \hlt{Thermal Specific Energy}}&\multicolumn{3}{c|}{}&\multicolumn{4}{c|}{}&\multicolumn{3}{c|}{}&\multicolumn{2}{c|}{}&\multicolumn{2}{c|}{}&\multicolumn{2}{c}{}\\
Th6e42-B4-m2e-2       &1.5    &29 & slight $\downarrow$ &5.8e40 &\hlt{5.8e42} &3e39-2e41&0      &0.02   &3e3   &3.9e32 &3e10 &1e-4 (t) &1   &N/A &N/A    &N/A \\ 
Th6e42-B4-m2e-1       &1.0   &29 & strong CF          &\hlt{5.8e42} &1.9e40 &5e40-1e42&0     &0.2    &3e3   &3.9e33 &3e9  &1e-4 (t) &1   &N/A &N/A    &N/A \\ 
Th6e43-B4-m2e-1       & 0.6   &0 & overheated            &5.8e41 &\hlt{5.8e43} &2e42-9e42&0      &0.2   &3e3   &3.9e33 &3e10 &1e-4 (t) &1   &N/A &N/A    &N/A \\ 
{\bf Th6e43-B4}    &{\bf 1.5} &4.3 & {\bf strong $\downarrow$}         &\hlt{\bf 5.8e43} &{\bf 1.9e41} &{\bf 2e40}     &{\bf 0}      &{\bf 2.0}    &{\bf 3e3}   &{\bf 3.9e34} &{\bf 3e9}  &{\bf 1e-4 (t)} &{\bf 1}   &{\bf N/A} &{\bf N/A}    &{\bf N/A} \\ 
Th6e44-B4          & 0.5   &0 & overheated            &5.8e42 &\hlt{5.8e44} &2e41     &0      &2.0    &3e3   &3.9e34 &3e10 &1e-4 (t) &1   &N/A &N/A    &N/A \\ 
\hline
\multicolumn{1}{c}{\bf \hlt{CR Specific Energy}}&\multicolumn{3}{c|}{}&\multicolumn{4}{c|}{}&\multicolumn{3}{c|}{}&\multicolumn{2}{c|}{}&\multicolumn{2}{c|}{}&\multicolumn{2}{c}{}\\
{\bf CR6e43-B4-m2e-1}  &{\bf 1.5} & {\bf 0}& {\bf quenched}          &{\bf 5.8e41} &{\bf 1.9e40} &{\bf 1e42-6e42}     &\hlt{\bf 5.8e43} &{\bf0.2} &{\bf 3e3}   &{\bf 3.9e33} &{\bf 1e7}  &{\bf 1e-4 (t)} &{\bf 1}   &{\bf N/A} &{\bf N/A} &{\bf N/A} \\ 
{\bf CR6e43-B4}    &{\bf 1.5} & {\bf 0.12}& {\bf quenched}          &{\bf 5.8e42} &{\bf 1.9e41} &{\bf 1e40}     &\hlt{\bf 5.8e43} &{\bf 2.0} &{\bf 3e3}   &{\bf 3.9e34} &{\bf 1e7}  &{\bf 1e-4 (t)} &{\bf 1}   &{\bf N/A} &{\bf N/A} &{\bf N/A} \\ 
\hline
\multicolumn{1}{c}{\bf \hlt{Duty Cycle}}&\multicolumn{3}{c|}{}&\multicolumn{4}{c|}{}&\multicolumn{3}{c|}{}&\multicolumn{2}{c|}{}&\multicolumn{2}{c|}{}&\multicolumn{2}{c}{}\\
Th6e44-B4-t$_d$10     & 0.6   &0 & overheated          &5.8e42 &\hlt{5.8e44} &8e41-2e43&0      &2.0    &3e3   &3.9e34 &3e10 &1e-4 (t) &1   &N/A &1/10   &N/A\\ 
{\bf Th6e44-B4-t$_d$100}&{\bf 1.5} & 0.064 & {\bf quenched}          &{\bf 5.8e42} &\hlt{\bf 5.8e44} &{\bf 2e41-6e41}&{\bf 0}      &{\bf 2.0}    &{\bf 3e3}   &{\bf 3.9e34} &{\bf 3e10} &{\bf 1e-4 (t)} &{\bf 1}   &{\bf N/A} &{\bf 10/100} &{\bf N/A}\\ 
\hline 
\hline
\end{tabular}
}
\end{center}
\begin{flushleft}
This is a partial list of simulations studied here: each was run using halo {\bf m14}, systematically varying the jet parameters. Columns list: 
(1) Model name:  The naming of each model starts with the primary form of energy flux and the energy flux value in erg s$^{-1}$ used. A run with `B4' in the name means the initial jet magnetic field has a toroidal geometry with a maximum field strength of $10^{-4} \mu{\rm G}$. The number after the `m' label is the mass flux in ${\rm M_\odot\,yr}^{-1}$. The numbers after the `w' and `pr' labels are the initial opening-angle and precession angle, respectively. The numbers after the `$t_p$' and `$t_d$' label the precession period and duty cycle period.  If a specific quantity is not labeled in the name, the jet model is launched with a constant mass flux of 2 ${\rm M}_\odot$ yr$^{-1}$, toroidal magnetic field, with a maximum field strength of $10^{-3} \mu{\rm G}$, $1^{\circ}$ opening-angle, no precession, and 100\% duty cycle.
(2) $\Delta T$: Simulation duration. All simulations are run to $1.5\,$Gyr, unless either the halo is completely ``blown out'' or completely unaffected. 
(3) The SFR averaged over the last 50 Myr.
(4) Summary of the results. `strong CF', `slight $\downarrow$', `significant  $\downarrow$' , and `quenched' correspond respectively to a SFR of $\gtrsim 20$, $\sim5-20$, $\sim1-5$ and $\lesssim 1 {\rm M}_\odot\,{\rm yr}^{-1}$. `Overheated' means the jet explosively destroys the cooling flow in $<500$ Myr, leaving a core with much lower density and high entropy and temperature (e.g. $\gg10^9$ K), violating observational constraints.
(5) $\dot{E}_{\rm Kin}$, $\dot{E}_{\rm Th}$, $\dot{E}_{\rm Mag}$, and $\dot{E}_{\rm CR}$ tabulate the total energy input of the corresponding form. The dominant energy form is highlighted in blue.
(6) $\dot{M}$, v, and $\dot{P}$ tabulate the mass flux, jet velocity and momentum flux.
(7) T: The initial temperature of the jet.
(8) B: The maximum initial magnetic field strength of the jet; (t) and (p) mean toroidal and poloidal respectively.
(9) $\theta_{\rm op}$: The opening angle of the jet.
(10) $\theta_{p}$: The precession angle of the jet.
(11) $T_{p}^{\rm on}$ and $T_{p}^{\rm all}$: The time that the jet is on in each duty cycle and the period of the duty cycle.
(12)  $T_{p}$: Precession period. 
\end{flushleft}
\end{table*}
\setlength{\tabcolsep}{6pt}

\section{Results} \label{S:results}
In this section, we summarize the results of all our simulations before turning to a more detailed analysis of individual mechanisms. \sref{S:sfr} describes the star formation and cooling-flow properties of all the runs. In \sref{S:success}, we further show the key X-ray observational properties of the quenched runs labeled as `strong $\downarrow$' (SFR $\sim1-5\, {\rm M_\odot\,yr}^{-1}$ ) or `quenched' (SFR $<1\,{\rm M_\odot\,yr}^{-1}$ ) in \tref{tab:run}, which do not ``overheat'' the halo (high entropy and low density within the cooling radius inconsistent with  X-ray observations).

\subsection{Star formation history and baryonic inflow} \label{S:sfr}
\begin{figure*}
\centering
 \includegraphics[width=16cm]{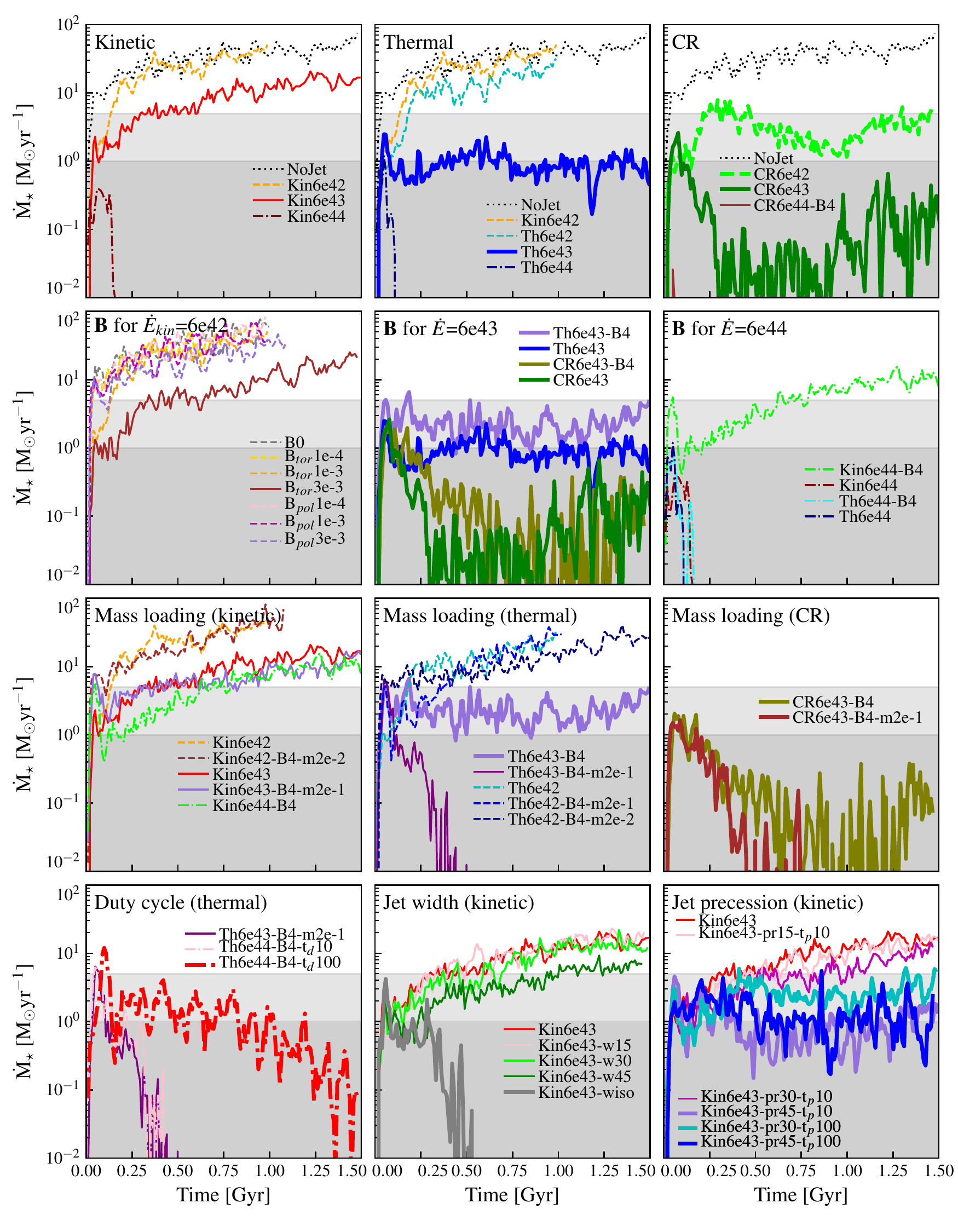}
\caption{The SFR as a function of time for all runs. Each panel labels the corresponding parameters we vary, and we also show the NoJet (no jet feedback) case in the top three panels. The thick lines indicate the quiescent and ``non-overheated'' cases. The gray areas at the bottom indicates an SFR below $\sim 5 {\rm M}_\odot$ yr$^{-1}$ and (darker grey) $\sim 1 {\rm M}_\odot$ yr$^{-1}$ , which we defined as `strong $\downarrow$' and `quenched' respectively in Table~\ref{tab:run}. Kinetic jets are the least effective in suppressing the SFR unless the opening-angle or precession angle is set to be wider than 30-45$^{\circ}$, or is significantly widened by magnetic fields. Thermal jets  can more effectively quench the galaxy when the energy input reaches $\sim6\times10^{43}$ erg s$^{-1}$. CR jets quench the most efficiently, requiring an energy input of only $\sim6\times10^{42}$ erg s$^{-1}$. In the sampled parameter space, magnetic fields cause less than a factor $\sim2$ effect in most cases. For the same thermal or cosmic ray input, lower mass flux and the higher specific energy produces more effective quenching. With the same averaged mass flux and thermal energy flux, duty cycles with periods $ \ll 100 \, {\rm Myr}$ are effectively the same as ``continuous'' jets at these $ \gg {\rm kpc}$ scales, while models with $\sim10\%$ duty cycles spread over $\gtrsim 100$ Myr periods are less effective than continuous or short period jets.}
\label{fig:sfr}
\end{figure*}

\begin{figure*}
\centering
 \includegraphics[width=16cm]{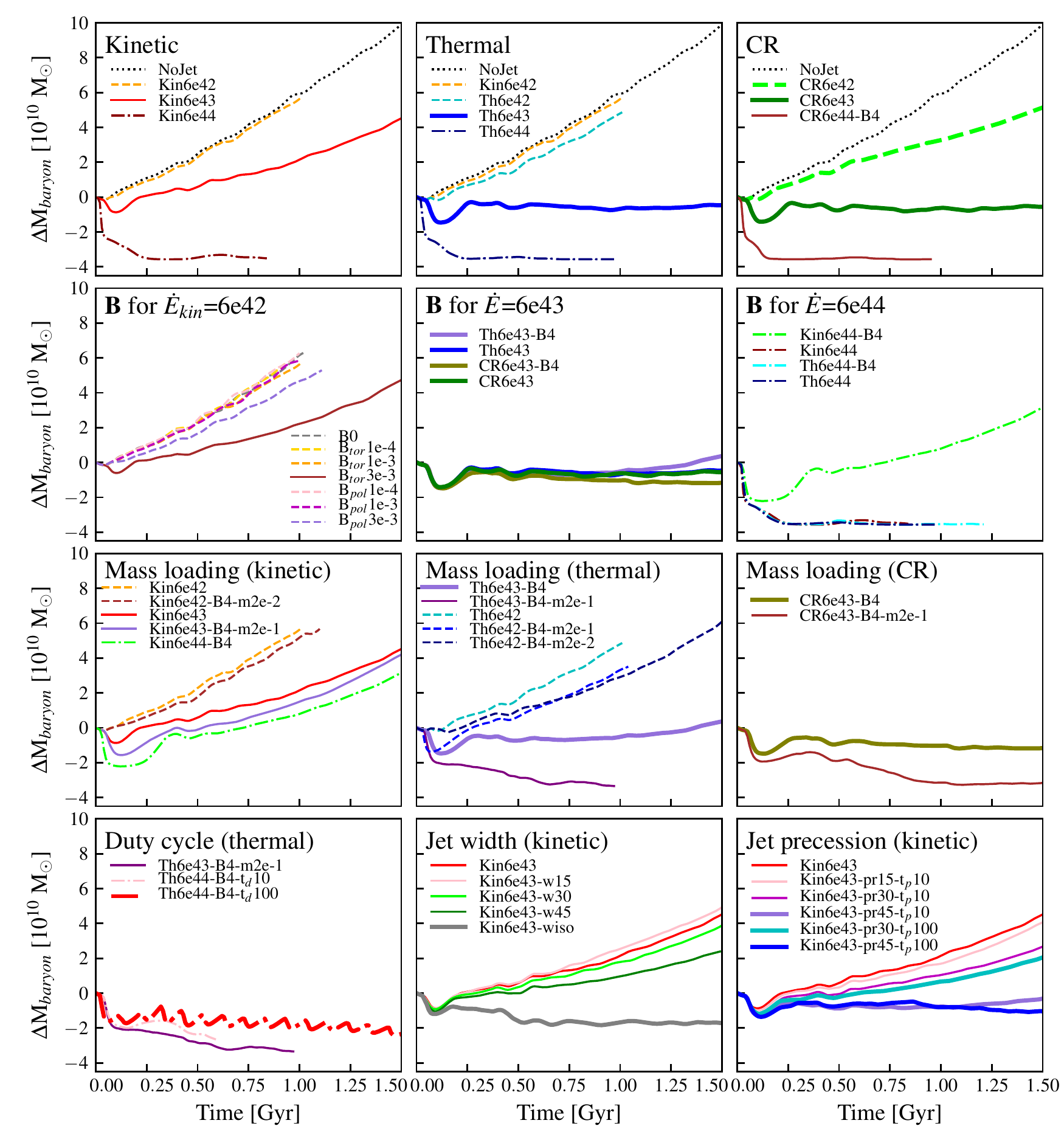}
\caption{The evolution of the core baryonic mass (total star + gas mass within 30 kpc) as a function of time. The panels and runs follow the same grid and style as in \fref{fig:sfr}. Most of the quiescent non-overheated runs (indicated with thick lines) have a core baryonic mass that remains almost constant over the duration of the run, indicating an explicit suppression of the cooling flow. The only exception is the lower flux cosmic ray jet run (`CR6e42'), where there is still a non-negligible increase of core baryonic mass even while the SFR is significantly suppressed.}
\label{fig:hotgas}
\end{figure*}

\fref{fig:sfr} and \fref{fig:hotgas} show the star formation rate and baryonic mass within 30 kpc of all the runs. Each panel shows a subset of simulations, selected to explore one parameter. The runs which are both ``not-overheated'' (consistent with X-ray observations) and ``quenched'' (labeled `strong $\downarrow$' or `quenched' in \tref{tab:ic},  defined as those with SFR $\lesssim 5 {\rm M}_\odot {\rm yr}^{-1}$ or sSFR $\lesssim 5 \times10^{-11} {\rm yr}^{-1}$ ) are highlighted with thicker lines. The averaged SFRs of the last 50 Myr of the simulations are also summarized in \tref{tab:run}.

Jets with most of their energy in a thermalized component (``thermal jets'', for the sake of brevity) can stably quench the galaxies with energy injection rates $\dot{E}\gtrsim 6\times 10^{43}$ erg s$^{-1}\sim6\times10^{-4}L_{\rm Edd}$ and $\dot{M}= 2{\rm M}_\odot$ yr$^{-1}$.  If the energy input is higher than $\sim 6\times 10^{44}$ erg s$^{-1}$, the results become explosive -- the gas is strongly expelled, and the cores are overheated, inconsistent with the X-ray observations (see \aref{a:d_t}). The balance between mass flux and thermal energy loading also changes the results. With the same thermal energy flux, the lower the mass flux (the higher the specific thermal energy), the more efficiently the galaxy is quenched. 

With the same averaged thermal energy and mass fluxes, the ``thermal jet'' run with a duty cycle reaching 100 Myr, `Th6e44-B4-t$_d$100',  tends to be less explosive than the runs with a shorter duty cycle (10 Myr) or with continuous jets with the same average energy and mass flux. The latter two cases essentially produce the same results.

Cosmic ray dominated jets (``CR jets''), on the other hand,  quench much more efficiently. With an order of magnitude lower energy input than is required for thermal jets, $\dot{E}\gtrsim 6\times10^{42}$ erg s$^{-1}\sim6\times10^{-5}L_{\rm Edd}$, the SFR is significantly suppressed to $\lesssim 3 {\rm M}_\odot$ yr$^{-1}$. With $\dot{E}\gtrsim 6\times10^{43}$ erg s$^{-1}$, the CR jets also quench more efficiently than thermal jets.

 Narrow jets with most of their energy in kinetic form (``kinetic jets'') quench much less efficiently than CR or thermal jets. With an energy input of $\dot{E}\lesssim 6\times10^{43} $ erg s$^{-1}$,  the star formation rate is only marginally suppressed. With an even higher energy input $\dot{E}\lesssim 6\times10^{44}$ erg s$^{-1}$, kinetic jets also do very little to the SFR (`Kin6e44-B4') unless they are significantly widened by strong magnetic fields (`Kin6e44), where the result becomes explosive. This will be discussed in \sref{s:kinetic}.

Making the kinetic jet wider by construction (at injection) or precessing with a wider angle can potentially help kinetic jets quench more efficiently. However, these effects only matter when the angles are wider than the solid angle affected by an initially narrower jet (30-45$^{\circ}$ in the case of `Kin6e43')  as we can see in `Kin6e43-wiso', `Kin6e43-pr45-t$_p$10', `Kin6e43-pr30-t$_p$100', and `Kin6e43-pr45-t$_p$100'. Interestingly, longer precession-period jets require a smaller opening-angle to be effective. We will discuss these effects in detail in \sref{dis:pres}.

Within the investigated parameter space, making the jet magnetic fields stronger generally has very weak effects, and the most powerful ``magnetic energy dominated'' jet we consider has quite a small (factor $\lesssim 2$) effect on the SFR. The only exception is  for the very fast kinetic jets, `Kin6e44', and `Kin6e44-B4', where the magnetic fields can non-linearly alter the jet propagation and make a difference between having minor effects and causing explosive quenching. The reason for this will be discussed in \sref{S:magnetic}.

Out of all the investigated runs, `Th6e43', `CR6e42',  `CR6e43',  `CR6e43-B4', `Kin6e43-wiso', `Kin6e43-pr45-t$_p$10', `Kin6e43-pr45-t$_p$100', `Th6e43-B4', and `Th6e44-B4-t$_d$100' are the runs that both have their SFRs significantly suppressed and avoid being overheated (i.e., are plausibly consistent with observational gas density and entropy profiles). Interestingly, among the nine runs, the run with $\dot{E}_{\rm CR} = 6\times10^{42}$ erg s$^{-1}$ (`CR6e42') is the only run with a significantly suppressed SFR which also maintains a  steady cooling flow (i.e., growth of the core baryonic mass). The rest of the runs all have roughly constant core baryonic mass, indicating an explicit suppression of the cooling flow and some switching between ``cool core'' and ``non cool core'' halos. The reason for this will be discussed in \sref{s:CR_th}.

\subsection{The resulting halo properties of the (non-overheated) quiescent runs} \label{S:success}

\subsubsection{X-ray luminosities} \label{S:luminosity}

\begin{figure}
\centering
 \includegraphics[width=8.2cm]{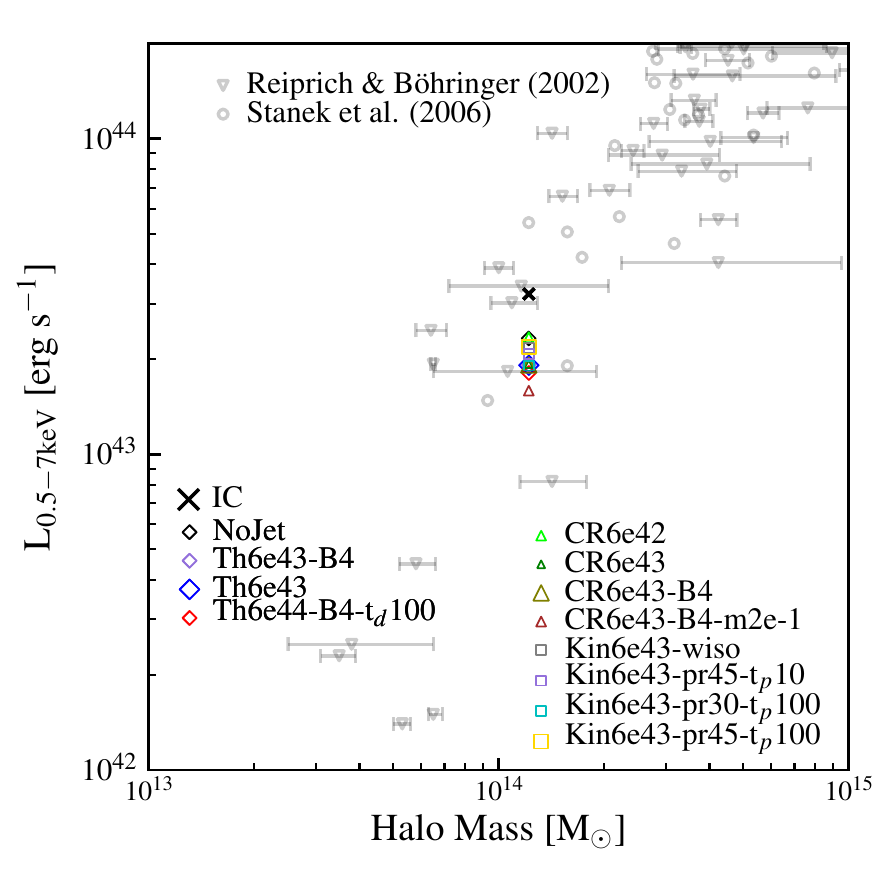}
\caption{The X-ray luminosity in the 0.5 - 7 keV band at the end of all the non-overheated quiescent runs. We use $M_{\rm 200m}$ as the halo mass for our simulations. The lighter markers and the error bars denote the observed values from \citep{2002ApJ...567..716R,2006ApJ...648..956S}. We observe very little evolution of the total X-ray luminosity in these runs (compared to the values for the initial conditions). All of the listed runs have X-ray luminosity within the observational range. Note that unless a run is very overheated, its cooling luminosity does not deviate from the initial condition by much. This is true even in the  `NoJet' run, and other `strong CF' runs. `Overheated' runs generally have slightly lower X-ray luminosity due to a lower core gas mass, ranging from 1-1.5 erg s$^{-1}$ except for 'Th6e44-B4-t$_d$10', which has $\sim2$ erg s$^{-1}$.}
\label{fig:luminosity}
\end{figure}

The resulting X-ray luminosity of the halo gas is an important constraint for AGN feedback models \citep[e.g.][]{2015MNRAS.449.4105C,2010MNRAS.406..822M}. \fref{fig:luminosity} shows the predicted X-ray cooling luminosity at the end of all the  ``non-overheated'' quiescent runs, which are runs with a strong suppression of cooling flows but which do not generate overheated entropy or density profiles (labeled `strong $\downarrow$' or `quenched' in \tref{tab:run}), integrated over all gas in the halo, from $0.5-7$\,keV. The luminosity is calculated using the same methods as in \cite{2009A&A...508..751S,2018MNRAS.478.3544R}, in which the cooling curve is calculated for photospheric solar abundances \citep{2003ApJ...591.1220L}, using the spectral analysis code SPEX \citep{1996uxsa.conf..411K} and scaled according to the local hydrogen, helium, and metal mass fractions.

 None of the non-overheated quiescent runs show a significant drop in the X-ray luminosity, indicating that they do not expel a significant amount of the virialized gas. The X-ray luminosity of all runs is above $\sim 2\times 10^{43}$ erg s$^{-1}$, within the observational range \citep{2002ApJ...567..716R,2006ApJ...648..956S,2006MNRAS.366..624B,2013ApJ...776..116K,2015MNRAS.449.3806A}. Among the runs, the `CR6e42' run has the highest X-ray luminosity (still within the constraint), consistent with the build up of a large core baryonic mass, as previously noted.

\subsubsection{Temperature, density, and entropy}

\begin{figure*}
\centering
 \includegraphics[width=16cm]{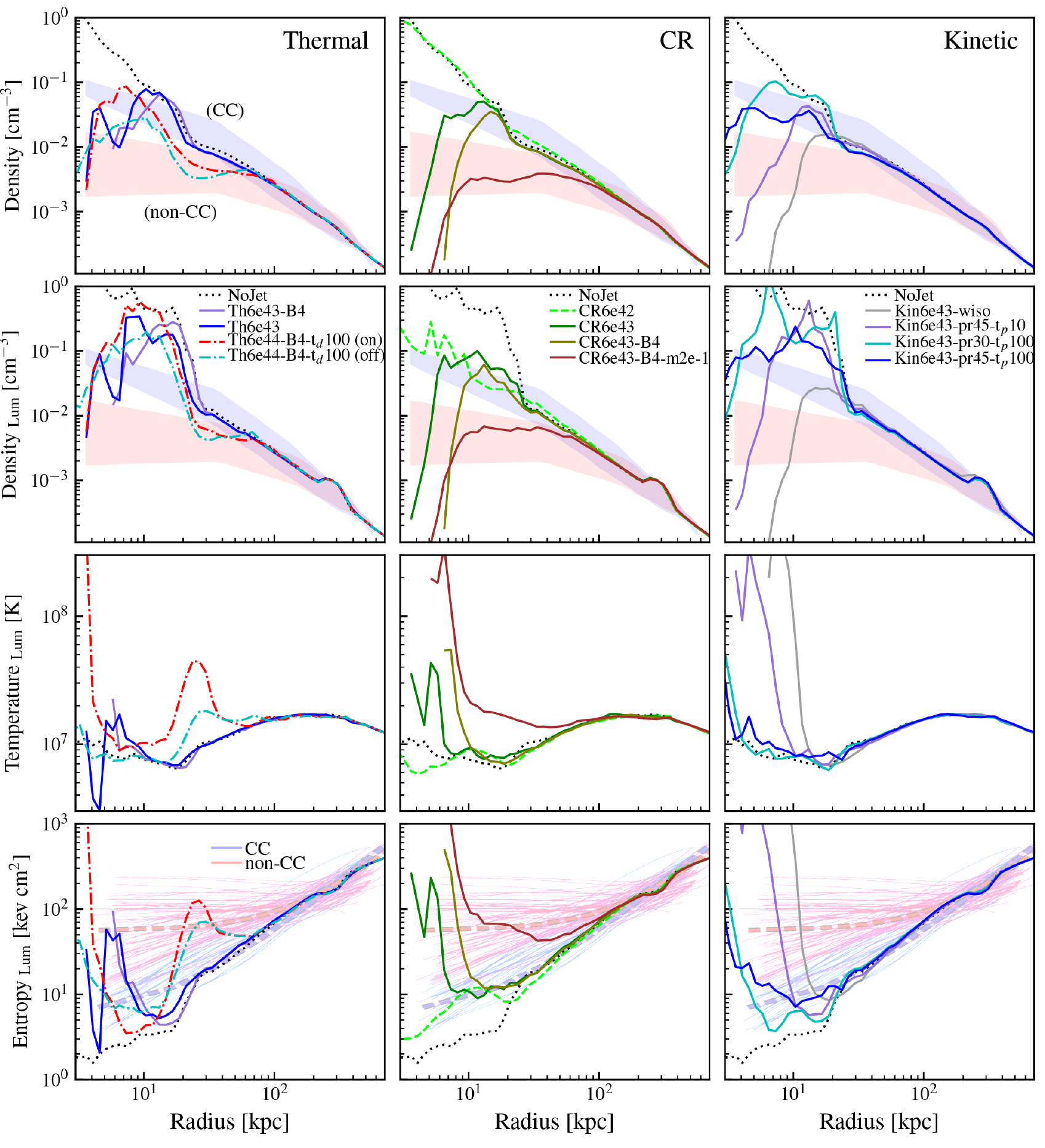}
\caption{Mean gas  density ({\em top row}), X-ray cooling luminosity-weighted density ({\em second row}),  luminosity-weighted temperature ({\em third row}), and luminosity-weighted entropy ({\em bottom row}) versus radius averaged over the last $\sim 50\,$Myr in the non-overheated quiescent  runs from \fref{fig:sfr}. 
The shaded regions in the first and second row and the light curves in the bottom row indicate the observational density and entropy profiles (scaled) for cool-core (blue) and non-cool-core (red) clusters \citep{2013ApJ...774...23M} (scaled to account for the halo mass differences). Almost all of the plotted runs with an energy input of $6\times10^{43}$ erg s$^{-1}$ have a heated core but are mostly within the observational range. 
The isotropic kinetic input run (`Kin6e43-wiso') has a larger heated region, possibly in tension with the observations. The density and entropy profiles of the run with low-energy cosmic ray jets (`CR6e42') and the runs with a long-precession-period kinetic jet (`Kin6e43-pr30-t$_p$100' and `Kin6e43-pr45-t$_p$100') resemble observed cool-core clusters. The thermal jet with a 100 Myr duty cycle (`Th6e44-B4-t$_d$100') results in a negative temperature and entropy profile when the jet is at the maximum energy flux, in tension with the observations. The run with a high specific energy CR jet (`CR6e43-B4-m2e-1') resembles the non-cool-core population.  The other non-overheated quiescent runs with  $6\times10^{43}$ erg s$^{-1}$ (`Th6e43', `Th6e43-B4', `CR6e43', and `CR6e43-B4') fall between the cool-core and non-cool-core populations.}
\label{fig:etd_sus}
\end{figure*}

\fref{fig:etd_sus} shows the average density, luminosity-weighted density, temperature, and entropy as a function of radius, averaged over the last $50\,$Myr for each simulations. We excluded the jet cells themselves from the calculation. The shaded regions in the top and second row indicate the observational density profiles (scaled) for cool-core (blue) and non-cool-core (red) clusters \citep{2013ApJ...774...23M}. To account for the difference in halo mass between the observations and our simulations, we use the panel for $z<0.1$ in  Fig. 9 of \cite{2013ApJ...774...23M} and assume $\rho_{\rm crit}\sim9.2\times 10^{30} {\rm g \,cm}^{-3}$ and $r_{\rm 500}=650$ kpc (our m14 initial condition).
The lightened curves in the bottom row indicate the observational entropy  profiles for cool-core (blue) and non-cool-core (red) clusters \citep{2013ApJ...774...23M}. We note that the halos in \cite{2013ApJ...774...23M} have a mass range of $\sim2 \times 10^{14} < M_{\rm 500} < 20 \times 10^{14} {\rm M}_\odot/h_{\rm 70}$.  We use their Fig. 2 and scale the average entropy at $r=700$ kpc to 500 kev cm$^2$ given that our halo is lower mass (cooler). \footnote{We also note that the luminosity-weighted density, temperature, and entropy are not precisely the same as the observed quantities, so the comparison should be viewed qualitatively. }

Most of the runs with a significantly suppressed SFR also have a lower gas density, and show the presence of heated gas in the core region of the galaxy (r$<$10 kpc).  `CR6e42' is, again, the only exception in that neither the temperature nor the density is significantly altered. For the run with a 100 Myr duty cycle, `Th6e44-B4-t$_d$100', we plotted the profiles both when the jet is on and when it is off. When the jet is on, we see a quick rise of temperature in the core region of the galaxy.  However, there is a delay in density suppression at around 10 kpc as the density there is the highest when the jet is off.

Most of the non-overheated quiescent runs with thermal or CR jets agree reasonably with the observations.  The density and entropy profiles in `CR6e42' end up very much  resembling those of observed cool-core clusters \citep[compare][]{2006MNRAS.372.1496S,2009MNRAS.395..764S,2010A&A...513A..37H,2013ApJ...774...23M}, while `CR6e43-B4-m2e-1' resembles a non-cool-core cluster. The other runs (`Th6e43', `Th6e43-B4', `CR6e43', and `CR6e43-B4') fall between the cool-core and non-cool-core profiles.  For the thermal jet run with a 100 Myr duty cycle (`Th6e44-B4-t$_d$100'), when the jet is on, and at a maximum energy flux, a negative temperature and entropy gradient is observed around 20-60 kpc, in possible tension with the observations. Such features die down when the jet is off, and then the run  resembles the cool-core clusters.

Kinetic jets with long precession period (100 Myr) also agree reasonably well with the observed cool-core clusters, while
the kinetic jet runs with a shorter precession period (10 Myr) have a slightly more extended heated core region and fall in between the cool-core and non-cool-core populations.
Isotropic kinetic input causes a very dramatic density suppression and a sharp temperature increase in the core region, resulting in the most significant tension with the observations.

\subsubsection{Turbulent Mach number} \label{S:mach}
\begin{figure*}
\centering
 \includegraphics[width=16cm]{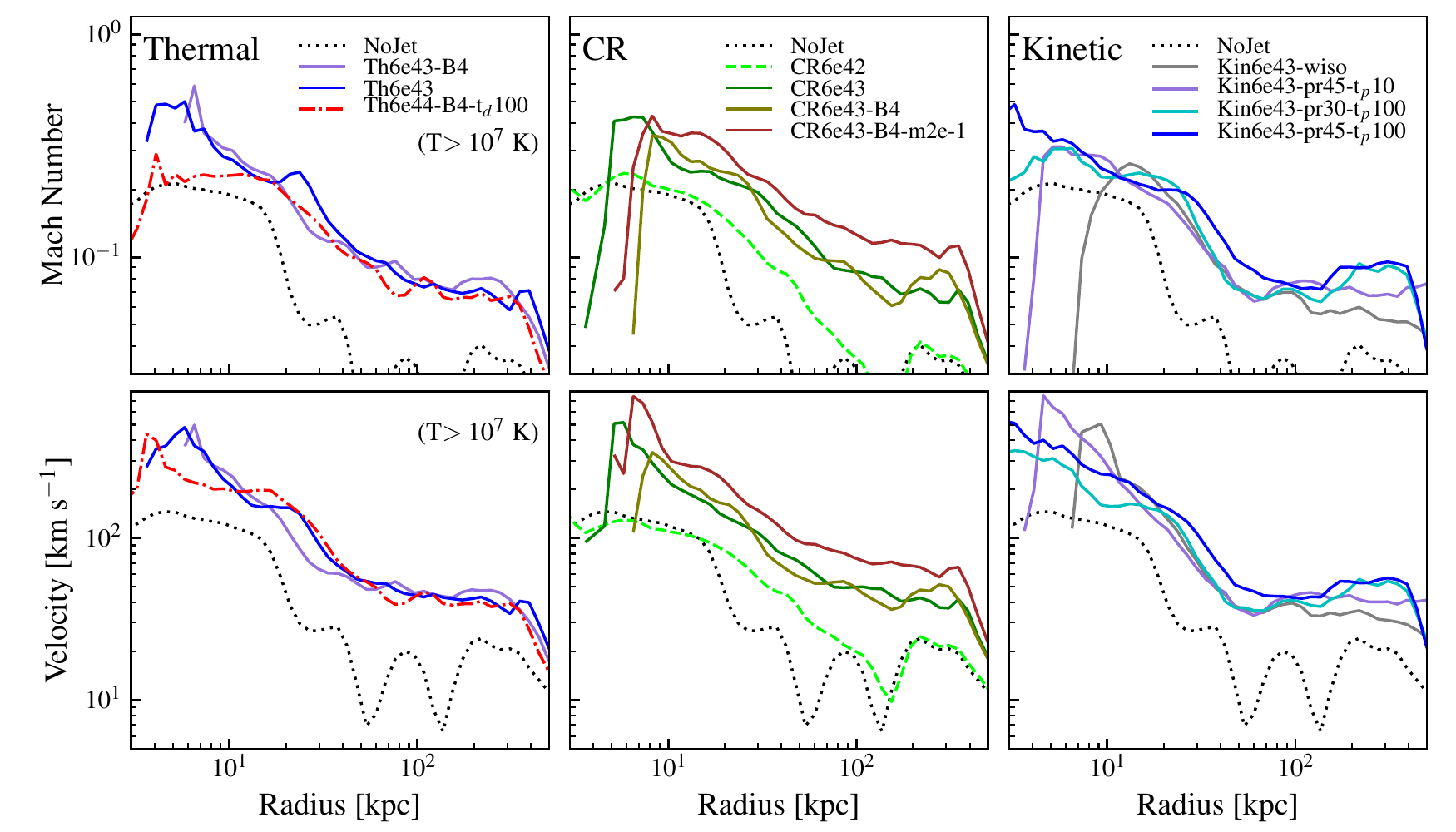}
\caption{ {\em Top row:} 1D rms Mach number in gas with $T>10^{7}\,$K (averaged over the last 50\,Myr of the runs) as a function of radius for the non-overheated quiescent runs. {\em Bottom row:} 1D rms velocity dispersion for the same gas.  All thermal and CR jets with $6\times 10^{43}$ erg s$^{-1}$ result in a similar boost of the turbulent velocity, reaching $\lesssim 400$ km s$^{-1}$ in the core region, which is slightly higher than the (limited) observational constraints. Kinetic jets with wide opening-angle or precession also boost the turbulent velocities to a similar value, and the velocity boosts extend to larger radii.  Long-precession-period jets  (100 Myr) further boost the turbulent velocity around 10-70 kpc when the opening-angle reaches 30-40$^{\circ}$, but only by a factor of $\lesssim 2$. The `CR6e42' run has slightly lower turbulent velocities, roughly 100 km s$^{-1}$,  which is broadly consistent with observations of the Perseus cluster \citep{2016Natur.535..117H,2018PASJ...70....9H}.}
\label{fig:turb}
\end{figure*}

\fref{fig:turb} shows the rms 1D turbulent velocity, defined as $v_{\rm turb}\equiv((v_{\theta}^2+v_{\phi}^2)/2)^{1/2}$, and the 1D Mach number ($v_{\rm turb}/v_{\rm thermal}$) for gas hotter than $10^7$K as a function of radius, averaged over the last 50 Myr of the runs. We exclude the radial velocity in the calculation due to the contamination of radial outflows and inflows. We exclude the spawned jet cells as well since they, by construction, can have velocities in excess of $3000$ km s$^{-1}$. 

All thermal jets with $6\times 10^{43}$ erg s$^{-1}$ result in a similar boost in velocity for radii larger than $\sim 30$ kpc, despite the different duty cycle, energy loading, and mass flux. Beyond this radius, the dynamical time is long enough that the duty cycle's effect can be averaged out.  We see a very small boost of the turbulent velocity for gas at small radii in the thermal jet run with a duty cycle of 100 Myr. The maximum velocity reaches roughly 200 km s$^{-1}$, broadly consistent with observations of the Perseus cluster \citep{2016Natur.535..117H,2018PASJ...70....9H}.
On the other hand, the continuous thermal jet with the same average energy flux boosts the turbulent velocity in the core region of the galaxy to $\lesssim 400$ km s$^{-1}$ , slightly higher than the observations.  

The cosmic ray jet with $6\times 10^{43}$ erg s$^{-1}$ results in a very similar turbulent velocity boost as the corresponding thermal jet run, and the maximum velocity reaches $\lesssim 400$ km s$^{-1}$.
The run with lower CR input, `CR6e42', has a much lower turbulent velocity ($\lesssim 200 $km s$^{-1}$) at all radii, consistent with the Perseus observation.

Kinetic jets with a wide opening-angle or with precession also boost the turbulent velocity to $\lesssim 400$ km s$^{-1}$, again, slightly higher than the observations.  Comparing with the thermal or CR jet runs with a similar energy input, the velocity boosts of kinetic jets extend to larger radii. Long-precession-period jets  (100 Myr) further boost the turbulent velocity around 10-70 kpc when the opening-angle reaches 30-40$^{\circ}$, but only by a factor of $\lesssim 2$.

\section{How jets quench}
\label{s:energy_form}
As shown in the first three panels of \fref{fig:sfr}, a cosmic ray jet quenches more efficiently than a thermal jet with the same energy flux --- i.e., CR jets require only about one-tenth of the energy flux as a thermal jet to quench the same galaxy. 
Kinetic jets are the least efficient at stopping the cooling flow, and only marginally suppress the SFR unless the opening-angles or procession angles are quite wide.
In this section, we discuss how different jet parameters affect jet propagation and galaxy quenching. We first provide a simple model in \sref{s:scaling} for the jet propagation and cocoon expansion, which helps us to interpret the results of our numerical experiments. Then we discuss how it applies to each of the cases in the following sections.

\subsection{A simple model for the cocoon expansion and how it impacts quenching}
\label{s:scaling}

\begin{figure*}
\includegraphics[width=17.7cm]{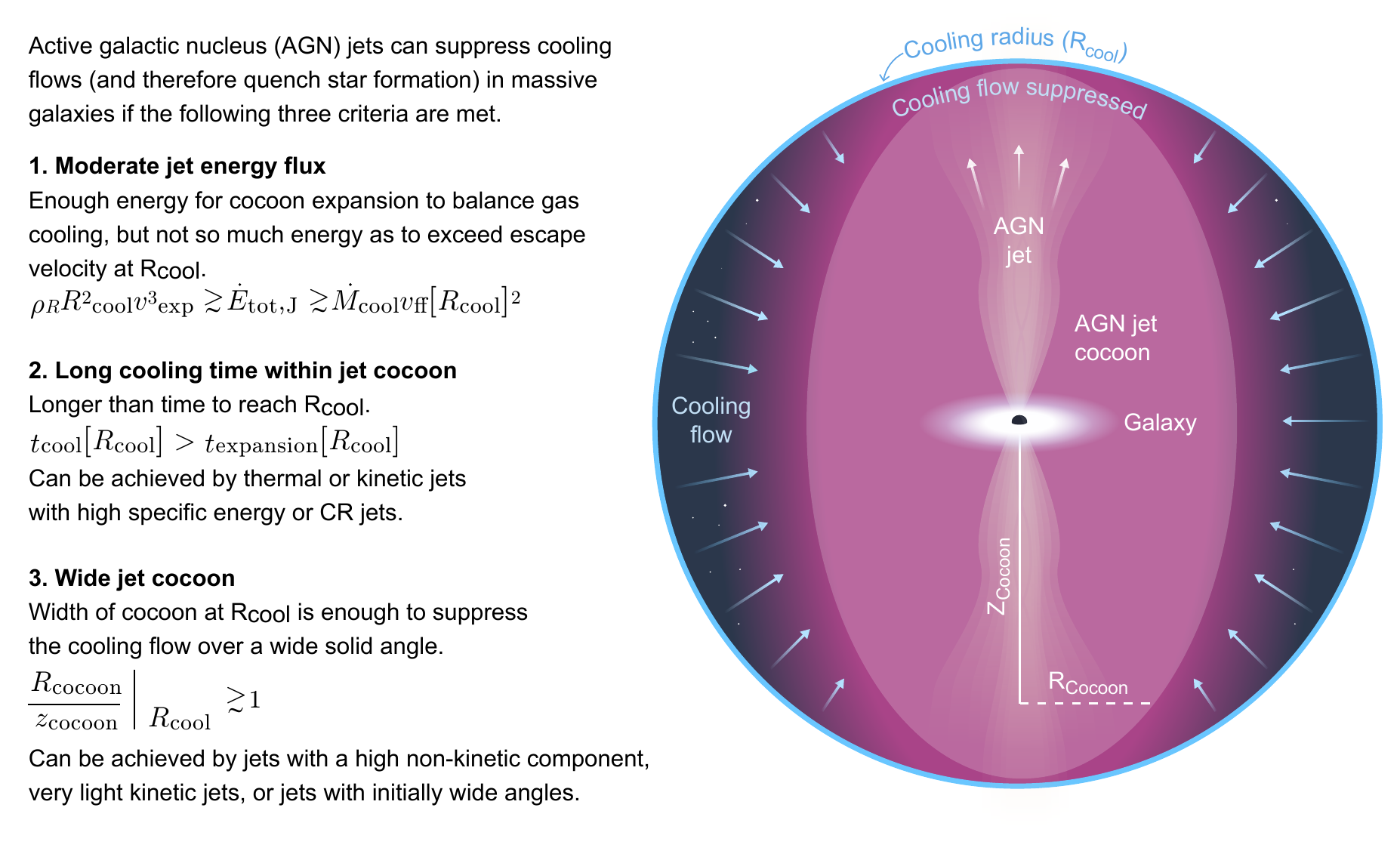}
\label{fig:cartoon}
\vspace{-0.7cm}
\caption{A cartoon picture of the criteria for successful jet models. }
\end{figure*}

Despite the different energy forms, the propagation of a jet builds up a pressurized region (cocoon) with the thermal, CR, or magnetic energy it carries at launch or gains through converting its kinetic energy through shocks. This both heats up gas within the cocoon, reversing the cooling and suppressing cooling instabilities \citep{2015Natur.519..203V}, as well as building up a pressure gradient, slowing down the gas inflow.
We found three criteria to successfully quench a halo, which are summarized in \fref{fig:cartoon}. First, the input energy of any form should, at a minimum, offset the gravitational collapse of the cooling gas:
\begin{align}
\label{eq:e_min}
\dot{E}_{\rm min}&\sim\dot{M}_{\rm cool}v_{\rm ff}[R_{\rm cool}]^2\notag\\
&\sim10^{43}\, {\rm erg\,s}^{-1} \left(\frac{\dot{M}_{\rm cool}}{100\,{\rm {\rm M}_\odot\, yr}^{-1}}\right) \left(\frac{v_{\rm ff}[R_{\rm cool}]}{300\,{\rm km\, s}^{-1}}\right)^2,
\end{align}
where $\dot{M}_{\rm cool}$ is the cooling rate and $v_{\rm ff}[R_{\rm cool}]$ is the free fall velocity at the cooling radius ($R_{\rm cool}$).\footnote{Here we define the cooling radius ($R_{\rm cool}\sim 30-50$ kpc) as the radius within which the cooling time is shorter than our simulation time ($\sim1-2$ Gyr).}

However, if the energy flux is too high, the result will be explosive. This happens at an energy flux of $\dot{E}_{\rm max}$, in which case the jet bubble expands quasi-isotropically at the escape velocity ($v_{\rm esc}[R_{\rm cool}]$) at the cooling radius. It may expand until it has swept out most of the ICM gas. Supposing the energy is purely advective, and the energy loss is negligible, then, the injected energy ($\dot{E}_{\rm tot,\,J}$) is related to the outer shell velocity ($v_{\rm exp}$) as
\begin{align}
\label{eq:t_exp}
\dot{E}_{\rm tot,\,J}\sim 4\pi R_{\rm cool}^2 v_{\rm exp} \times \frac{1}{2}\rho_R v_{\rm exp}^2 \propto\rho_R R_{\rm cool}^2v_{\rm exp}^3,
\end{align}
where $\rho_R$ is the density around the cooling radius.
Therefore,
\begin{align}
\label{eq:e_max}
\frac{\dot{E}_{\rm max}}{\dot{E}_{\rm min}} \propto\left(\frac{v_{\rm esc}[R_{\rm cool}]}{v_{\rm ff}[R_{\rm cool}]}\right)^3\sim 10-50,
\end{align}
 given that $\dot{M}_{\rm cool}\propto \pi R_{\rm cool}^3\rho_R/ t_{\rm ff} \propto R_{\rm cool}^2\rho_R v_{\rm ff}[R_{\rm cool}]$. This gives a roughly one-order-of-magnitude range for the allowed energy flux.

Another important criterion is that the cooling time within the jet cocoon/bubble has to be long enough such that the energy will not be lost before the cocoon reaches the cooling radius, otherwise, the bulk of the gas within the cooling radius will not be affected and the cooling flows persist. The cooling time is roughly $t_{\rm cool}\sim kT/\bar{n}\Lambda(T)$, where $T$ is the temperature within the jet cocoon/bubble, $\bar{n}$ is the average number density within the cooling radius, and $\Lambda$ is the cooling function. From \Eqref{eq:t_exp}, the expansion time is roughly $t_{\rm exp}\sim R_{\rm cool}/v_{\rm exp}\propto\rho_R^{1/3}\dot{E}_{\rm tot,\,J}^{-1/3}R_{\rm cool}^{5/3}$. The cooling time is long enough when $t_{\rm cool}\gg t_{\rm exp}$ or
\begin{align}
T& \gg \frac{\bar{n}\Lambda(T)\rho_R^{4/3}R_{\rm cool}^{5/3}\dot{E}_{\rm tot,\,J}^{-1/3}}{k}\notag\\
&\sim 10^8 {\rm K} \left(\frac{\bar{n}}{0.01 {\rm cm}^{-3}}\right)^{7/3}\left(\frac{R_{\rm cool}}{30 {\rm kpc}}\right)^{5/3}\left(\frac{\dot{E}_{\rm tot,\,J}}{10^{45} {\rm erg\,s}^{-1}}\right)^{-1/3}.
\end{align}

The third criterion is that the solid angle affected by the jet cocoon should be wide enough that cooling can be suppressed over a significant fraction of the volume.    The propagation of the jet cocoon qualitatively follows  momentum conservation in the z-direction \citep[e.g.,][]{1989ApJ...345L..21B},
\begin{align}
\label{eq:mom_conserve}
 A_c v_z \bar{\rho} v_z= \bar{\rho} \pi R_{\rm cocoon}^2 v_z^2 = \frac{1}{2}\dot{M}_{\rm J} v_{\rm J},
\end{align}
and the balance of energy flux in the perpendicular directions,
\begin{align}
\label{eq:e_conserve}
2A_{\rm tot}v_R\left(\frac{1}{2}\bar{\rho} v_R^2\right)= (2\pi R_{\rm cocoon} z_{\rm cocoon})v_R \bar{\rho} v_R^2\beta=\frac{\gamma}{2}(\dot{M}_{\rm J}\,v_{\rm J}^2),
\end{align}
where $A_c$ is the cross section of the whole pressurized region, $A_{\rm tot}$ is the total surface area of the same region, $R_{\rm cocoon}$ is the radius of $A_c$, z is the height to which the jet reaches, $v_R\equiv dR_{\rm cocoon}/dt$ and $v_z\equiv dz_{\rm cocoon}/dt$ are the expansion velocities of the pressurized region in the mid-plane and polar directions, $\bar{\rho}$ is the averaged density within the radius out to which the jet reaches, $\dot{M}_{\rm J}$ is the jet initial mass flux, $v_{\rm J}$ is the initial jet velocity, $\beta$ is an order-of-unity geometric factor for the surface area of the pressured region, and $\gamma\equiv \dot{E}_{\rm expansion}/\dot{E}_{\rm kin}\propto \dot{E}_{\rm tot,\,J}/\dot{E}_{\rm kin}\equiv f_{\rm kin}^{-1}$ is the ratio of the energy flux in the perpendicular direction 
(proportional to the total injected energy $\dot{E}_{\rm tot,\,J}$) to the injected kinetic energy flux.

From the equations above, we can solve for the time dependence of $R_{\rm cocoon}$ and $z_{\rm cocoon}$. In particular, for a fixed  $\bar{\rho}$, $\dot{M}_{\rm J}$, $\dot{E}_{\rm tot,\,J}$ and time, the opening-angle of the resulting cocoon scales as 
\begin{align}
\label{eq:angle_t}
\frac{R_{\rm cocoon}}{z_{\rm cocoon}}\propto \left(\frac{\dot{E}_{\rm tot,\,J}}{\dot{E}_{\rm kin}}\right)^{5/12}.
\end{align}
On the other hand, for a fixed propagation height $z_{\rm cocoon}=z_0$, the opening-angle scales as
\begin{align}
\label{eq:angle_z}
\frac{R_{\rm cocoon}}{z_{\rm cocoon}}&= \frac{\gamma z_0}{16\beta}\left(\frac{2\pi \bar{\rho} v_{\rm J}}{\dot{M}_{\rm J}}\right)^{1/2}\notag\\
           &\propto \begin{cases}
      \left(\frac{\dot{E}_{\rm tot,\,J}}{\dot{E}_{\rm kin}}\right)^{3/4}  & \text{for a fixed  $\bar{\rho}$, $\dot{M}_{\rm J}$, $\dot{E}_{\rm tot,\,J}$}\\
       \left(\frac{\dot{E}_{\rm tot,\,J}}{\dot{M}_{\rm J}}\right)^{3/2}   & \text{for a fixed  $\bar{\rho}$, $v_{\rm J}$, $\dot{E}_{\rm tot,\,J}$}\\
       \left(\frac{v_{\rm J}}{\dot{M}_{\rm J}}\right)^{1/2}   & \text{for a fixed  $\bar{\rho}$, $\dot{E}_{\rm kin}$, $\dot{E}_{\rm tot,\,J}$}
    \end{cases} 
\end{align}
Therefore, to have a wider cocoon, the jet needs to have either  a) a smaller kinetic component for fixed total energy and mass flux, b) a higher non-kinetic specific energy for fixed jet velocity and total energy flux, or c) a lighter but faster jet with fixed total and kinetic energy flux.

At the cooling radius, $R_{\rm cool}\sim 30\,{\rm kpc}$, the opening-angle for the jet cocoon is wide enough to be considered quasi-isotropic when $R_{\rm cocoon}/z_{\rm cocoon}\gtrsim 1$.\footnote{$R_{\rm cocoon}/z_{\rm cocoon}= 1$ corresponds to a polar angle of $45^{\circ}$.}  As the jet cocoon evolves, it gradually widens and becomes quasi-isotropic (if ever) at a certain height $z_{\rm cocoon}=z_{\rm iso}$.
If $z_{\rm iso} \gg R_{\rm cool}$, the jet can only suppress the inflows within a small solid angle and will not quench the cooling flow, no matter how high the injected energy is. On the contrary, if $z_{\rm iso} \ll R_{\rm cool}$, the cooling flow can be suppressed if the energy flux is sufficiently high.

For a purely kinetic jet launched with a very high Mach number, which is always the case, 
\begin{align}
\gamma\sim \frac{\rho_{\rm after-shocked} v^2_{\rm after-shocked}}{\rho_{\rm pre-shocked} v^2_{\rm pre-shocked}} \sim1/4.  
\end{align}
The scaling of $z_{\rm iso}$ generally follows 
\begin{align}
\label{eq:z_iso}
z_{\rm iso}=&\left(\frac{\dot{M}_{\rm J}}{2\pi\rho v_{\rm J}}\right)^{1/2}\left(\frac{16\beta}{\gamma}\right)\notag\\
           \sim& 23\, {\rm kpc}\,\, f_{\rm kin}\notag\\
           &\times\left(\frac{\dot{M}_{\rm J}}{2\,{\rm M}_\odot {\rm yr}^{-1} }\right)^{1/2}\left(\frac{\bar{n}}{0.01\,{\rm cm}^{-3} }\right)^{-1/2}\left(\frac{v_{\rm J}}{10^4\,{\rm km\, s}^{-1} }\right)^{-1/2} .
\end{align}

Therefore, $z_{\rm iso}<R_{\rm cool}$ when
\begin{align}
\label{eq:vj_kin}
v_{\rm J}\gtrsim& 2\times10^4\, {\rm km\,s}^{-1} f_{\rm kin}^2\notag\\
          &\times\left(\frac{\dot{M}_{\rm J}}{2\,{\rm M}_\odot {\rm yr}^{-1} }\right) \left(\frac{v_{\rm ff}}{300\, {\rm km\,s}^{-1} }\right)
           \left(\frac{\dot{M}_{\rm cool}}{100\,{\rm M}_\odot {\rm yr}^{-1} }\right)^{-1},
\end{align}
which corresponds to a purely kinetic jet with a very high velocity, or with
\begin{align}
\label{eq:fkin_nonkin}
f_{\rm kin}\lesssim& 0.4 \left(\frac{v_{\rm J}}{3000\, {\rm km\,s}^{-1}}\right)^{1/2}\left(\frac{\dot{M}_{\rm J}}{2\,{\rm M}_\odot {\rm yr}^{-1} }\right)^{-1/2}\notag\\
          &\times \left(\frac{v_{\rm ff}}{300\, {\rm km\,s}^{-1} }\right)^{-1/2}
           \left(\frac{\dot{M}_{\rm cool}}{100\,{\rm M}_\odot {\rm yr}^{-1} }\right)^{1/2},
\end{align}
which corresponds to the jet models with more than half of the energy in a non-kinetic form. In the following sub-sections, we discuss how the above scaling relations help explain which jet models quench most efficiently.

\subsection{Kinetic jets transfer energy to larger distances within smaller solid angles}
Kinetic jets quench  galaxies mostly through heating and reversing the inflow in the affected region. By shock heating the gas within a specific solid angle $\Omega$ (and also transferring momentum/kinetic energy to the gas), they halt the inflows in that cone and suppress cooling flows roughly proportional to $\Omega/4\pi$. 

\begin{figure*}
\includegraphics[width=12cm]{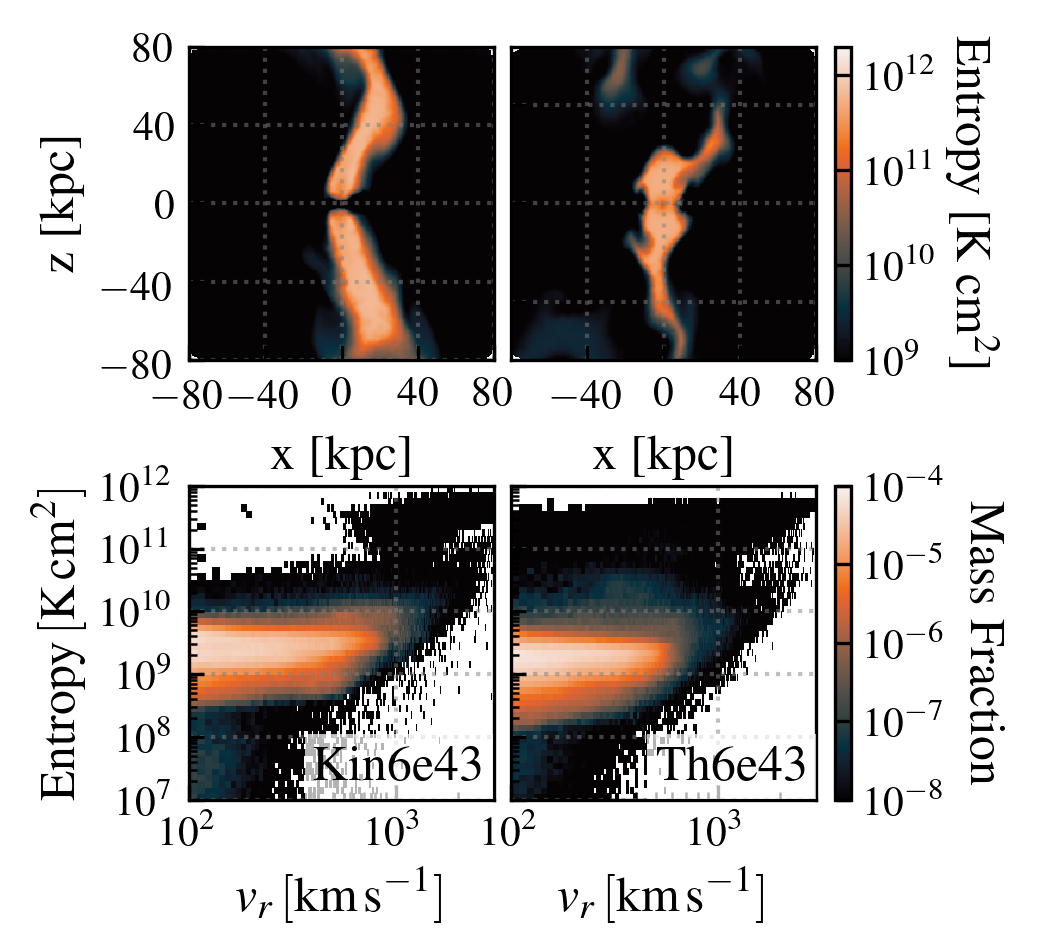}
\caption{{\bf Top row:} Entropy distribution of the kinetic jet and thermal jet runs with an equal energy flux of $6\times10^{43}{\rm erg\,s}^{-1}$ (a $\delta y=10$ kpc slice).  {\bf Bottom row:} Entropy and radial velocity distribution. The thermal jet causes a wider heated region in the core ($\lesssim 20$ kpc), while the kinetic jet propagates more efficiently to a larger radius. Despite the difference in jet velocity and jet temperature, both thermal and kinetic jets show a similar positive correlation between entropy and radial velocity in outflowing gas. }
\label{fig:morph_kin_thermal}
\end{figure*}

In fact, many properties of the outflows in both thermal and kinetic jet runs are very similar despite the very different initial jet temperature and jet velocity. The bottom row of \fref{fig:morph_kin_thermal} shows entropy vs. radial velocity plots of the thermal and kinetic runs with identical energy fluxes (`Kin6e43' and `Th6e43'). In both runs, the uppermost disconnected part (at high entropy) consists entirely of spawned jet elements. The kinetic jet elements are shock-heated to a similar entropy and temperature as the thermal jet elements, which are initialized with these values.  As expected, the post-shock velocity of the kinetic jet cells is slightly higher than that of the thermal jet cells, but it is important to note that even the ``thermal jets'' are moving with very large bulk velocity $>3000{\rm km\,s}^{-1}$ on kpc scales. Because the energetics and ram+thermal pressure of the jets are similar in both cases,  the bulk outflow velocity of the affected ICM is similar in both runs. 

Although the properties of entrained, out-flowing gas around the polar direction are similar in these runs, the distribution of inflow vs. outflow is quite different away from the poles.  Injecting energy in a predominantly  kinetic form generally transfers energy out to a large distance from the BH, in a narrower solid angle. As shown in the entropy slices (the top row of \fref{fig:morph_kin_thermal}), the kinetic jet (`Kin6e43') is narrower, especially at smaller radius, while a thermal jet with the same energy flux  (`Th6e43') more effectively heats the core region.  This is consistent with \Eqref{eq:vj_kin} and \Eqref{eq:fkin_nonkin}: an initially narrow kinetic jet ($v_{\rm J}=10^4\, {\rm km\,s}^{-1}$), has an effective opening-angle less than $45^{\circ}$ at the cooling radius ($\sim 30$ kpc), while the thermal jet with a small $f_{\rm kin}$ widens very quickly.  `Kin6e43' also generates a more continuous chimney, indicating a more efficient propagation to large radius. This is expected since the kinetic jets initially have a much higher velocity. Accordingly, kinetic jets mostly invert the inflow within a confined solid angle and only reduce the cooling flow proportionally.  On the other hand, thermal jets more uniformly heat up the core region $r\lesssim 30 {\rm kpc}$  and more effectively quench. Whether the substantially less-collimated thermal jets are theoretically and observationally realistic remains to be determined.

\subsection{Width of the jet cocoon determines how efficiently it quenches }
\label{s:kinetic}

Given that the effects of the kinetic jets are limited to a relatively small solid angle, the value of the solid angle can determine how effective the jet is at quenching. Keeping a similar solid angle but increasing the energy (or momentum) flux, on the other hand, has much smaller effects.
To produce a kinetic-jet-inflated cocoon with a wide enough solid angle to suppress the cooling flows, we either need a jet with $v_{\rm J} \gg 2\times 10^4 {\rm \, km\,s}^{-1}$ (following \Eqref{eq:vj_kin}), or a jet that is initialized with a wide opening-angle. Exploring the former possibility is limited by the maximum jet velocity we can adopt in our MHD simulations.

As we can see in the ``Mass loading (kinetic)'' panel of \fref{fig:sfr} and \fref{fig:hotgas}, despite very different jet velocities, specific energies and initial jet opening-angles, `Kin6e43',  `Kin6e44-B4', `Kin6e43-w15',  `Kin6e43-w30', and `Kin6e43-w45' share very similar core baryonic mass growth and SFRs.   `Kin6e44-B4' and `Kin6e43-wiso', on the other hand, are quenched and have lower core baryonic mass due to the larger opening-angle of the jet-inflated cocoon.

\begin{figure*}
\includegraphics[width=17.7cm]{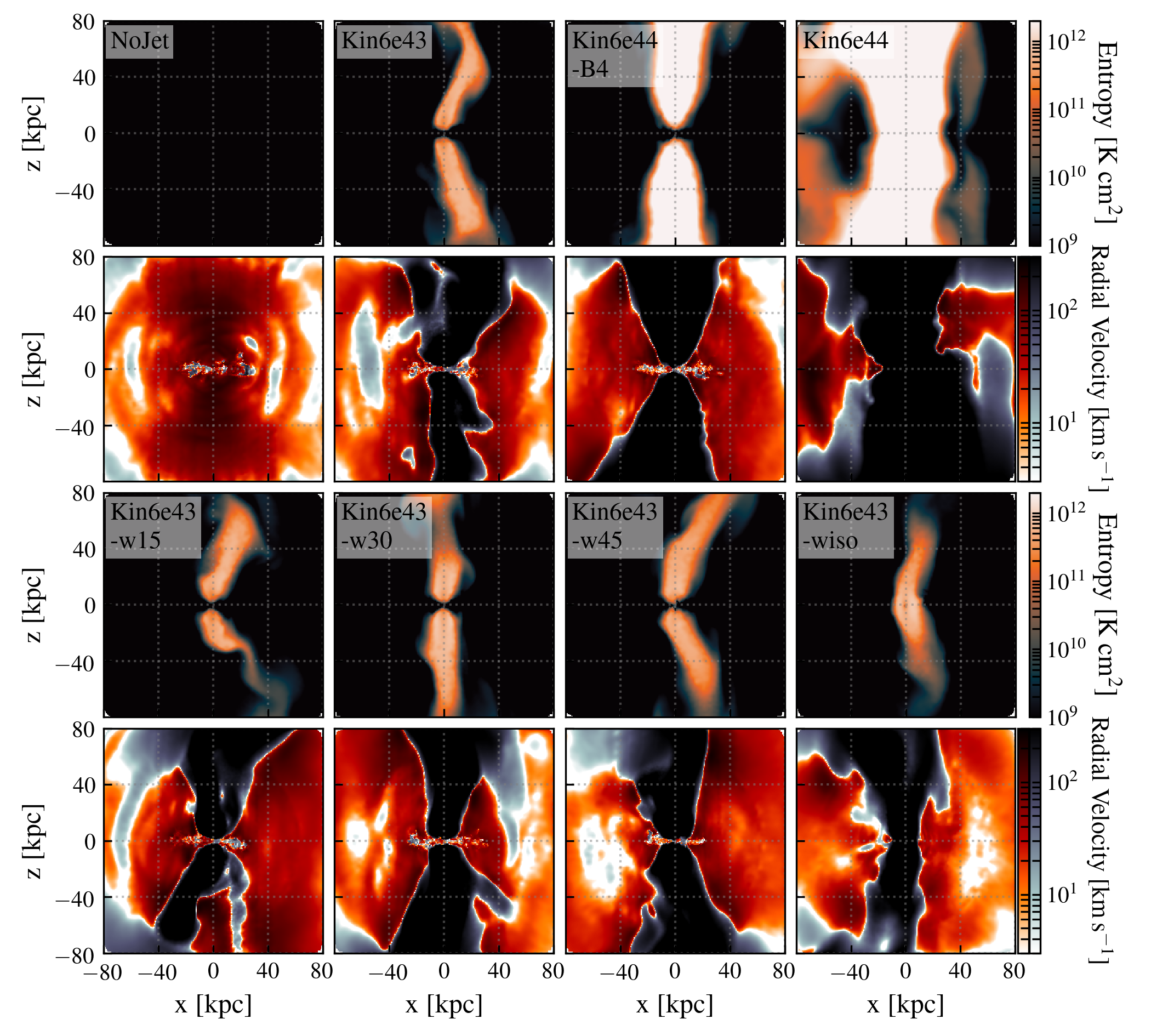}
\label{fig:morph_kin}
\caption{Entropy and radial velocity of some representative kinetic jet runs shown in a $\delta y=10$ kpc slice. The radial velocity is indicated separately for inflowing gas (red color scale) and outflowing gas (gray color scale).  Despite the different opening-angles,  and kinetic energy fluxes,  `Kin6e43', `Kin6e44-B4', `Kin6e43-w15', and `Kin6e43-w30' all have outflowing gas confined within a very similar solid angle. The entropy distribution shows larger differences. 
`Kin6e44' has a significantly wider opening-angle due to the magnetic fields, which is explained in \sref{S:magnetic}.}
\end{figure*}

\begin{figure}
\centering
 \includegraphics[width=8.2cm]{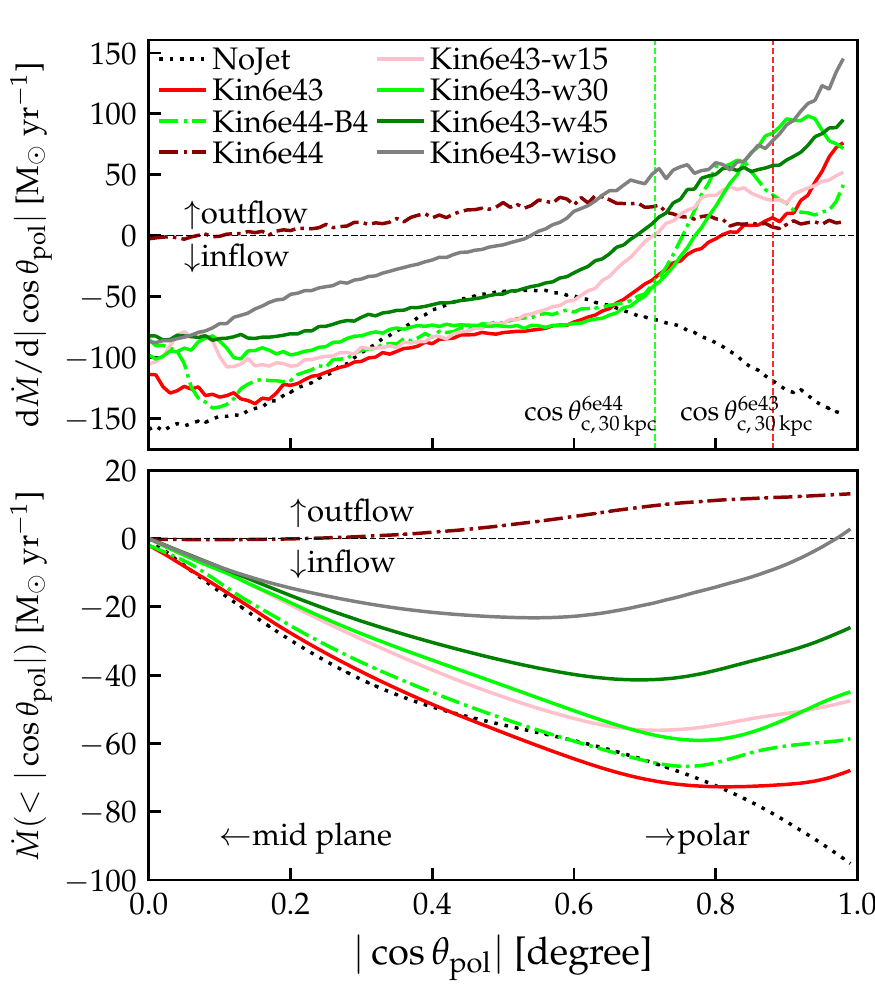}
\caption{{\em Top:} The distribution of radial gas inflow ($\dot{M}<0$) and outflow ($\dot{M}>0$) as a function of the cosine of the polar angle ($|\cos\theta_{\rm pol}|$) calculated in a shell of 25-35 kpc averaged over the last 50 Myr of each run.  
{\em Bottom:} The integrated value of the mass flux from  $|\cos\theta_{\rm pol}|=0$ to  $|\cos\theta_{\rm pol}|$. The vertical lines label the corresponding cocoon opening-angle at 30 kpc for the `Kin6e43' and `Kin6e44-B4' runs according to \Eqref{eq:angle_z}. Without jets, there is net inflow in all directions. With kinetic jets, we generically see outflow along the polar directions and inflow in the mid-plane, with net (angle-integrated) inflow. The outflow opening-angle increases as we increase the effective jet opening-angle, but even isotropic jets  (`Kin6e43-wiso') are collimated by the inflowing halo gas and feature inflows within $\sim30^{\circ}$ of the mid-plane (only the most violent jet here produces isotropic outflows).
}
\label{fig:inoutflow_kin}
\end{figure}

\fref{fig:morph_kin} shows the distributions of the entropy (1st and 3rd row) and the radial velocity (2nd and 4th row) of these seven runs and the `NoJet' run. Inflowing and outflowing velocities are indicated by red and gray color scales, respectively.  \fref{fig:inoutflow_kin} shows the inflow/outflow rates along different angles as a function of polar angle.
It is clear that although the entropy distribution differs from run to run, the velocity structures of the runs with similar baryonic inflows (`Kin6e43',  `Kin6e44-B4', `Kin6e43-w15', `Kin6e43-w30', and `Kin6e43-w45') look quite similar. In the `NoJet' case, gas is inflowing in all directions. In the five aforementioned runs, on the other hand, the velocity is all outflowing at small polar angle (closer to the jet), while at large polar angle (closer to the mid-plane), the velocity is all inflowing. This transition happens at 30$^{\circ}$-40$^{\circ}$ for all of the four cases. As labeled on the plot, the angle is roughly consistent with the angle derived from \eqref{eq:angle_z} at $r=30\,{\rm kpc}$ assuming $n\sim 0.01\, {\rm cm}^{-3}$. Even the `Kin6e43-wiso' run, despite the isotropic injection, is still collimated by the surrounding over-density in the midplane of the galaxy. However, the inflow at the core region of this run is shut down isotropically, consistent with the zero SFR at a later time.
In contrast, `Kin6e44' has a much wider region that is dramatically outflowing. The reason for this is that the jet has a non-thermal component  (in this case, magnetic) with comparable energy to the kinetic component, producing rapid broadening of the cocoon.


In \fref{fig:inoutflow_kin}, the inflows and outflows are calculated in a shell of 25-35 kpc, somewhat inside the cooling radius.\footnote{We note that the plotted values will vary if we choose a different radius, but the conclusions remain the same.}  We clearly see that runs which produce a wider opening-angle outflow  ``cocoon'' or bubble in \fref{fig:morph_kin}, and thus have outflow along a broader range of angles, have lower integrated inflow rates, which is consistent with their accreted baryonic mass and SFRs in \fref{fig:sfr} and \fref{fig:hotgas}. 

In brief, the jet drives shocks which inflate a cocoon, directly impacting gas within an opening angle that depends on the form of the jet.
Beyond this effective opening-angle, the cooling flows are much less affected. As a result, increasing the kinetic energy input does not necessarily mean more effective quenching if the affected solid angle is not enlarged because the maximum effect is expelling all the gas in that cone. Consistent with this, the `Kin6e44-B4' run has very similar inflow and outflow structures as `Kin6e43' in \fref{fig:inoutflow_kin} despite having an order of magnitude higher kinetic energy flux.
However, enlarging the initial kinetic jet opening-angle while keeping the total energy flux the same can potentially make the jet quench the galaxy more efficiently.

Remarkably, if we use \Eqref{eq:z_iso} to estimate the scaling of the cocoon opening-angle with the jet $\dot{M}$ and $\dot{E}$ or $v_{\rm J}$, we obtain the same scaling for opening-angle as seen in \fref{fig:inoutflow_kin} for the initially ``narrow'' jets (vertical dashed lines).

\subsection{Why the CR jet quenches more efficiently}
\label{s:CR_th}

\begin{figure}
\centering
 \includegraphics[width=8.4cm]{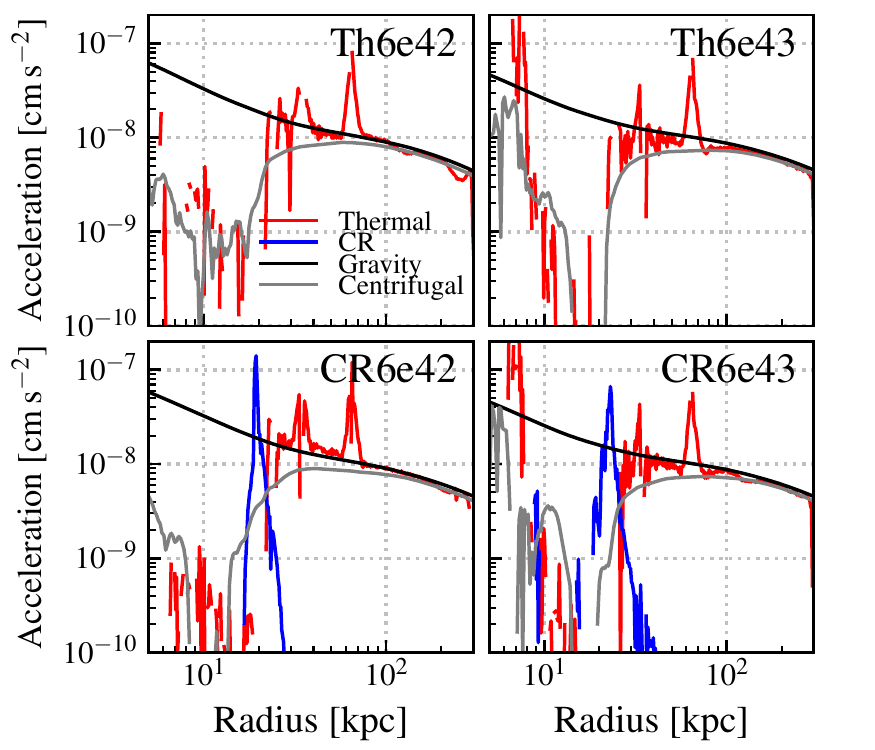}
\caption{Comparison of gravitational, rotational, thermal pressure and CR pressure gradient acceleration. The centrifugal acceleration is defined as $GM_{\rm enc}/r^2-v_{\rm rot}^2/r$. In the core region, where cooling is rapid, the thermal pressure gradient is not outward and support is lost. In our CR runs, the CR pressure gradient predominantly balances gravity in the core region. }
\label{fig:pres_gr}
\end{figure}

\pfh{We find that cosmic ray (CR)-dominated jets quench more efficiently, and potentially more stably, than thermal, kinetic, and magnetic jets. We argue that this is due to three factors: (i) CR pressure support, (ii) modification of the thermal instability, and (iii) CR propagation. Injected CRs provide pressure support to the gas and have long cooling times, which leads to the formation of a CR pressure-dominated cocoon. Because the CR energy density is much larger than kinetic energy density, the CR jet cocoon covers a wider angle (as expected), and can therefore more efficiently suppress inflow. If the CR losses are negligible and CRs become quasi-isotropic, with an effective isotropically-averaged diffusivity $\tilde{\kappa}$ (which includes streaming+advection), then as shown in various studies \citep{Buts18,hopkins:cr.mhd.fire2,2020MNRAS.496.4221J,2021MNRAS.501.3640H,2021MNRAS.501.3663H,2021MNRAS.501.4184H} the CR pressure for steady-state injection is $P_{\rm CR}(r) \sim \dot{E}_{\rm CR} /12\pi\,\tilde{\kappa}\,r$. Comparing the outward acceleration $\rho^{-1}\nabla P$ to gravity ($\sim v_{c}^{2}/r$), where $v_c$ is the circular velocity,} CR pressure alone can support the gas if
\begin{align}
    \label{eqn:EdotCR} \dot{E}_{\rm CR} &\gtrsim  10^{43}\,{\rm erg\,s^{-1}}\,\notag\\
    &\left( \frac{10^{29}}{\tilde{\kappa}}\right)\,\left( \frac{n_{\rm gas}}{0.01\,{\rm cm^{-3}}}\right)\,\left(\frac{r}{30\,{\rm kpc}}\right)\,\left(\frac{v_{c}}{500\,{\rm km\,s^{-1}}} \right)^{2}
\end{align}
\pfh{(where we recall the cooling radius is $\sim 30\,$kpc here and the diffusivity used here is $10^{29}\,{\rm cm^{2}\,s^{-1}}$; see \citealt{chan:2018.cosmicray.fire.gammaray,hopkins:cr.mhd.fire2,2021MNRAS.501.4184H}). This roughly explains the required CR energetics we find in our simulations, as well as the radii/densities where CR pressure dominates for a given $\dot{E}_{\rm CR}$.} 
This is also consistent with the comparison of the total centrifugal acceleration and the acceleration due to CR and thermal pressure gradient as shown in \fref{fig:pres_gr}.\footnote{We used the median pressure (weighted by mass) in each radial bin to calculate the pressure gradient in \fref{fig:pres_gr} to better show the difference.}

\begin{figure}
\centering
 \includegraphics[width=8.2cm]{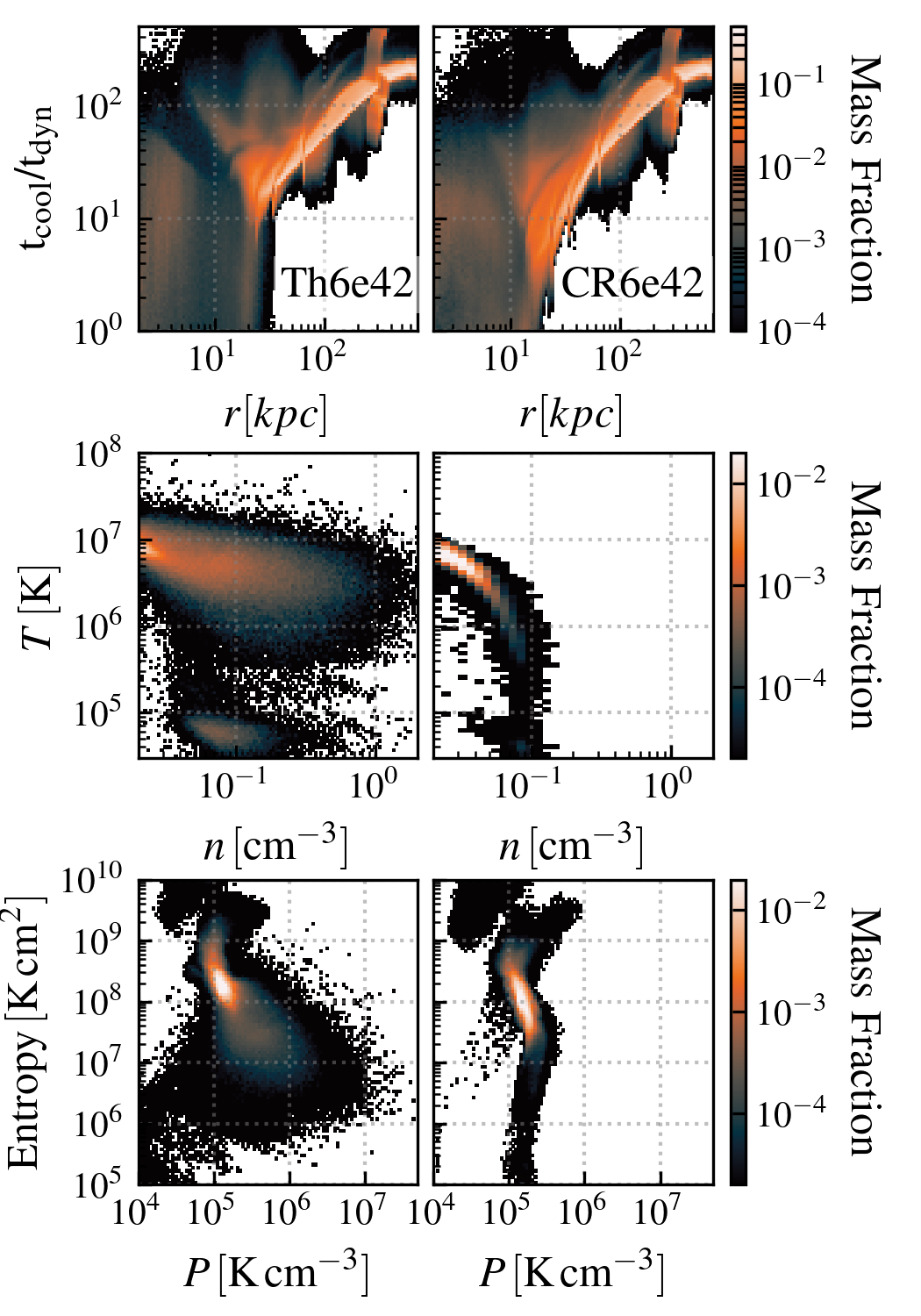}
\caption{ Ratio of the cooling time to dynamical time ($t_{\rm cool}/t_{\rm dyn}$) (top row) and the phase structure for 15-30 kpc in the temperature-density plane (middle row) and entropy-pressure (thermal + CR) plane (bottom row) for the thermal and CR jet runs. The plots are averaged over 0.95-1 Gyr of each run. 
In the CR  runs (`CR6e42'), there is more warm, thermally-unstable gas residing at intermediate densities (and at $\sim 10-30$ kpc) without accreting (e.g., our `CR6e42' run) and the formation of the dense, cold phase appears somewhat delayed.
 The cosmic ray jet run also has gas following a constant total pressure path with narrow density variation while cooling. In the thermal jet run, there is generally a wider density distribution.}
\label{fig:phase_CRth}
\end{figure}

\pfh{CR-dominated jets can also help quench by modifying the non-linear behavior of the thermal instability, as suggested in \citet{2020MNRAS.496.4221J} and shown rigorously in \citet{2020ApJ...903...77B}. In brief: {\em if} CR pressure balances gravity and dominates over thermal pressure in a thermally-unstable medium, then cooling gas follows total pressure equilibrium (not just thermal pressure equilibrium) and cooling gas can remain diffuse (rather than being compressed to high densities; i.e.,\ the cooling changes from isobaric to isochoric), as shown in \fref{fig:phase_CRth}.} This, in turn, slows the ``precipitation'' of dense, cold gas from the cooling flow that would otherwise accrete \citep{2017ApJ...845...80V}, allowing it to instead remain diffuse and supported by CR pressure. We see indirect evidence for this (in \fref{fig:phase_CRth}) in our CR runs as more warm thermally-unstable gas resides at intermediate densities (and at $\sim 10-30$ kpc radii) without accreting (e.g., our `CR6e42' run) and formation of the dense, cold phase appears somewhat delayed. There is also a slight thermal instability difference in the cosmic ray jet runs with different fluxes. We discuss this further in \aref{a:cr}.

Heating from cosmic rays plays a smaller role in quenching. For all of our CR jet runs,  CR collisional heating and streaming heating contribute at most $1/20$ and $1/6$ of our CR injection rate, respectively. This amount of heating should not have a major effect on quenching, as we can see in the thermal jet runs with such a corresponding energy flux.

\pfh{The ability of CRs to stream or diffuse (i.e.,\ large $\tilde{\kappa}$ above) relative to the gas is crucial to these behaviors. If CRs were purely advected with the gas, they would simply represent slowly-cooling internal energy. But because CRs can stream through gas, their pressure profile operates akin to a fixed background, which means that if we increase $\dot{E}_{\rm CR}$ further, the behavior is not ``explosive.'' Specifically, as shown in \citet{2021MNRAS.501.3640H}, even if $|\nabla P_{\rm CR}| \gg \rho\,|{\bf a}_{\rm grav}|$, then although CR pressure is sufficient to drive gas outflows, these outflows are weak. {\em Independent} of $P_{\rm CR}$ or $\dot{E}_{\rm CR}$, they rapidly accelerate gas up to a terminal velocity $\sim V_{c}$, i.e.,\ trans-sonic with respect to the hot halo gas, which then ``coasts'' in the outer halo and beyond. In comparison, a conventional continuous-injection pressure-driven blastwave generally produces hypersonic outflows (and accelerates more rapidly in a declining density profile as is typical of outer halos).} 
CR diffusion is more effective than thermal conduction (except at high temperature),  making both `CR6e44' and `CR6e43-B4-m2e-1'  less overheated than the corresponding `Th6e44' and `Th6e43-B4-m2e-1', as shown in \fref{fig:etd_sus1} and \fref{fig:etd_sus2}.

\subsection{The effects of Magnetic fields}
\label{S:magnetic}
Magnetic fields usually only have limited effects (factor $\lesssim 2$) in quenching the galaxy or suppressing cooling flows. 
The  exception is `Kin6e44', where the galaxy is  quenched while an otherwise identical simulation with an order of magnitude lower magnetic field strength (`Kin6e44-B4') has strong cooling flows and high SFRs.

In `Kin6e44', the magnetic energy input is $\sim 2-3\times 10^{44}$ erg s$^{-1}$, similar to the kinetic energy input.  We therefore expect the non-kinetic pressure to broaden the jet cocoon (\Eqref{eq:fkin_nonkin}) and indeed, we see a much wider jet cocoon which produces a wide-angle outflow.   
In other runs, the highest magnetic energy input rate is $\sim 10^{43}$ erg s$^{-1}$, insufficient to strongly broaden the cocoon. %

\begin{figure}
\includegraphics[width=8.2cm]{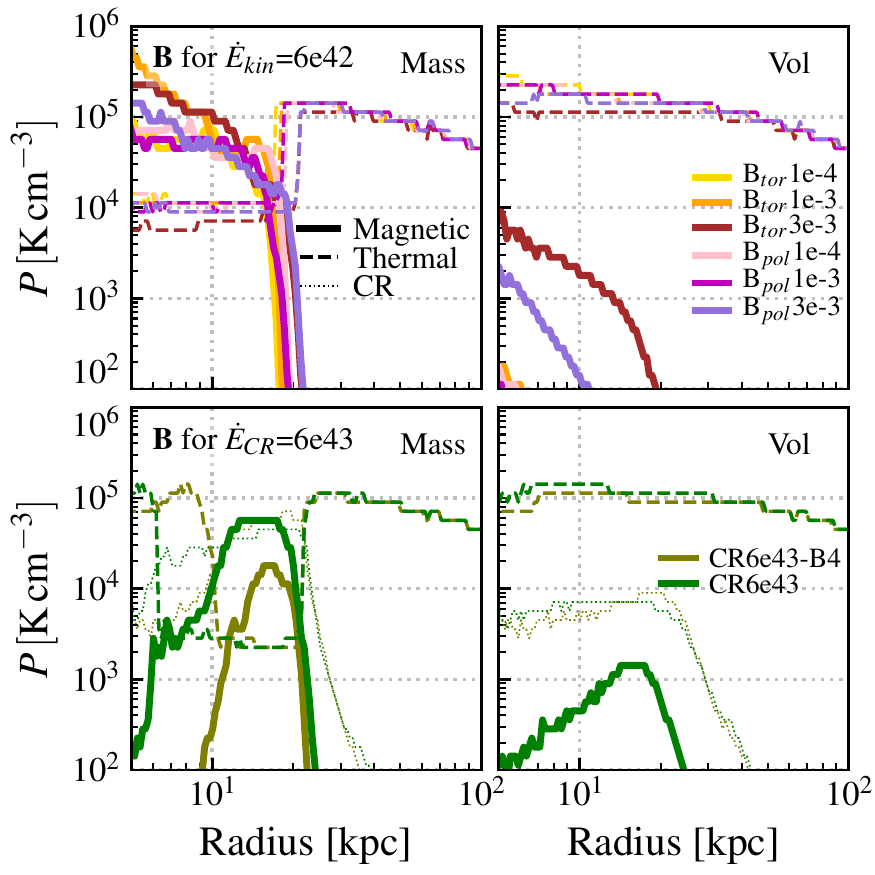}
\label{fig:pres_profile}
\caption{The volume and mass-weighted thermal (thin dashed), magnetic (thick solid) and CR (thin dotted if applicable) pressure profiles for runs with different jet magnetic field variations. The magnetic pressure can only be comparable to the thermal pressure in the 10-30 kpc range and then, only in dense cool gas, given that the volume-weighted values are much lower than the mass-weighted ones. 
At radius $\gtrsim100$ kpc (where jet effects are weak), or wherever the B-fields are a large fraction of the total pressure (i.e., appear to have saturated), $|\bar{B}|$ is weakly sensitive to the injected fields.}
\end{figure}

The direct effect of gas acceleration by magnetic pressure is weak. Magnetic pressure is only high in the core region and dense structures, as shown in \fref{fig:pres_profile}. The magnetic pressure can be higher than the thermal pressure within $\sim20$ kpc if mass-weighted but is always subdominant to thermal pressure if volume-weighted. This indicates that magnetic field strengths  are only high in the dense cooler gas, and in that gas the B-fields appear saturated (independent of injected field strengths).  Moreover, in those dense structures, the cooling is already effective, and the extra magnetic pressure cannot do much. In the regions where the density is relatively low, the magnetic pressure support is less important.

Outside of these dense regions, we do see slightly broader cocoons (hence more efficient quenching) for jets with toroidal (vs. poloidal) fields, at the same initial strength (compare e.g., `B$_{\rm tor}$1e-4 ' and `B$_{\rm pol}$1e-4 ' in \fref{fig:pres_profile}). This is most likely due to toroidal fields being less of an impediment to perpendicular cocoon expansion.
This difference may also partially result from enhanced amplification, as the galactic field is also toroidal near the injection site.

\begin{figure*}
\includegraphics[width=16cm]{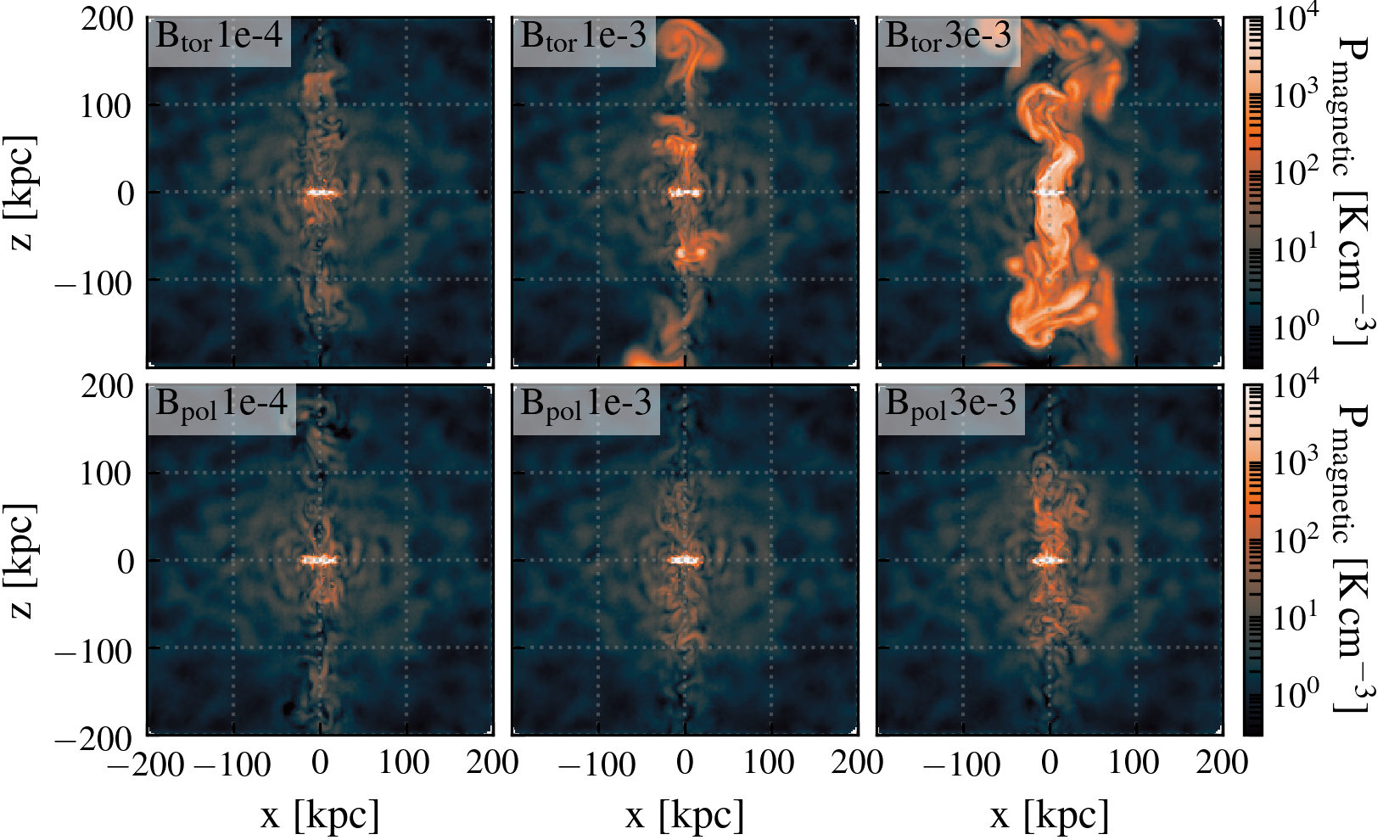}
\label{fig:morph_kin_mag}
\caption{The magnetic pressure morphology shown in a $\delta y=10$ kpc slice of the runs with the same kinetic jet, $6\times 10^{42} {\rm erg\, s}^{-1}$, but with different jet magnetic fields. Beyond $\sim30$ kpc, the magnetic pressure is high only close to the z-axis, where the gas is directly affected by the jet. With similar initial magnetic field strengths, a jet with initially toroidal magnetic fields is able to maintain larger  magnetic field strengths  further along the jet axis, compared to  a jet with poloidal magnetic fields.}
\end{figure*}

\subsection{Mass flux and Energy Loading}
We also tested how the cooling flows and SFRs differ when using the same total energy in a given form but with different mass flux and energy loading.  Given a fixed total energy, the lower the mass flux (i.e., the higher the specific energy of the jet), the more effective the quenching is, as expected from simple analytic predictions of how rapidly the cocoon can inflate (\Eqref{eq:angle_z} and \Eqref{eq:fkin_nonkin}).

\begin{figure*}
\centering
 \includegraphics[width=16cm]{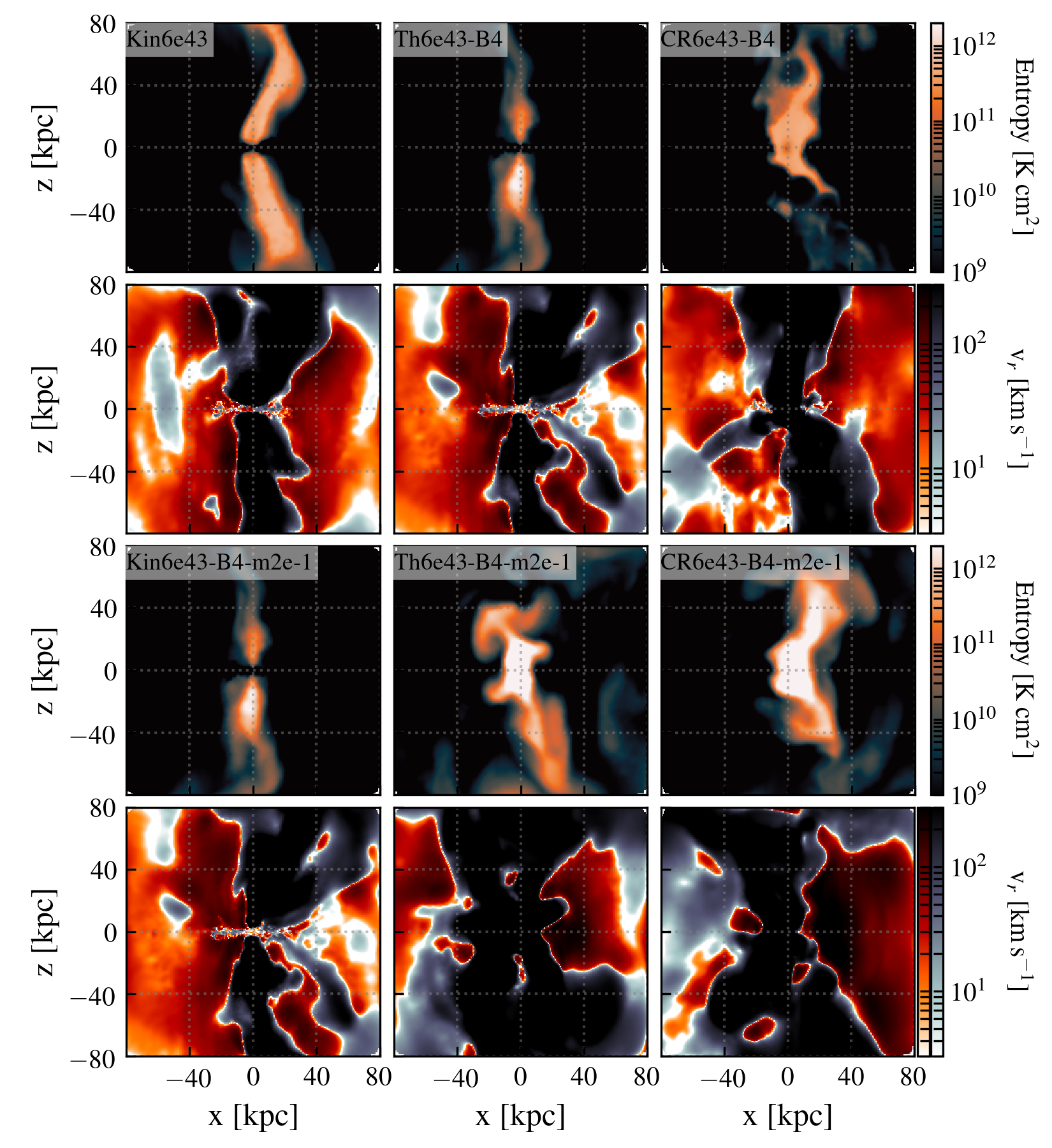}
\caption{Entropy and radial velocity morphology for runs with different mass fluxes while keeping the same energy flux, shown in a $\delta y=10$ kpc slice. With the same thermal or CR energy flux of $6\times10^{43}$ erg s$^{-1}$, a run with lower mass flux but higher energy per unit mass results in a wider jet cocoon. The situation is not as evident in the cases with lower energy thermal jets or kinetic jets.}
\label{fig:morph_massload}
\end{figure*}

As we can see in \fref{fig:sfr} ,`Th6e43-B4-m2e-1'  quenches more effectively than `Th6e43-B4'. The former jet has an order of magnitude lower mass flux than the latter despite having the same thermal energy flux. `Th6e42-B4-m2e-2' also quenches slightly more efficiently than `Th6e42-B4-m2e-1' and `Th6e42'.
The major difference between jets with different specific thermal energies and mass fluxes is the cocoon width, as shown in \fref{fig:morph_massload}, especially in the 6$\times10^{43}$ erg s$^{-1}$ runs. `Th6e43-B4-m2e-1' has a wider solid angle for which the gas is outflowing than for the `Th6e43-B4' run.
We see the same comparing CR runs (`CR6e43-B4-m2e-1' and `CR6e43-B4' ). In all of these cases, the width of the low-density evacuated cocoon scales roughly $\propto (\dot{E}_{\rm tot,\,J}/\dot{M}_{\rm J})^{3/2}$, as  expected from \Eqref{eq:angle_z}. The effects in the low-energy thermal jet runs or pure kinetic runs are weaker, as the cooling flows are not strongly suppressed (the cocoon has insufficient energy to grow). As shown in \Eqref{eq:vj_kin}, a purely kinetic jet with $\dot{M}=2 \, {\rm {\rm M}_\odot\,yr}^{-1}$ needs a velocity $ > 2\times10^4 \, {\rm km\,s}^{-1}$ to have a sufficient width at the cooling radius.

\subsection{Duty cycle}
\label{dis:duty}
\begin{figure*}
\centering
 \includegraphics[width=17.7cm]{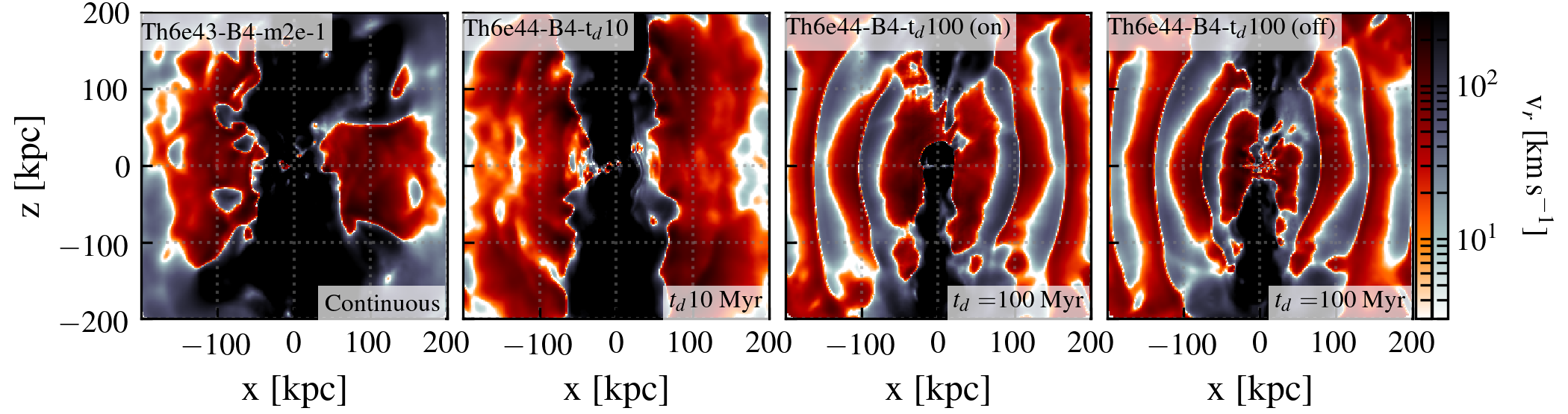}
\caption{The radial velocity field for thermal jet runs with the same time-averaged energy flux $6\times 10^{43}$ erg s$^{-1}$ within a $\delta y=10$ kpc slice, but with different duty cycles (continuous, versus ``on'' for $\sim10\%$ of the time with a period of 10 Myr or 100 Myr, as labeled). The results are shown at the end of each simulation. The run with a 10 Myr period looks effectively similar to the continuous run. Both of them eventually shut down the inflows at the core region of the galaxy completely. The run with a 100 Myr period has concentric shells of inflows and outflows, showing the previous cycles. The inflows at the core region rebuild again when the jet is off.}
\label{fig:morph_duty}
\end{figure*}

We also tested the difference between a jet with constant  mass and energy flux and jets with various duty cycles but the same averaged mass and energy flux.  We found that the run with a  $\sim100$ Myr episodic period and a $\sim10\%$  duty cycle (`Th6e44-B4-t$_d$100')  is less effective than the run with a shorter period and the same duty cycle (`Th6e44-B4-t$_d$10') or the continuous run `Th6e43-B4-m2e-1'.  
The two latter runs turn out to be very similar, as shown in \fref{fig:morph_duty}.
`Th6e44-B4-t$_d$100', on the other hand, has layers of inflows and outflows indicating different episodes of the jet. 
As shown in \fref{fig:hotgas}, the core baryonic mass of `Th6e44-B4-t$_d$100' takes roughly 100 Myr  to recover in each duty cycle while remaining overall constant on average. In comparison, the dynamical time at the region where multiphase gas starts to form ($\sim 30$ kpc) is roughly 10 Myr and the cooling time for the hot gas at the same radius is $\gtrsim 100$ Myr, so the cooling flow has time to recover when the jet is off. On the contrary, in the run with a 10 Myr period, both the cooling time and dynamical time of the gas around the same region are larger than or equal to the period, so the effect is approximately the same as a continuous jet.


The jet with visible duty cycles (`Th6e44-B4-t$_d$100') has a weaker effect than a continuous jet with the same average flux (`Th6e43-B4-m2e-1') most likely owing to threshold effects. 
The continuous run with 10x higher energy (`Th6e44-B4') quenches similarly to the continuous run (`Th6e43-B4-m2e-1'), so increasing the flux in the ``on'' state does not increase the efficiency of quenching dramatically. This is likely due in part to the fact that the perpendicular expansion of the cocoon is sub-linear in time and $\dot{E}$. Additionally, since cooling times are not much longer than 100 Myr, some of the injected energy is lost in each ``off'' cycle. 

\subsection{Jet Precession}
\label{dis:pres}

We experimented with different precession angles and precession periods and find that the dominant effect of the precessing kinetic jet is still shock heating the surrounding gas and suppressing the inflows or pushing the gas outward within a specific solid angle. Thus, making an otherwise narrow cocoon ``efficient'' requires precession angles $\gtrsim 30-45^{\circ}$ so that the cocoon can become effectively quasi-isotropic.  

\begin{figure*}
\centering
 \includegraphics[width=16cm]{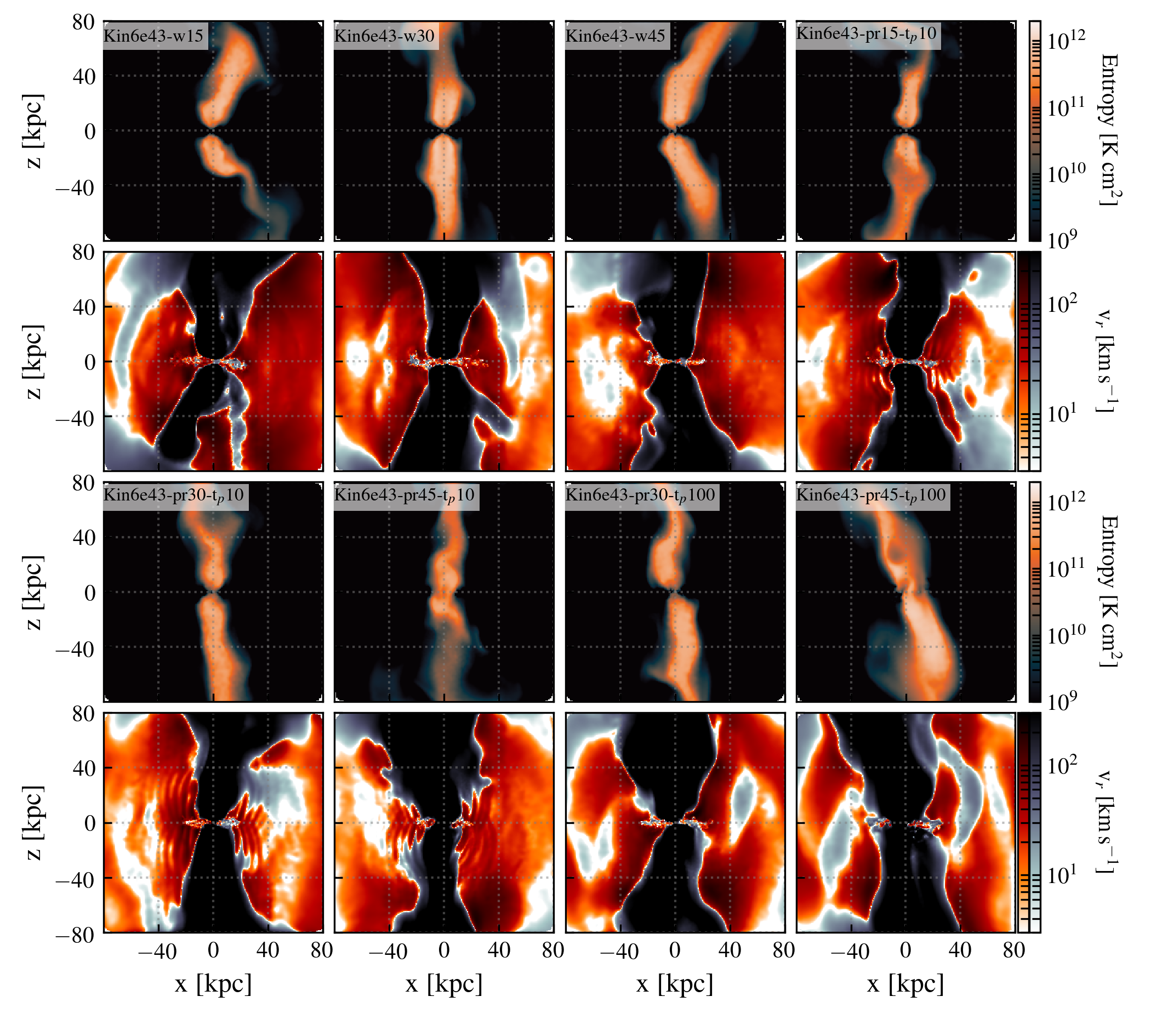}
\caption{The entropy and radial velocity morphology for runs with different jet opening-angles, precession periods, and precession angles, shown in a $\delta y=10$ kpc slice. The runs with a wider opening-angle and/or  more extended precession (`Kin6e43-pr30-t$_p100$', `Kin6e43-pr45-t$_p100$', `Kin6e43-pr45-t$_p10$', and `Kin6e43-w45') have a slightly wider solid angle for outflowing material at the core region of the galaxy, consistent with their lower cooling flows and SFR.}
\label{fig:width_pres}
\end{figure*}

Specifically, we see that when the precession period is around 10 Myr, the jet becomes more effective only after the precession angle reaches $\gtrsim 45^{\circ}$ ('Kin6e43-pr45-t$_p$10'), where the SFR is lower (see \fref{fig:sfr}) and the cocoon becomes effectively wider than `Kin6e43' (\fref{fig:width_pres}).  
When the period is 100 Myr, a slightly smaller 30$^{\circ}$ precession ('Kin6e43-pr30-t$_p$100') is sufficient to suppress the SFR instead of 45$^{\circ}$. Consistently, as shown in \fref{fig:width_pres}, 'Kin6e43-pr30-t$_p$100' has a wider ``effective'' solid angle than 'Kin6e43-pr30-t$_p$10' in the core region.

A non-precessing wide jet (`Kin6e43-w15', `Kin6e43-w30', and `Kin6e43-w45') can be viewed as a high-speed precessing jet, since the spawned cells are sampling the opening-angle, with an effective period $\ll$ Myr. Consistent with the cocoon behaviour discussed above, although the SFR starts to drop when the opening-angle reaches 45$^{\circ}$, `Kin6e43-w45' has a higher SFR and stronger cooling flows than the precessing jet with $45^{\circ}$ precession angle. Therefore, an opening-angle between $45^{\circ}$ and isotropic should be required to reach a similar level of quenching effect.

We also see a factor of $\lesssim 2$ boost in the turbulent velocity at $\sim$ 10-70 kpc in the runs with a 100 Myr precession period when the opening-angle reaches 30-40$^{\circ}$. So precessing jets can stir some turbulence, but this, by itself is nowhere near the level of turbulence required to quench (see \citetalias{2020MNRAS.491.1190S}).  

The reason for these ``second order'' trends is that, all else equal, a more slowly precessing jet can more efficiently expand  to a sufficiently large radius. Since it stays in each direction for a longer time, it `clears out' a path before moving to another direction. 

\begin{figure}
\centering
 \includegraphics[width=8.2cm]{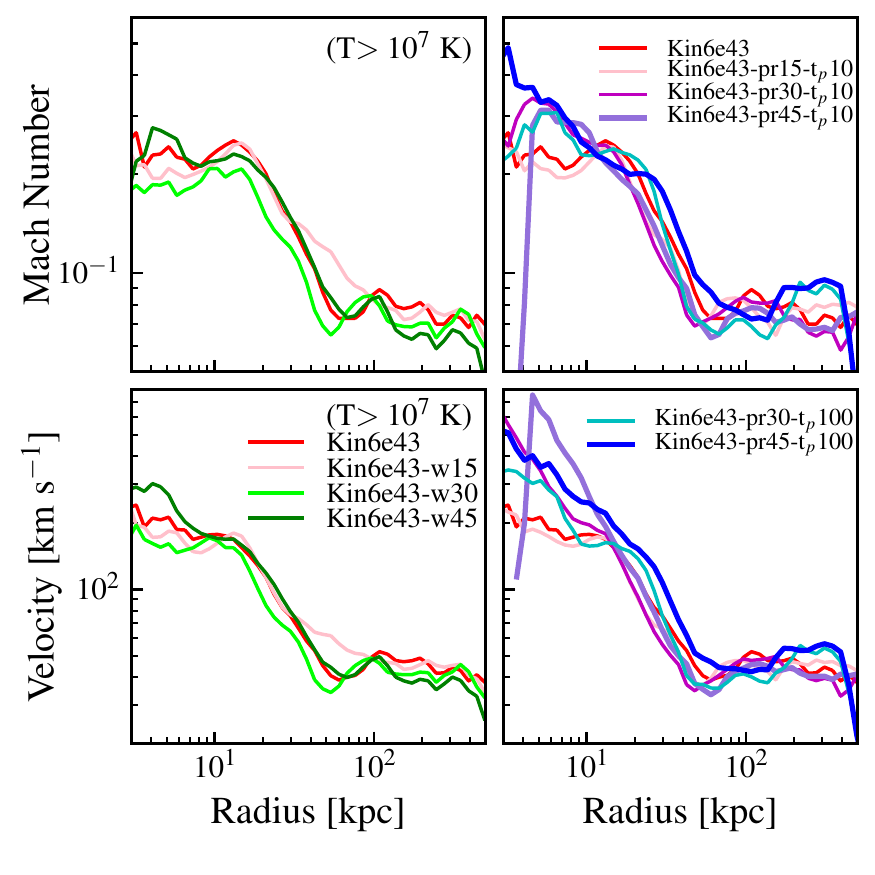}
\caption{As \fref{fig:turb}, 1D rms Mach number ($v_{\rm turb}^{\theta,\phi}/\sqrt{2} v_{\rm thermal}$; top), and dispersion ($v_{\rm turb}^{\theta,\phi}/\sqrt{2}$; below) in gas with $T>10^{7}\,$K, averaged over the last 50\,Myr of the runs, as a function of radius for the runs with different widths, precession periods and precession angles from \fref{fig:width_pres}. Jet models with a long precession period and a sufficiently wide precession angle can slightly boost the turbulent velocity at 10-70 kpc by a factor of $\lesssim 2$, indicating more efficient turbulent ``stirring''.}
\label{fig:turb_pres}
\end{figure}

\section{Comparisons and Caveats} 
\label{S:discussion}

\subsection{Comparison with \citetalias{2020MNRAS.491.1190S} }

\pfh{In \citetalias{2020MNRAS.491.1190S} \citep{2020MNRAS.491.1190S}, we considered the same ICs, with AGN toy models using 4 different mechanisms: CR injection, thermal heating, radial momentum injection, and turbulent stirring with varied $\dot{E}$ and radial distribution of the injection (i.e.,\ these did not follow any physical propagation model). The CR-dominated jets here produce broadly similar results to the simpler CR-injection runs in \citetalias{2020MNRAS.491.1190S}, as the CR-dominated cocoon becomes quasi-isotropic and dominates the dynamics. The thermal jets here are qualitatively different from the thermal heating runs in \citetalias{2020MNRAS.491.1190S}: here, extremely hot, low-density, slow-cooling plasma is injected, which inflates a quasi-isotropic cocoon providing pressure and bouyancy support to halt/reverse inflows inside $R_{\rm cool}$. In \citetalias{2020MNRAS.491.1190S}, the ``thermal heating'' was applied as a direct heating term to the {\em pre-existing} gas in the galaxy or halo: this requires far more energy to offset cooling ``directly'' (since it requires re-heating dense cool-phase gas where cooling is rapid) and is far more unstable, as either the applied heating is less than cooling (in which case the gas still cools) or is greater than cooling (in which case a Sedov-Taylor-type explosion immediately results in ``overheating''). The kinetic jet models here are also distinct from, although in some ways in-between, the momentum-injection and turbulent-stirring runs from \citetalias{2020MNRAS.491.1190S}. Our widest-angle kinetic jets are somewhat akin to the isotropic radial momentum injection runs in \citetalias{2020MNRAS.491.1190S}, but less explosive owing to the cocoon dynamics that occur here (and not in \citetalias{2020MNRAS.491.1190S} owing to the isotropy and effectively large mass-loading of the coupling). None of our models here produces turbulent power approaching the level identified in \citetalias{2020MNRAS.491.1190S} as required for quenching from ``pure turbulence'' effects.}

\subsection{Comparison with other simulations }
We place our results in context through several comparisons as follows (although there is too much previous work to make the comparison comprehensive):

The fiducial model in \cite{2014ApJ...789..153L,2014ApJ...789...54L} has the energy flux equally distributed between thermal and kinetic energy. They also tested various jet models with a varied balance between kinetic and thermal energy  and different  efficiencies. They found, like us, that feedback stronger than the favored value causes overheating in the core region. 
They also found that with a pure kinetic model, the cocoons are narrower and less pressurized.  However, they suggested that the exact kinetic fraction versus thermal fraction does not significantly alter galaxy evolution, while we found that thermal jets are more effective in quenching. 
This likely owes to the fact that they do not fix the jet energies: in their runs, when a specific jet model is not as efficient in quenching at the same energy flux, the accretion rate and energy flux rise to compensate.  
 Kinetic jets are also shown to be able to quench the galaxy, reach self-regulation, and maintain the cool-core properties across a range of halo mass in   \cite{2011MNRAS.411..349G,2011MNRAS.415.1549G,2012MNRAS.424..190G,2012ApJ...746...94G}. This probably occurs for the same reason discussed above, and also the frequently higher jet velocity in their model. 

In \cite{2017MNRAS.472.4707B}, the authors  discussed the CGM turbulence caused by kinetic jets and compared with the CGM turbulence caused by substructures. The authors found that the jet is mostly responsible for the smaller scale turbulence around its edges, but not as effective in inducing larger-scale turbulence. We find a qualitatively similar result that even our most widely precessing jets boost the turbulence at a large radius at most by a factor of 2, insufficient to quench on its own \citepalias{2020MNRAS.491.1190S}.

Cosmic ray jets have been studied by \cite{2020MNRAS.493.4065W} (in a $4\times 10^{13} {\rm M}_\odot$ halo) and \cite{2017ApJ...844...13R} and  \cite{2019ApJ...871....6Y} in a more massive Perseus-mass cluster. The cosmic ray energy flux we find to stably quench  a $10^{14} {\rm M}_\odot$ halo ($\sim 6\times10^{42}$-$\sim 6\times10^{43} {\rm erg s}^{-1}$) in this work is roughly consistent with the energy range suggested in  \cite{2020MNRAS.493.4065W}, given their slightly less massive system. Also consistent with \cite{2019ApJ...871....6Y}, we find that a cosmic ray dominated jet is generally more efficient at quenching and results in a wider bubble compared to a kinetic jet.
Like \cite{2017ApJ...844...13R}, we find that CR pressure plays a key role regulating the cooling flows, and that including CR motion relative to the gas is the key. Otherwise, if CRs were purely advected, they would behave explosively like thermal energy.  Note that we parametrized the CR motion relative to the gas as diffusion, but within our approximations, this is mathematically identical to their super-Alfv\'en streaming.

\subsection{Possible further observational probes}
\label{s:observation}
To further constrain the models here, a more detailed analysis of X-ray properties and further comparisons with the multi-phase observations will be required. We leave this for future work, but briefly comment on directions we think would be fruitful. Although various models in this work are broadly consistent with the X-ray inferred radially averaged density, temperature, and entropy profiles, the detailed spatial distribution of these properties may vary, especially between the region closer to and further from the jet axis. These can be further constrained by more extensive X-ray map comparisons. 
Likewise mapping the kinetic properties (inflow/outflow and turbulent velocities vs. polar angle) near the jet can further constrain the models. Given that the gas properties at very large radii are dominated by the initial conditions in isolated galaxy simulations, thermal and kinetic Sunyaev–Zeldovich \citep{1970Ap&SS...7....3S} properties might not be as sensitive to the jet model. As shown in \fref{fig:phase_CRth}, the thermal properties of lower temperature gas can differ within $\sim30$ kpc between runs with cosmic-ray and thermal jets of different energy flux. These will predict different column densities of various ions in different phases.

We have verified that, in our CR jets, the predicted $\sim$GeV gamma-ray luminosity from hadronic interactions is below the current observational upper limits \citep{2016ApJ...819..149A,2019MNRAS.488..280W}. `CR6e43' and `CR6e43-B4-m2e-1' have $L_\gamma\sim1-3\times 10^{41}{\rm erg\,s}^{-1}$. In `CR6e42' $L_\gamma$ grows from $\sim10^{41}{\rm erg\,s}^{-1}$  to $\sim1-2\times 10^{42}{\rm erg\,s}^{-1}$, roughly at the upper limit. The values in the latter case are higher due to the denser core that develops. Likewise, the estimated $\sim$ GHz radio luminosity  from secondary CR electrons is well within the observational constraint from the radio flux assuming all the secondary CR electrons decay via synchrotron emission \citep[e.g.,][]{2014ApJ...781....9G,2016arXiv160300368B}.\footnote{The gamma-ray energy flux per volume is roughly $\dot{e}_\gamma\sim5/6*\Lambda_{\rm had}e_{\rm CR} n$, where $\Lambda_{\rm had}\sim 7.51\times10^{-16} s^{-1}$ is the hadronic coupling coefficient, $e_{\rm CR} $ is the CR energy density, and $n$ is the number density. The radio flux per volume in GHz from secondary electrons is roughly $\dot{e}_{\rm radio}\sim f_{\rm GHz}(2/3)(1/4)(5/6)\Lambda_{\rm had}e_{\rm CR} n$, where $f_{\rm GHz}\sim 0.01$ is the fraction of energy flux in the GHz band.  } 
To make more detailed predictions for radio emission, explicitly modeling the cosmic ray electrons will be required.

\subsection{Limitation of our models}
\label{s:caveat}
We emphasize that we are, by design, testing jet models with constant flux in a fixed initial cluster configuration. Our model does not include dynamically-variable black hole accretion, so is not ``self-regulating''. Although we explore the effect of duty cycles, we expect a self-consistently fueled jet will have a more complicated duty cycle. With these limitations in mind, the less dramatic ``overheated'' models we considered  (like `Th6e43-B4-m2e-1' or `Kin-6e43-iso') may be allowed if the jet only lasts for a  shorter duration.  The most overheated models like `Kin6e44' or `Th6e44', on the other hand, may still result in tension with observations even if the jet is on only for a short time episodically, like what we see in our duty cycle test `Th6e44-B4-t$_d$100'.

Another limitation of this work is the lack of a cosmological environment in our simulations. 
For example, there are no satellites/substructures in our halo, which can alter the large scale turbulence and other ICM gas properties.

We also note that we try to test each form of energy flux broadly within the plausible range instead of attempting to match any specific theoretically or observationally motivated model for the energy composition at the jet launching scale.
Finally, our experiments are limited because we use a Newtonian MHD code, and cannot consider truly relativistic jets like those in black hole scale simulations.

\section{Conclusions}\label{sec:conclusions}

\begin{table*}
\begin{center}
 \caption{Executive summary of our experiments with different jet models in a $10^{14} {\rm M}_\odot$ halo}
 \label{tab:exe}
 \begin{tabular*}{\textwidth}{@{\extracolsep{\fill}}cccccc}
\hline
\hline
\vspace{-0.2cm}\\
 Form   & Variations & $\dot{E}_{q}$ at $2{\rm M}_\odot {\rm yr}^{-1}$& Other Criteria & Higher $\dot{E}/\dot{m}$ &Problem \\
 \hline
\mrrr{Kinetic} & non-precessing, narrow & none               & needs broader cocoon                                    &wider cocoon & inefficient \\
               & non-precessing wide    & $\sim$ 6e43       & $\theta_{\rm op}\gg45^{\circ}$ & not tested              &extended heated core \\
               & \mr{precessing narrow} & \mr{$\sim$ 6e43}  & $\theta_{\rm p}\gtrsim 45^{\circ}$ if $t_p\sim$ 10 Myr & \mr{not tested} &\mr{extremely large precession}\\
               &                        &                   & $\theta_{\rm p}\gtrsim 30^{\circ}$ if $t_p\sim$ 100 Myr & &\\
\hline              
\mr{Thermal}  &constant $\dot{E}$       &$\sim$ 6e43        & $T_{\rm jet\,\, material}\gtrsim10^9$ K                                   & wider cocoon  &narrow $\dot{E}$ range  \\
              & 10\% duty cycle         &$\langle\dot{E}(t)\rangle=$6e44      & $t_{\rm d}\gtrsim100$ Myr                  & not tested &  $dT/dr<0$ when ``on''\\
\hline              
{Cosmic ray}    & {constant $\dot{E}$}       & {$\sim$ 6e42-6e43+}  & {no other criteria}                                    &wider cocoon & none \\
\hline
Magnetic      &constant ${\bf B}_{\rm ini}$                  &none                &none quenches                                    &wider cocoon & inefficient by itself  \\
\hline 
\hline
\end{tabular*}
\end{center}
\begin{flushleft}
This is a  summary of all the parameter space we explore. Each column is as follow: (1) ``Form'': The dominant energy form in the jet at launch. (2) ``Variations'': Qualitative model variations we considered.  (3) $\dot{E}_{\rm q}$ at $2{\rm M}_\odot {\rm yr}^{-1}$: The required energy flux to stably quench the galaxy (for our default IC) when the mass flux is $2{\rm M}_\odot {\rm yr}^{-1}$. (4) ``Other criteria'': Additional requirements for this model group to stably quench. $\theta_{\rm op}$: opening-angle. $\theta_{\rm p}$: precessing angle. (5) Higher $\dot{E}/\dot{m}$: Effect of increased specific energy in the jet. (6) ``Problems'': Physical problems or major inconsistencies with observations common to all runs in a given ``group''.
\end{flushleft}
\end{table*}

In this paper, we have attempted a systematic exploration of different AGN jet models that inject energy into massive halos, quenching galaxies and suppressing cooling flows. We specifically considered models with pure kinetic jets, thermal energy dominated jets, and cosmic ray jets. We also systematically varied the mass loading, jet width, jet magnetic field strength and field geometry, precession angle and period, and jet duty cycle. These were studied in full-halo-scale but non-cosmological simulations including radiative heating and cooling, self-gravity, star formation, and stellar feedback from supernovae, stellar mass-loss, and radiation, enabling a truly ``live'' response of star formation and the multi-phase ISM to cooling flows. We used a hierarchical super-Lagrangian refinement scheme to reach $\sim 10^{4}\,{\rm M}_{\sun}$ mass resolution, much higher than many previous global studies.

We summarize our key results in the following points and in \tref{tab:exe}:
\begin{itemize}

\item \pfh{All our successful models quench via the (initially narrow) jet inflating a quasi-isotropic (large solid-angle at $r\lesssim R_{\rm cool}$) cocoon in which pressure (ram or thermal or CR) is able to balance gravity and ``loft'' and heat gas within most of the solid angle inside $R_{\rm cool}$. Narrow-angle cocoons fail to quench regardless of energetics, as inflow continues near the midplane. We stress that the mode with which the jet delivers energy is important and it is not enough to simply directly dump in thermal energy or induce turbulent dissipation to offset cooling.  
The qualitative behaviors of these cocoons in our kinetic+thermal+magnetic+CR runs are well-described by simple similarity solutions (\S~\ref{s:energy_form} \&\ \S~\ref{s:CR_th}).}

\item \pfh{This implies three necessary criteria for jet quenching, which we find are sufficient to identify all our quenched runs. {\bf (1)} A mean energy input rate sufficient to reverse the cooling flow dynamics (sustain pressure that balances gravity), $\langle \dot{E}_{\rm tot,\,J} \rangle \gtrsim 3\times 10^{43}\,{\rm erg\,s^{-1}}\,(\dot{M}_{\rm cool}/100\,{\rm M_{\odot}\,yr^{-1}})\,(v_{c}[R_{\rm cool}]/500\,{\rm km\,s^{-1}})^{2}$. {\bf (2)} A specific energy of jet material large enough that the direct (for thermal/CR) or post-shock (for kinetic) cocoon cooling time is always much longer than the cocoon expansion time, $\dot{E}_{\rm tot,\,J}/\dot{M}_{\rm J} \gtrsim 10^{17}\,{\rm erg\,g^{-1}}$ (e.g.\ $T>10^{9}\,$K, for thermal jets, or $v\gtrsim 5000\,{\rm km\,s^{-1}}$ for kinetic). {\bf (3)} A means to ensure the cocoon can expand to fill broad solid angles (so effectively suppress inflows) before the jet breaks through $\sim R_{\rm cool}$. This can be accomplished by either \textbf{(a)} the jet having a dominant fraction of its injection energy in non-kinetic (thermal, CR, or magnetic) form (with the relevant solid angle scaling as $(\dot{E}_{\rm tot,\,jet}/\dot{E}_{\rm kin,\,J})^{3/2}$); \textbf{(b)} an extremely ``light'' kinetic jet having a high specific-energy at $\sim 10\,$pc (our coupling radius), with jet velocity at this radius $\gtrsim 10^{4}(\dot{M}_{\rm J}/{\rm M}_\odot {\rm yr}^{-1})(v_{\rm ff}/300\, {\rm km\,s}^{-1} )(\dot{M}_{\rm cool}/ 100\,{\rm M}_\odot{\rm yr}^{-1})^{-1}$;
or \textbf{(c)} a large kinetic jet opening or precession angle. 
}

\item \pfh{For thermal+kinetic+magnetic jets (provided the above conditions are met), the criterion for this quenching to become ``explosive'' is a larger mean $\langle \dot{E}_{\rm tot} \rangle$ by only a factor $\sim 10$ or so. Beyond this energy flux, the jet violently expels the inner halo gas, leaving a remnant which is too hot and has an inverted temperature/entropy gradient compared to observations. Modifying precession or opening-angles or duty cycles can shift the ``preferred'' energies slightly but does not appreciably widen this range. Thus there is a rather narrow range of energetics where such jets quench without violating observations. But it remains possible that jets ``self-regulate'' to this range in models where accretion and jet power scale self-consistently with nuclear gas properties.}

\item \pfh{For CR-dominated jets, the fact that CRs can diffuse or stream through the gas provides a sort of ``pressure valve,'' making the induced outflows less ``overheated'' at high energies. At lower energies, the combination of efficient CR diffusion isotropizing the CR cocoon, efficient CR pressure support of cool gas, and the modified nature of thermal instability in a CR-pressure dominated medium allows CR jets to quench at order-of-magnitude lower energetics. Together this means the allowed dynamic range of energetics for CR-dominated jets is much larger (factor $\sim 100$). Moreover, the lower-energy CR jets are the only successfully quenched models here which do not strongly alter the core density and therefore retain observed cool-core features and lower turbulent velocities $\lesssim 100 \,{\rm km\, s}^{-1}$.}
    
\end{itemize}

In summary, our study supports the idea that quenching -- at least of observed $z\sim0$ massive halos -- can be accomplished within the viable parameter space of AGN jets. But with this study and \citetalias{2020MNRAS.491.1190S}, we show the viable parameter space which produces successful quenching and does not violate observational constraints is rather narrow, and points to specific jet/cocoon processes and quite possibly a role for CRs. Many caveats remain (see \sref{s:caveat}) to explore in future work, alongside more detailed comparisons with observations (\sref{s:observation}). 


\acknowledgments
We thank Eliot Quataert for useful discussion. 
KS acknowledges financial support from the Simons Foundation.
Support for PFH and co-authors was provided by an Alfred P. Sloan Research Fellowship, NSF Collaborative Research Grant \#1715847 and CAREER grant \#1455342, and NASA grants NNX15AT06G, JPL 1589742, 17-ATP17-0214.  
GLB acknowledges financial support  from the NSF (grant AST-1615955, OAC-1835509, AST-2006176) and computing support from NSF XSEDE.   
RSS, CCH and DAA were supported by the Simons Foundation through the Flatiron Institute. 
DAA was supported by NSF grant AST-2009687. 
CAFG was supported by NSF through grants AST-1715216 and CAREER award AST-1652522; by NASA through grant 17-ATP17-0067; and by a Cottrell Scholar Award and a Scialog Award from the Research Corporation for Science Advancement. 
DK was supported by NSF grant AST-1715101 and the Cottrell Scholar Award from the Research Corporation for Science Advancement. 
TKC  is  supported  by  Science  and  Technology  Facilities  Council  (STFC)  astronomy  consolidated  grant ST/T000244/1.
Numerical calculations were run on the Flatiron Institute cluster ``popeye'' and ``rusty'', Caltech  cluster ``Wheeler'', allocations from XSEDE TG-AST120025, TG-AST130039 and PRAC NSF.1713353 supported by the NSF, and NASA HEC SMD-16-7592. 
This work was carried out as part of the FIRE project and in collaboration with the SMAUG collaboration. SMAUG gratefully acknowledges support from the Center for Computational Astrophysics at the Flatiron Institute, which is supported by the Simons Foundation.

\vspace{0.5cm}

\bibliographystyle{mnras}
\bibliography{mybibs}

\appendix
\normalsize
\section{Density and Entropy Profiles for All Runs}
\label{a:d_t}
\begin{figure*}
\centering
 \includegraphics[width=16cm]{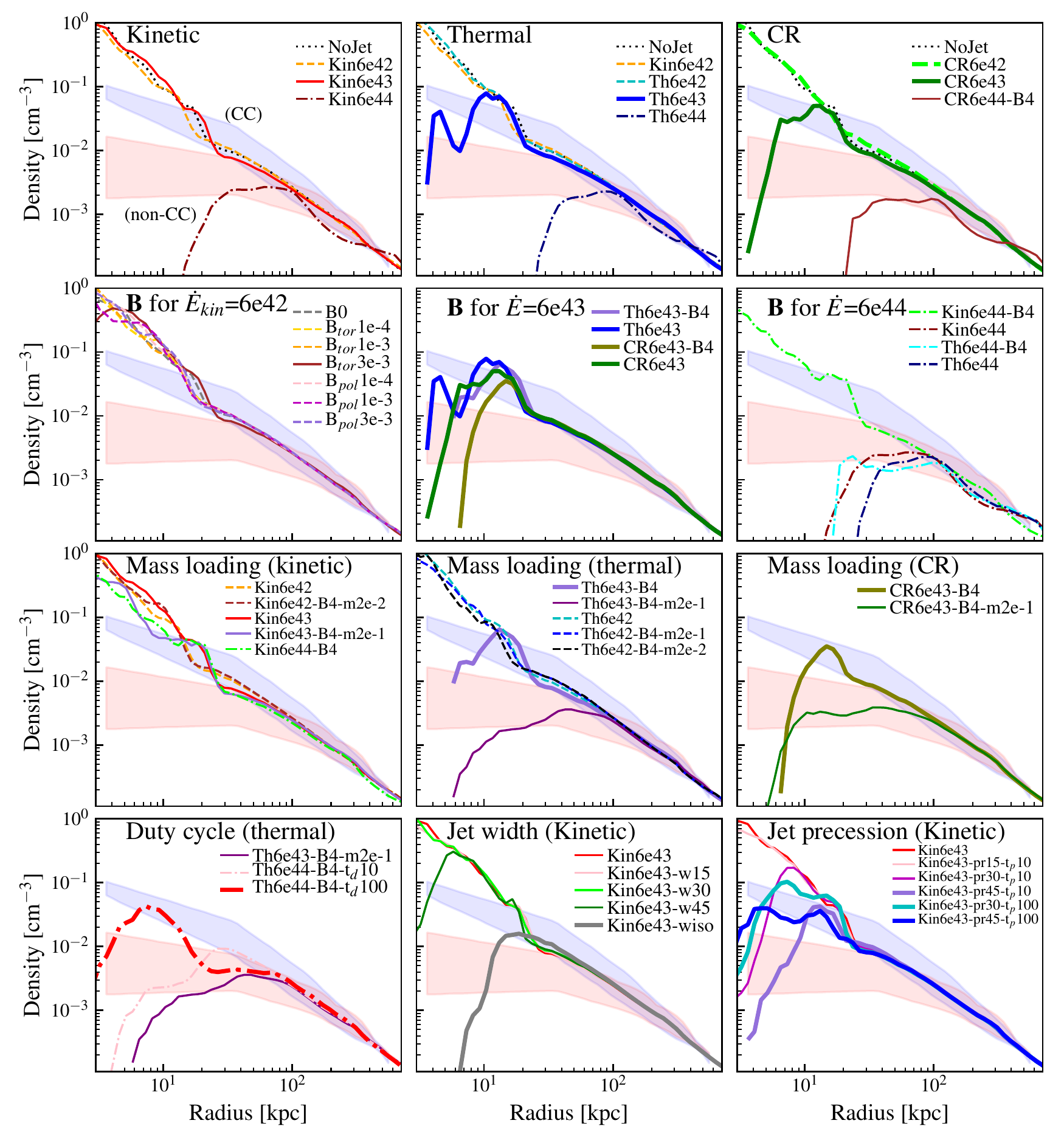}
\caption{Density versus radius averaged over the last $\sim 50\,$Myr in the all runs from \fref{fig:sfr}. 
The shaded regions indicate the observational density profiles (scaled) for cool-core (blue) and non-cool-core (red) clusters \citep{2013ApJ...774...23M}, scaled according to the halo mass differences. Runs labeled `overheated' in \tref{tab:run} generally have very low density core regions. Runs labeled `strong CF' or `slight$\downarrow$' in \tref{tab:run} generally have an over-dense core region. The other runs agree reasonably well with the observations.}
\label{fig:etd_sus1}
\end{figure*}

\begin{figure*}
\centering
 \includegraphics[width=16cm]{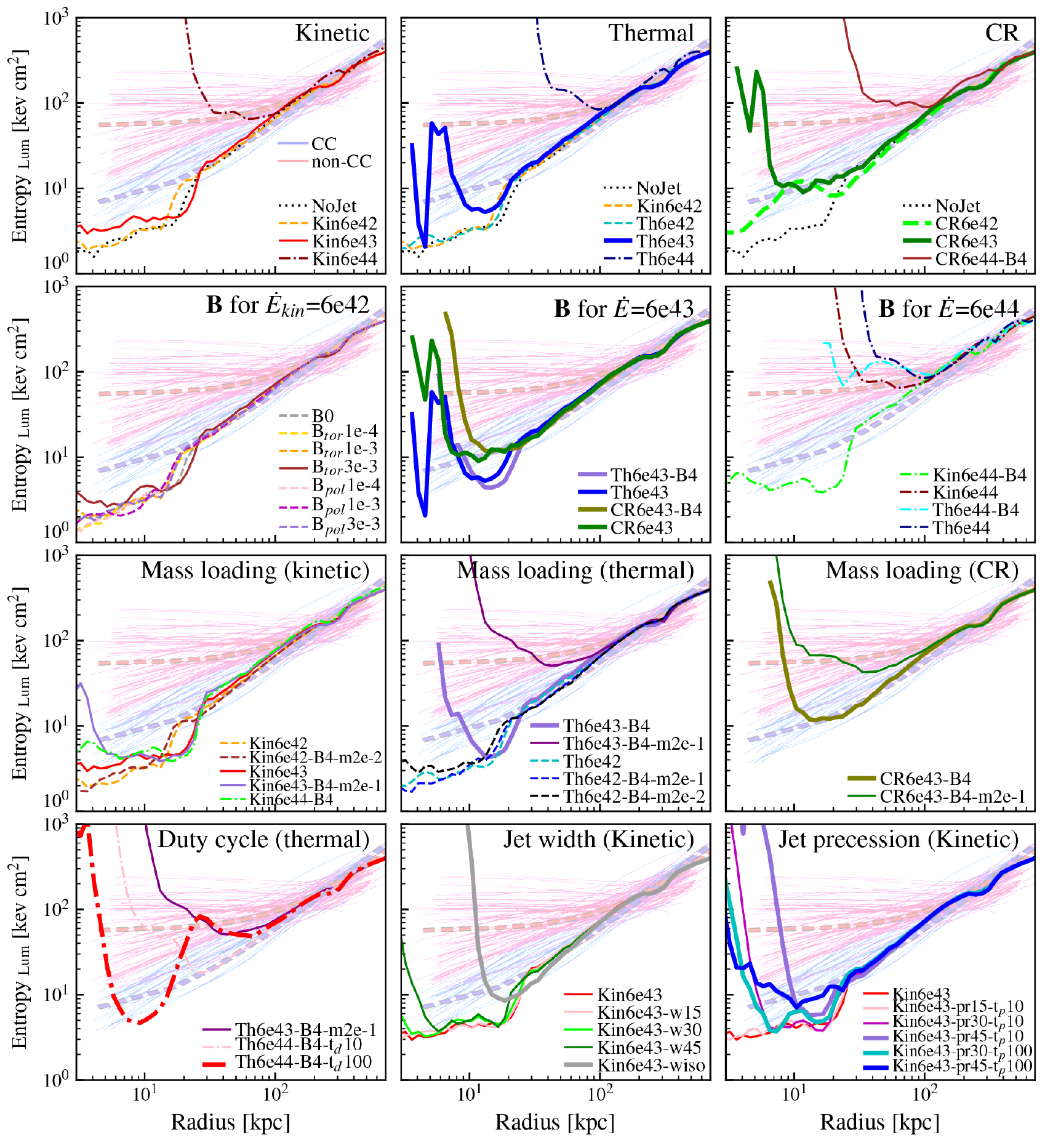}
\caption{Luminosity-weighted entropy versus radius averaged over the last $\sim 50\,$Myr in the non-overheated quiescent runs from \fref{fig:sfr}. 
The light curves in the bottom row indicate the observational entropy profiles (scaled) for cool-core (blue) and non-cool-core (red) clusters \citep{2013ApJ...774...23M} (scaled according to the halo mass differences). 
Runs labeled `overheated' in \tref{tab:run} generally have very high entropy in the core regions. The other runs have entropy profiles resembling the observed cool-core populations.}
\label{fig:etd_sus2}
\end{figure*}

In \fref{fig:etd_sus1} and \fref{fig:etd_sus2} we provide the density and luminosity-weighted entropy profiles of all our runs averaged over the last $\sim 50\,$ Myr. The runs labeled `overheated' in \tref{tab:run} generally have very low density, high entropy core regions within $\sim 500$ Myr. The runs labeled `strong CF' or `slight$\downarrow$' in \tref{tab:run} generally have an over-dense core region. The other runs agree more reasonably with the observations.

\section{Thermal stability for CR jets with different flux}
\label{a:cr}
\begin{figure*}
\centering
 \includegraphics[width=16cm]{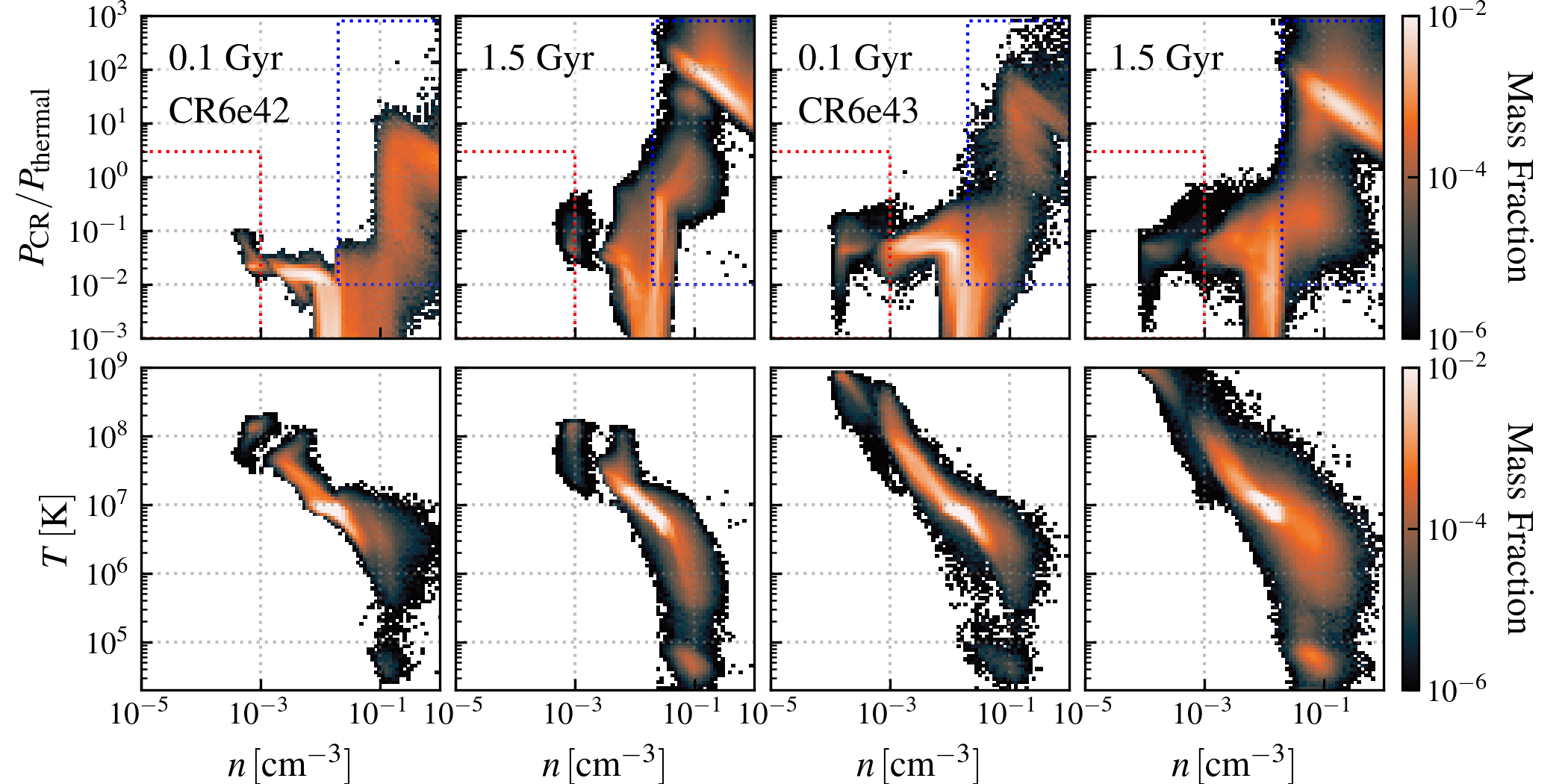}
\caption{The distribution of $P_{\rm CR}/P_{\rm thermal}$ as a function of gas density (top), and the phase distribution in the temperature-density plane (bottom) for the `CR6e42' and `CR6e43' runs at 100 Myr and 1.5 Gyr, for the gas above $10^4$ K and from 10-30 kpc. At 1.5 Gyr, the run with a lower CR flux jet (`CR6e42') reaches a higher $P_{\rm CR}/P_{\rm thermal}$ than the run with higher CR flux (`CR6e43') for the gas that is cooling (the region outlined with a blue dotted line in the plot). The latter run has more cosmic rays distributed to the lower density phase at a larger radius (the red region in the plot). Only when $P_{\rm CR}/P_{\rm thermal}$ builds up to a sufficiently high value does the density distribution tighten.}
\label{fig:phase_CRth_t}
\end{figure*}

CRs stabilize the gas more effectively in the runs with lower CR energy injection. The reason is due to the balance between CR energy and thermal energy, $f_{\rm CR}=P_{\rm CR}/P_{\rm thermal}$. The gas will only follow an isochoric and constant thermal+CR pressure process when CRs are the dominant energy form (high $f_{\rm CR}$).

The first row of \fref{fig:phase_CRth_t} shows such a ratio of the two cosmic ray injection runs at the beginning (100 Myr) and end (1.5Gyr) of the simulations. In `CR6e42', initially, the ratio, $f_{\rm CR}$, for the gas that is cooling (the blue square region in \fref{fig:phase_CRth_t}) is not sufficiently high, so the density in that phase shows a broader distribution resembling that in the `Th6e42' run. After the CR energy builds up as the energy injection continues and $f_{\rm CR}$ increases, the density distribution becomes narrow. On the other hand, in the higher CR flux run, `CR6e43', initially, the ratio $f_{\rm CR}$ is slightly higher than the `CR6e42' run. However, at a later time, the CR energy of the gas with density $n>10^{-2}$ cm$^{-3}$ does not increase much due to gas expansion (because of the suppressed density) and the advection of CR rich gas. Instead, the CR energy goes into the lower density phase at larger radii (the red square region in \fref{fig:phase_CRth_t}).

\label{lastpage}

\end{document}